\documentclass[portrait]{cernsem}
\usepackage{graphicx,epsfig,a4,fancybox,fancyhdr,epic,rotating,color}
\slidesmag{2}
\renewcommand{\Semlabel}{Wien2003}

\begin{document}
\begin{PSlide}{}
%\begin{Slide}{}

\color{blue}

\begin{center}

{ \Large \bf \underline{Super Yang-Mills theories and the structure}}\\
{ \Large \bf \underline{of anomalies and spontaneous parameters} }
\\

\vspace{1.0cm}

{\large \bf

\color{blue}
Peter Minkowski
\\
University of Bern, Switzerland
\\
\vspace{0.5cm}
\vspace{0.5cm}

Seminar at the TU in Vienna\\
November 2003

}
\end{center}

\vspace{0.5cm}
\color{green}
in collaboration with Luzi Bergamin
\vspace{0.5cm}

\color{red}
\begin{center}
{\Large \bf \underline{Topics} }
\end{center}

%\color{blue}

{\large \bf

1 Logical possibilities in super Yang-Mills theories
\vspace*{0.2cm}

2 supercurrent algebra \\
\vspace*{0.2cm}

3 on the road of conserved susy\\
\vspace*{0.2cm}

4 the predominant role of\\
\hspace*{0.4cm} $F \ ( \ x \ ) \ = \ \frac{1}{4} 
\ \left ( \ F_{\ \mu \nu}^{\ A} \ F^{\ \mu \nu \ A} \ \right )
\ ( \ x \ )$\\
\hspace*{0.4cm} and ${\cal{B}}^{\ 2} \ = \ 
\ \left \langle \ \Omega \ \right |
\ F \ ( \ x \ ) \ \left | \ \Omega \ \right \rangle$
\vspace*{0.2cm}

5 susy breaking : through anomalies and\\ 
\hspace*{0.4cm} spontaneously
$\rightarrow$ mass gap

6 conclusions - outlook

}

\end{PSlide}

\begin{PSlide}{}

\color{red}

\color{red}

\begin{center}
{\Large\bf Contents}
\end{center}

{\large\bf

\noindent
\begin{tabular}{llr}
1 & Logical possibilities in &  \\ 
  & super Yang-Mills theories & {\color{blue} 8}
\vspace*{0.2cm} \\
2 & Supercurrent algebra & {\color{blue} 9}
\vspace*{0.2cm} \\ 
3 & On the road of conserved susy - & \\
  & the minimal source extension & {\color{blue} 14}
\vspace*{0.2cm} \\
4 & Grassmann variables and & \\
  & base for susy algebra & {\color{blue} 20}
\vspace*{0.2cm} \\
4a & Insertion on the & \\
   & $SL \ ( 2 , C )$ matrices ${\cal{A}} \ , \ \cdots$ & {\color{blue} 22}
\vspace*{0.2cm} \\
4b & Base susy algebra & {\color{blue} 25}
\vspace*{0.2cm} \\
4c & Base susy & \\
   & differential representation & {\color{blue} 28}
\vspace*{0.2cm} \\
4d & Explicit construction of & \\
   & chiral (scalar) superfield & {\color{blue} 29}
\vspace*{0.2cm} \\
4e & Chiral invariants from chiral & \\
   & (scalar) superfields (superpotentials) & {\color{blue} 33}
\vspace*{0.2cm} \\
4f & The chiral chain & \\
   & $\Phi \ \rightarrow \ \Phi^{\ '}$ or
     $\Phi_{\ n} \ \rightarrow \ \Phi_{\ n + 1}$ & \hspace*{-0.3cm} {\color{blue} 38} \hspace*{0.0cm}
\end{tabular}

}

\newpage

\color{red}

{\large \bf

\noindent
\begin{tabular}{llr}
5 & The N = 1 super-Yang-Mills structure & {\color{blue} 39}
\vspace*{0.2cm} \\
5a & The ${\cal{D}}$ valued gauge connection & \\
   & from a hermitian vectorfield & \\
   & in the Wess-Zumino gauge & {\color{blue} 39}
\vspace*{0.2cm} \\
5b & The ( SL2C- ) spin one matrices & \\
   &  $\left ( \ \sigma^{\ \mu \nu} \ \right )_{\ \left \lbrace \ \gamma \alpha \ \right \rbrace}$ 
     & {\color{blue} 46}
\vspace*{0.2cm} \\
5c & The chiral field strengths & \\
   & multiplets $w_{\ \alpha} \ ,  \ \overline{w}_{\ \dot{\alpha}}$ & \\
   & and the chiral Lagrangean multiplet & \\ 
   & $\begin{array}{lll}
      {\cal{L}} & = & {\cal{N}} \ tr \ w^{\ \alpha} \ w_{\ \alpha} \ , \ {\cal{N}}^{\ -1} \\
                & = & 4 \ C \ ( \ {\cal{D}} \ ) \ g^{\ 2}
     \end{array} $ & {\color{blue} 48}
\vspace*{0.2cm} \\
5d & Some numbers for the & \\
   & simple compact Lie groups & {\color{blue} 51}
\vspace*{0.2cm} \\
5c & continued & \hspace*{-0.3cm} {\color{blue} 57} \hspace*{0.0cm}
\end{tabular}

}

\newpage

\color{red}

{\large \bf

\noindent
\begin{tabular}{llr}
6 & The minimal source extension & \\
  & of N=1 super-Yang-Mills structure & {\color{blue} 60}
\vspace*{0.2cm} \\
6a & The minimal source extension  & \\
   & of the $D^{\ A}$-eliminated & \\
   & Lagrangean multiplet & \\
   & for N=1 super-Yang-Mills & {\color{blue} 67}
\vspace*{0.2cm} \\
6b & The minimal source extension & \\
   &  $D^{\ A}$-eliminated & \\ 
   & $\Phi$ multiplet ( results ) & {\color{blue} 71}
\vspace*{0.2cm} \\
6c & The minimal source extension & \\
   &  $D^{\ A}$-eliminated & \\ 
   & Lagrangean multiplet $\color{blue} {\cal{L}} \ ( \ \color{red} J \ \color{blue} )$ 
       & \hspace*{-0.3cm} {\color{blue} 72} \hspace*{0.0cm}
\end{tabular}

}

\newpage

\color{red}

{\large \bf

\noindent
\begin{tabular}{llr}
7 & The Legendre transform from & \\
  & (minimal) sources to & \\
  & classical fields representing & \\
  & the Lagrangean multiplet & {\color{blue} 78}
\vspace*{0.2cm} \\
7a & Susy transformations of the & \\
   &  Lagrangean multiplet -- $\delta_{\ 2}$ & {\color{blue} 91}
\vspace*{0.2cm} \\
7b & Susy transformations of the & \\
   & Lagrangean multiplet -- & \\ 
   & $\delta_{\ 1} \ = \ \color{magenta} \tau^{\ \delta} \ \color{red} q_{\ \delta}$ 
     & {\color{blue} 92}
\vspace*{0.2cm} \\
7c & Susy transformations of the & \\
   & Lagrangean multiplet -- & \\
 & $\delta \ = \ \delta_{\ 1} \ + \ \delta_{\ 2}
\ = \ \color{magenta} \tau^{\ \gamma} \ \color{red} q_{\ \gamma}
\ + \ \color{magenta} \overline{\tau}^{\ \dot{\delta}} \ \color{red} \overline{q}_{\ \dot{\delta}}$
& \hspace*{-0.3cm} {\color{blue} 97} \hspace*{0.0cm}
\end{tabular}

}

\newpage

\color{red}

{\large\bf

\noindent
\begin{tabular}{llr}
8 & Indirect derivation of the & \\
  & anomaly of the susy current & {\color{blue} 98}
\vspace*{0.2cm} \\
3 & ({\small of topics}) On the road of & \\
  & conserved susy & {\color{blue} 99}
\vspace*{0.2cm} \\
8a & Ad 1 , 2 : susy covariance and & \\
   & the fermionic arguments of $\Phi$ & {\color{blue} 102}
\vspace*{0.2cm} \\
8a 1 & Ad 3 : Trivial (in)dependence of $\Gamma$ on & \\
     & the fermionic arguments & {\color{blue} 106}
\vspace*{0.2cm} \\
8a 2 & Ad 3 : General dependence of $\Gamma$ on & \\
     & the fermionic arguments & {\color{blue} 106}
\vspace*{0.2cm} \\
8b & Special values for the & \\
   & arguments of $\Gamma$ & \hspace*{-0.3cm} {\color{blue} 108} \hspace*{0.0cm}
\end{tabular}

}

\newpage

\color{red}

{\large\bf

\noindent
\begin{tabular}{llr}
9 & Last but not least -- & \\
  & the main problem : & \\
  & inconsistency of & \\
  & only sponteneous susy breaking & {\color{blue} 109}
\vspace*{0.2cm} \\
9a & Connection with trace anomaly & \\
   & (N=1 susy) and the sign of & \\
   & the vacuum energy density & {\color{blue} 110}
\vspace*{0.2cm} \\
10 & Consequence and conjectured structure & \\
   & of the anomalous & \\
   & susy current divergence & {\color{blue} 112}
\vspace*{0.2cm} \\
11 & Outlook (conclusions) & {\color{blue} 115}
\vspace*{0.2cm} \\
   & References (partial) & \hspace*{-0.3cm} {\color{blue} 117} \hspace*{0.0cm}
\end{tabular}

}

\end{PSlide}

\begin{PSlide}{}

\color{red}

\begin{center}

{\Large \bf
1 Logical possibilities in super Yang-Mills theories 
}
\end{center}

\vspace{0.3cm}

{\large \bf

\color{blue}

\begin{displaymath}
\begin{array}{l}
\left \langle \ \Omega \ \right | \ \vartheta_{\ \mu \nu}
\ \left | \ \Omega \ \right \rangle \ = \ \varepsilon \ g_{\ \mu \nu}
\hspace*{0.1cm} ; \hspace*{0.1cm}
g_{\ \mu \nu} \ = \ diag \ ( \ 1 \ , \ -1 \ \otimes \ 3 \ )
\vspace*{0.3cm} \\
\varepsilon \ =
\ \left \lbrace
\ \begin{array}{l}
> 0 \ \leftrightarrow \ +
\vspace*{0.1cm} \\
= 0 \ \leftrightarrow \ 0
\vspace*{0.1cm} \\
< 0 \ \leftrightarrow \ -
\end{array}
\right .
\hspace*{0.3cm}
\mbox{ \begin{tabular}{l} \color{green} original focus was
\vspace*{0.1cm} \\ 
\color{green} on cases $b_{\ 0} \ , \ b_{\ +}$
\end{tabular}}
\end{array}
\end{displaymath}

\color{blue}

\begin{tabular}{|c|cccc|}
\hline 
 & \multicolumn{2}{c}{susy breaking } 
& \begin{tabular}[t]{c} massless \\
   goldstino
   \end{tabular} & $\varepsilon$ \\
cases & \begin{tabular}[t]{c} anomalous \\
        current \\
        divergences \\
        $j_{\ \mu \ \alpha} \ , \ j^{\ *}_{\ \nu \ \dot{\beta}}$
        \end{tabular} & spontaneous & & \\
 & & & & \\
\hline 
\color{magenta}
$a_{\ -}$ & \color{magenta} yes & \color{magenta} yes & \color{magenta} no 
& \color{magenta} - 
 \\
\hline 
  \multicolumn{5}{|c|}{\color{cyan} (to be) excluded} \\
\hline 
$a_{\ 0}$ & yes  & no  & no & 0 \\
$a_{\ +}$ & yes  & yes & no & + \\
$b_{\ +}$ & no   & yes & yes & + \\
$b_{\ -}$ & no   & yes & yes & - \\
$b_{\ 0}$ & no   & no  & no  & 0 
\color{blue} \\
\hline
\end{tabular}
}
%\vspace*{0.5cm}

\color{red}

\end{PSlide}

%\newpage
\begin{PSlide}{}
\color{red}
\begin{center}

{\Large \bf
2 Supercurrent algebra [1]
}
\end{center}

\color{blue}
\vspace{0.3cm}
{\Large \bf
supercharges $Q_{\ \alpha} \ , \ Q^{\ *}_{\ \dot{\beta}}$

supercurrents $j_{\ \mu \ \alpha} \ , \ j^{\ *}_{\ \nu \ \dot{\beta}}$

we follow the path $b_{\ +}$ 
\color{green}
\ $\leftrightarrow$ susy broken {\it only} \\
\hfill spontaneously

\color{blue}
\begin{displaymath}
\begin{array}{l}
Q_{\ \alpha} \ = \ \displaystyle{\int}_{\ t} \ d^{\ 3} x
\ j_{\ 0 \ \alpha} \ ( \ t \ , \ x \ )
\hspace*{0.2cm} ; \hspace*{0.2cm}
Q_{\ \alpha} \ \rightarrow \ Q^{\ *}_{\ \dot{\alpha}}
\vspace*{0.3cm} \\
P_{\ \mu} \ = \ \displaystyle{\int}_{\ t} \ d^{\ 3} x
\ \vartheta_{\ 0 \ \mu} \ ( \ t \ , \ x \ )
\end{array}
\end{displaymath}

the 'once local' form of the susy relation is

\vspace{-0.4cm}
\begin{equation}
\label{eq:1}
\begin{array}{l}
\hspace*{0.4cm}
\displaystyle{\int} \ d^{\ 3} x
\ \left \lbrace 
\ j_{\ 0 \ \alpha} \ ( \ t \ , \ x \ ) \ , 
\ j^{\ *}_{\ 0 \ \dot{\beta}} \ ( \ t \ , \ y \ )
\ \right \rbrace
\vspace*{0.3cm} \\
\ = \ \displaystyle{\int} \ d^{\ 3} x
\ \sigma^{\ \mu}_{\ \alpha \dot{\beta}}
\ \vartheta_{\ 0 \ \mu} \ ( \ t \ , \ y \ ) \ \delta^{\ 3} 
\ ( \ \vec{x} \ - \ \vec{y} \ )
\end{array}
\end{equation}
}

\vspace{-0.4cm}
{\small
\color{magenta}
[1] L. Bergamin and P. Minkowski,
{\it Spontaneous Susy Breaking in N=2 Super-Yang-Mills Theories},
  Contribution to 30th International Conference on High-Energy Physics (ICHEP
  2000), Osaka, Japan, 27 Jul - 2 Aug 2000,
  Published in *Osaka 2000, High energy physics, vol. 2* 1387-1388,
  hep-ph/0011041.
}

\newpage

\color{blue}
{\Large \bf the relation in eq. (\ref{eq:1}) can be written in
covariant form, using general space-time positions 
$x \ = \ ( \ t \ , \vec{x} \ )$ and $x \ \rightarrow \ y$

\begin{equation}
\label{eq:2}
\begin{array}{l}
 \displaystyle{\int} \ d^{\ 4} x
\color{red}
\ e^{\ i \ q \ x}
\color{blue}
\ \partial^{\ \mu}_{\ x}
\ T \ \left ( 
\ j_{\ \mu \ \alpha} \ ( \ x \ ) \ j^{\ *}_{\ \nu \ \dot{\beta}} \ ( \ y \ )
\ \right )
\vspace*{0.3cm} \\
= \ \sigma^{\ \varrho}_{ \alpha \dot{\beta}} 
\ \vartheta_{\ \nu \varrho} \ ( \ y \ )
\color{red}
\hspace*{0.5cm} \mbox{for} \hspace*{0.2cm} \nu \ = \ 0
\end{array}
\end{equation}

\color{blue}
In eq. (\ref{eq:2}) $T$ denotes time ordereing. Irrespective of eventual
Schwinger terms the once integrated relation (\ref{eq:2}) can be
extended to all values of $\nu$. Introducing the
Fourier transform of the time ordered susy current correlation
function eq. (\ref{eq:2}) becomes

\vspace*{-0.5cm}
\begin{equation}
\label{eq:3}
\begin{array}{l}
\tau_{\ \mu \alpha \ ; \ \nu \dot{\beta}} \ ( \ q \ )
\ = \ \displaystyle{\int} \ d^{\ 4} x
\color{red}
\ e^{\ i \ q \ x} \ \times
\vspace*{0.3cm} \\
\color{blue}
\times \ \left \langle \ \Omega \ \right | 
\ T \ \left (
\ j_{\ \mu \ \alpha} \ ( \ x \ ) \ j^{\ *}_{\ \nu \ \dot{\beta}} \ ( \ 0 \ )
\ \right )
\ \left | \ \Omega \ \right \rangle
\vspace*{0.3cm} \\
\frac{1}{i} \ q^{\ \mu}
\ \tau_{\ \mu \alpha \ ; \ \nu \dot{\beta}} \ ( \ q \ )
\ = \ \sigma^{\ \varrho}_{ \alpha \dot{\beta}}  
\ \left \langle \ \Omega \ \right |
\ \vartheta_{\ \nu \varrho} \ ( \ 0 \ )
\ \left | \ \Omega \ \right \rangle
\end{array}
\end{equation}

\newpage

\color{blue}
Eqs. (\ref{eq:2}) , (\ref{eq:3}) are precisely valid if
the susy currents are conserved 
$\partial^{\ \mu} \ j_{\ \mu \ \alpha} \ ( \ x \ ) \ = \ 0$ .

We can evaluate eq. (\ref{eq:3}) for $q \ \rightarrow \ 0$
using the Lorentz covariant decomposition

\begin{equation}
\label{eq:4}
\begin{array}{l}
\tau_{\ \mu \alpha \ ; \ \nu \dot{\beta}} \ ( \ q \ )
\ = \ \begin{array}[t]{l}
\Gamma_{\ \mu \nu \varrho} \ ( \ q \ )
\ \sigma^{\ \varrho}_{ \alpha \dot{\beta}} \ \tau \ ( \ q^{\ 2} \ )
\ +
\vspace*{0.3cm} \\ 
\ + \ \mbox{transverse terms}
\end{array}
\vspace*{0.3cm} \\ 
\Gamma_{\ \mu \nu \varrho} \ ( \ q \ )
\ = \ g_{\ \mu \varrho} \ q_{\ \nu}
\ + \ g_{\ \nu \varrho} \ q_{\ \mu}
\ - \ g_{\ \mu \nu} \ q_{\ \varrho}
\vspace*{0.3cm} \\ 
q^{\ \mu} \ \Gamma_{\ \mu \nu \varrho} \ ( \ q \ ) 
\ = \ q^{\ 2} \ g_{\ \nu \varrho}
\end{array}
\end{equation}

and obtain

\begin{equation}
\color{blue}
\label{eq:5}
\begin{array}{l}
\mbox{for}
\ \left \langle \ \Omega \ \right |
\ \vartheta_{\ \nu \varrho} \ ( \ 0 \ )
\ \left | \ \Omega \ \right \rangle
\ = \ \varepsilon \ g_{\ \nu \varrho}
\vspace*{0.3cm} \\
\lim_{\ q \ \rightarrow \ 0} \ q^{\ 2} \ \tau \ ( \ q^{\ 2} \ )
\ = \ \varepsilon
\end{array}
\end{equation}

From eqs. (\ref{eq:4}) , (\ref{eq:5}) we deduce that for 
$\varepsilon \ \neq \ 0$ , $\tau$ must inherit a goldstino induced 

\newpage

\color{blue}
pole of the form

\begin{equation}
\label{eq:6}
\begin{array}{c}
\Gamma_{\ \mu \nu \varrho} \ ( \ q \ )
\ \sigma^{\ \varrho}_{ \alpha \dot{\beta}} \ Abs \ \tau \ ( \ q^{\ 2} \ )
\ =
\vspace*{0.3cm} \\
 =  \displaystyle{\int} \ d \ \varrho \ ( \ m^{\ 2} \ ) 
 \sum_{\ n}
\left (
\begin{array}{l}
 \left \langle \ \Omega \ \right |
\ j_{\ \mu \alpha} \ ( \ 0 \ ) 
\ \left | \ p \ ; n \ \right \rangle \times
\vspace*{0.3cm} \\
\times  \left \langle \ p \ ; n \ \right |
\ j^{\ *}_{\ \nu \dot{\beta}} \ ( \ 0 \ )
\ \left | \ \Omega \ \right \rangle
\end{array}
\right )
\vspace*{0.3cm} \\
\color{magenta}
\varrho \ ( \ m^{\ 2} \ ) \ > \ 0
\vspace*{0.3cm} \\
\color{magenta}
\downarrow
\vspace*{0.3cm} \\
\color{blue}
q \ \rightarrow \ 0 \ :
\ \tau \ \sim \ \left . | \ f_{\ g} \ |^{\ 2}
\ \right / \ ( \ q^{\ 2} \ + \ i \ \eta \ )
\end{array}
\end{equation}

In eq. (\ref{eq:6}) the positive measure $\varrho \ ( \ m^{\ 2} \ )$
refers to the Kallen-Lehmann representation for the two point
function $\tau_{\ \mu \alpha \ ; \ \nu \dot{\beta}} \ ( \ q \ )$
(\ref{eq:4}) and $f_{\ g}$ denotes the goldstino 'decay constant'
analogous to $f_{\ \pi}$ for pions, but of dimension $mass^{\ 2}$,
which is defined modulo an arbitrary phase.

$\eta \ \downarrow \ + \ 0$ is used for the infinitesimal 

\newpage

\color{blue}
positive
imaginary part of $q^{\ 2}$ to distinguish 
it from the vacuum energy density $\varepsilon$ .

Combining eqs. (\ref{eq:5}) and (\ref{eq:6}) we obtain

\begin{equation}
\label{eq:7}
\begin{array}{l}
| \ f_{\ g} \ |^{\ 2} \ = \ \varepsilon \ \geq \ 0
\vspace*{0.3cm} \\
 \left \langle \ \Omega \ \right |
\ j_{\ \mu \alpha} \ ( \ 0 \ )
\ \left | \ p \ ; s \ goldstino \ \right \rangle \ =
\vspace*{0.3cm} \\
\ = \ f_{\ g} \ \sigma_{\ \mu \ \alpha \dot{\beta}} 
\ u^{\ \dot{\beta}} \ ( \ p \ ; \ s \ )
\end{array}
\end{equation}

\color{magenta}
Eq. (\ref{eq:7}) eliminates the case $b_{\ -}$ in the table of section 1.

}

\color{red}

\vspace*{0.5cm}
\vspace*{0.5cm}
\color{blue}

\end{PSlide}

\begin{PSlide}{}

\begin{center}

\color{red}
{\Large \bf
3 On the road of conserved susy
}
\end{center}

\color{blue}
{\Large \bf
In this section we follow, for N=1 super Yang-Mills theory with 
(simple) gauge group G,
the road $b_{\ 0}$ of the table in section 1,
\color{magenta} i.e. $\varepsilon \ = \ 0$ ,
\color{blue}
as long as possible, ignoring the prejudice originating from
the trace anomaly :

\vspace*{-0.5cm}
\begin{equation}
\label{eq:8}
\begin{array}{l}
\vartheta^{\ \mu}_{\ \mu} \ = 
\ \left ( \ - \ 2 \ \beta \ ( \ g \ ) \ / \ g \ \right ) \ {\cal{L}}
\vspace*{0.3cm} \\
\ \hspace*{0.8cm}  =  \ - \ ( \ 2 \ \beta_{\ 0} \ ) 
\ \left \lbrack \ \frac{1}{4} 
\ F^{\ A}_{\ \mu \nu} \ F^{\ A \ \mu \nu} 
\ \right \rbrack_{\ ren.gr.inv.}
\vspace*{0.3cm} \\
\beta_{\ 0} \ = \ \color{red} - \ b_{\ 0} \ \color{blue} = \ 3 \ C_{\ 2} \ ( \ G \ ) 
\ / \ ( \ 16 \ \pi^{\ 2} \ )
\vspace*{0.3cm} \\
\rightarrow \ \varepsilon \ < \ 0
\end{array}
\end{equation}

In eq. (\ref{eq:8}) the suffix $_{ren.gr.inv.}$ 
-- dropped subsequently --
denotes the field strength bilinear operator renormalized in a 
renormalization group invariant way. This presents no (obvious) problems
for $\beta_{\ 0} \ > \ 0$. 

\newpage

\color{blue}
$C_{2} \ ( \ G \ )$ denotes the second 
Casimir operator of the gauge group, with the normalization
$C_{2} \ ( \ SU_{\ n} \ ) \ = \ n$ . 

The question we ask is : how does
\color{red}
the 
\color{blue}
(N=1) susy covariant effective
action for the composite operators of the Lagrangean multiplet
determine the vacuum - or spontaneous parameters

\vspace*{-0.5cm}
\begin{equation}
\label{eq:9}
\begin{array}{l}
\left \langle \ \Omega \ \right |
\ \vartheta_{\ \mu \nu}
\ \left | \ \Omega \ \right \rangle
\ = \ \varepsilon \ g_{\ \mu \nu}
\vspace*{0.3cm} \\
\left \langle \ \Omega \ \right |
\ \Lambda^{ \ A \ \alpha} \ \Lambda^{\ A}_{\ \alpha}
\ \left | \ \Omega \ \right \rangle
\end{array}
\end{equation}

In eq. (\ref{eq:9}) $\Lambda^{ \ A \ \alpha} \ ( \ x \ )$
denote the gaugino fields normalized in accord with the field strengths
$F^{\ A}_{\ \mu \nu} \ ( \ x \ )$ [2] .

}
%\vspace{-0.1cm}
{\small
\color{magenta}
[2] some selected references :

G. Veneziano and S. Yankielowicz,
Phys. Lett. B113 (1982) 231,

G. M. Shore, Nucl. Phys. B222 (1983) 446,

R. Dijkgraaf and C. Vafa, hep-th/0208048,

R. Dijkgraaf, M.T. Grisaru, C.S. Lam, C. Vafa and D. Zanon, \\
Phys.Lett. B573 (2003) 138, hep-th/0211017, 

L. Bergamin and P. Minkowski, hep-th/0301155. 
}

\newpage

\color{blue}
{\Large \bf
We called this effective action \\

\color{red}
\begin{center}
'the minimal source extension'

$\downarrow$
\end{center}

\color{blue}
a) chiral (base) superfield

First let me discuss a primary (base) chiral superfield
$\Phi$ subjected to the constraint 
$\overline{D}_{\ \dot{\beta}} \ \Phi \ = \ 0$

\vspace*{-0.5cm}
\begin{equation}
\label{eq:10}
\begin{array}{l}
\Phi \ = \ \left \lbrace
\begin{array}{ll}
\hspace*{0.5cm} \vartheta^{\ 2} \ H \ ( \ x^{\ -} \ ) & \color{red} +2
\vspace*{0.3cm} \\
\color{blue}
+ \ \vartheta^{\ \alpha} \ \eta_{\ \alpha} \ ( \ x^{\ -} \ ) & \color{red} -2 
\vspace*{0.3cm} \\
\color{blue}
+ \ \varphi \ ( \ x^{\ -} \ ) & \color{red} +2
\end{array}
\right .
\vspace*{0.3cm} \\
x^{\ - \ \mu} \ = \ x^{\ \mu} \ - \ \frac{i}{2}
\ \vartheta^{\ \alpha} \ \sigma^{\ \mu}_{\ \alpha \dot{\beta}}
\ \overline{\vartheta}^{\ \dot{\beta}}
\end{array}
\end{equation}
 
The signed number in the last column of the bracket in eq. (\ref{eq:10})
denotes the bosonic (+) and fermionic (-) number of degrees of freedom
per unit phase space pertaining to the given local fields
$H \ , \ \eta_{\ \alpha} \ , \ \varphi$ .

\newpage

\color{blue}
The complex scalar field $H$ (for 'Hilfsfeld') is thought to be an auxiliary
field, fully determined from the dynamical variables 

\vspace*{-0.5cm}
\begin{equation}
\label{eq:11}
\begin{array}{l}
H \ = \ H \ ( \ \eta_{\ \alpha} \ , \ \varphi \ )
\end{array}
\end{equation}

so that susy is to be achieved ignoring the highest 
$\vartheta^{\ 2}$ component in eq. (\ref{eq:10}) .

Expanding $x^{\ -}$ we obtain the full chiral scalar field

\vspace*{-0.5cm}
\begin{equation}
\label{eq:12}
\begin{array}{l}
\Phi \ = 
\vspace*{0.3cm} \\
\left \lbrace
\begin{array}{c}
\vartheta^{\ 2} \ \overline{\vartheta}^{\ 2} 
\ \left ( \ - \ \frac{1}{4} \ \mbox{\fbox{\rule{0mm}{1.0mm}}} \ \varphi
\ \right )
\vspace*{0.3cm} \\
\vartheta^{\ 2} \ \overline{\vartheta}_{\ \dot{\delta}}
\ \left ( \ \frac{-i}{2} \ \partial^{\ \dot{\delta} \alpha} \ \eta_{\ \alpha}
\ \right )
\ + \color{red} \ \overline{\vartheta}^{\ 2} \ \vartheta^{\ \alpha} \ 0
\vspace*{0.3cm} \\
\color{blue}
\vartheta^{\ 2} \ H \ + \  \vartheta^{\ \alpha} 
\ \sigma^{\ \mu}_{ \alpha \dot{\beta}} \ \overline{\vartheta}^{\ \dot{\beta}}
\ \left ( \ \frac{-i}{2} \ \partial_{\ \mu} \ \varphi \ \right )
\ + \ \color{red} \overline{\vartheta}^{\ 2} \ 0
\vspace*{0.3cm} \\
+ \ \color{red} \overline{\vartheta}^{\ \dot{\beta}} \ 0
\color{blue} \ + \ \vartheta^{\ \alpha} \ \eta_{\ \alpha}
\vspace*{0.3cm} \\
+ \ \varphi 
\end{array}
\right \rbrace
\end{array}
\end{equation}

\newpage

\color{blue}
For full transparency we exhibit the antichiral field $\overline{\Phi}$

\color{red}
\vspace*{-0.5cm}
\begin{equation}
\label{eq:13}
\begin{array}{l}
\color{blue} \overline{\Phi} \ = 
\vspace*{0.3cm} \\
\color{blue}
\left \lbrace
\begin{array}{c}
\vartheta^{\ 2} \ \overline{\vartheta}^{\ 2}
\ \left ( \ - \ \frac{1}{4} \ \mbox{\fbox{\rule{0mm}{1.0mm}}} \ \varphi^{\ *}
\ \right )
\vspace*{0.3cm} \\
\color{red}
\vartheta^{\ 2} \ \overline{\vartheta}_{\ \dot{\delta}} \ 0
\color{blue} \ + \ \overline{\vartheta}^{\ 2} \ \vartheta^{\ \alpha}
\ \left ( \ \frac{i}{2} \ \partial^{\ \dot{\gamma}}_{\ \alpha} 
\ \eta^{\ *}_{\ \dot{\gamma}}
\ \right )
\vspace*{0.3cm} \\
\color{red} \vartheta^{\ 2} \ 0
\color{blue} \ + \ \vartheta^{\ \alpha} 
\ \sigma^{\ \mu}_{ \alpha \dot{\beta}} \ \overline{\vartheta}^{\ \dot{\beta}} 
 \left ( \ \frac{i}{2} \ \partial_{\ \mu} \ \varphi^{\ *}  \right )
\ + \ \overline{\vartheta}^{\ 2}  H^{\ *}
\vspace*{0.3cm} \\
\color{red}
+ \ \vartheta^{\ \alpha} \ 0
\color{blue}
\ + \ \overline{\vartheta}_{\ \dot{\gamma}} 
\ \eta^{\ * \ \dot{\gamma}}
\vspace*{0.3cm} \\
 + \ \varphi^{\ *}
\end{array}
\right \rbrace
\end{array}
\end{equation}

\color{magenta}
{\small{Supersymmetry in the sense of manifestly equal number of 
bosonic and fermionic degrees of freedom is clearly not obvious 
through the chiral constraint (eqs. \ref{eq:12} and \ref{eq:13}) .}
} 

\newpage

\color{blue}
We note the structure of the highest component of $\overline{\Phi} \ \Phi$

\vspace*{-0.5cm}
\begin{equation}
\label{eq:14}
\begin{array}{l}
\left .
\overline{\Phi} \ \Phi
\ \right |_{\ \vartheta^{\ 2} \ \overline{\vartheta}^{\ 2}}
\ =
\ \left \lbrace
\begin{array}{c}
H^{\ *} \ H \ - \ \frac{1}{4} \ \mbox{\fbox{\rule{0mm}{1.0mm}}} 
\ \left ( \ \varphi^{\ *} \ \varphi \ \right )
\vspace*{0.3cm} \\
\ +
\ \left ( 
\ \partial^{\ \mu} \ \varphi^{\ *}
\ \right )
\ \left ( 
\ \partial_{\ \mu} \ \varphi
\ \right )
\vspace*{0.3cm} \\
\ + \ \eta^{\ *}_{\ \dot{\gamma}} \ \frac{i}{2}
\ \raisebox{0.5cm}{$\leftharpoondown \hspace*{-0.4cm} \rightharpoonup$}
\hspace*{-0.5cm} \partial_{\ \mu} \ \sigma^{\ \mu \ \dot{\gamma} \alpha} \ \eta_{\ \alpha}
\end{array}
\right \rbrace
\end{array}
\end{equation}

In eq. (\ref{eq:14}) we recognize the kinetic Lagrangean density
for the complex scalar field $\varphi$ and the irreducible (Majorana-)
spinor $\eta_{\ \alpha}$ .

$\ \mbox{\fbox{\rule{0mm}{1.0mm}}}
\ \left ( \ \varphi^{\ *} \ \varphi \ \right )$
is relevant for the construction of the energy momentum
tensor for the scalar field $\varphi$, whereas the term $H^{\ *} \ H$ 
interpreted as a negative potential, is not bounded from below,
signalling the special role of the 'auxiliary' field H.

}

\end{PSlide}

\begin{PSlide}{}

\color{red}

\begin{center}
{\large \bf
4 Grassmann variables and base for susy algebra
}

\end{center}

\color{blue}

{\large \bf
Grassmann variables $\vartheta \ , \ \eta \ , \ \cdots$ are chosen in connection 
with $SL \ (2,C)$ representations for spin
\\

\vspace*{-0.5cm}
\begin{equation}
\label{eq:15}
\begin{array}{l}
\mbox{\color{red} right chiral} 
\color{blue} \ :
\ \vartheta_{\ \alpha} \ , \ \eta_{\ \alpha} \ , \ \cdots
\hspace*{0.2cm} ; \hspace*{0.2cm}
_{\alpha} \ = \ 1,2
\vspace*{0.3cm} \\
\vartheta_{\ \alpha} \ \vartheta_{\ \beta} \ + 
\ \vartheta_{\ \beta} \ \vartheta_{\ \alpha} \ \equiv \ 0
\vspace*{0.3cm} \\
{\color{red} SL \ (2,C)} \color{blue} \ :
\ \vartheta_{\ \alpha} \ \rightarrow \ {\cal{A}}_{\ \alpha \beta} \ \vartheta_{\ \beta}
\vspace*{0.3cm} \\
\mbox{\color{green} left chiral} \color{blue} \ :
\ \overline{\vartheta}^{\ \dot{\gamma}} \ , \ \overline{\eta}^{\ \dot{\gamma}}
\ , \ \cdots \hspace*{0.2cm} ; \hspace*{0.2cm} ^{\dot{\gamma}} \ = \ 1,2 
\vspace*{0.3cm} \\
\overline{\vartheta}^{\ \dot{\gamma}} \ \overline{\vartheta}^{\ \dot{\delta}} \ +
\ \overline{\vartheta}^{\ \dot{\delta}} \ \overline{\vartheta}^{\ \dot{\gamma}}
\ \equiv \ 0
\vspace*{0.3cm} \\
{\color{red} SL \ (2,C)} \color{blue} \ :
\ \overline{\vartheta}^{\ \dot{\gamma}} \ \rightarrow 
\ \widetilde{{\cal{A}}}^{\ \dot{\gamma} \dot{\delta}} \ \overline{\vartheta}^{\ \dot{\delta}} 
\vspace*{0.3cm} \\
\widetilde{{\cal{A}}} \ = \  \left ( \ {\cal{A}}^{\ \dagger} \ \right )^{\ -1}
\ = \ \varepsilon \ \overline{{\cal{A}}} \ \varepsilon^{\ -1}
\end{array}
\end{equation}

\noindent
In eq. (\ref{eq:15}) $\varepsilon$ denotes the symplectic invariant
(\color{red} $SL \ (2,C) \ \equiv \ SP \ (1,C)$\color{blue} )

\vspace*{-0.5cm}
\begin{equation}
\label{eq:16}
\begin{array}{l}
\left ( \ \varepsilon \ \right )_{\ \alpha \beta} \ =
\ \left (
\ \begin{array}{rl}
0 & 1
\vspace*{0.1cm} \\
- 1 & 0
\end{array}
\ \right )_{\ \alpha \beta}
\ =
\ \left ( \ \overline{\varepsilon} \ \right )_{\ \dot{\alpha} \dot{\beta}}
\end{array}
\end{equation}

\noindent
with the inverse 

\vspace*{-0.5cm}
\begin{equation}
\label{eq:17}
\begin{array}{l}
\left ( \ \varepsilon^{\ '} \ \right )^{\ \alpha \beta} \ =
\ - \ \left (
\ \begin{array}{rl}
0 & 1
\vspace*{0.1cm} \\
- 1 & 0
\end{array}
\ \right )_{\ \alpha \beta}
\ =
\ \left ( \ \overline{\varepsilon}^{\ '} \ \right )^{\ \dot{\alpha} \dot{\beta}}
\end{array}
\end{equation}

}

\newpage

\color{blue}

{\large \bf
\noindent
The pair $\varepsilon \ , \ \varepsilon^{\ '}$ can be used to lower and raise
spinor indices respectively. Note however, that this process does not yield
$\varepsilon \ \leftrightarrow \ \varepsilon^{\ '}$ when applied to these invariants themselves,
rather leaves them {\it invariant} .

\color{magenta} 
\noindent
The $SL \ (2,C)$ representations ${\cal{A}} \ , \ \widetilde{{\cal{A}}}$
and equivalents 

\color{blue}
\noindent
These form the 'quadrangle' represented by 

\vspace*{-0.5cm}
\begin{equation}
\label{eq:18}
\begin{array}{l}
\begin{array}{ccc}
\theta_{\ \alpha} \ ; \ {\cal{A}} & \longleftrightarrow & \theta^{\ \alpha} \ =
\ \left ( \ \varepsilon^{\ '} \ \right )^{\ \alpha \beta} \ \vartheta_{\ \beta}
\ ; \ {\cal{A}}^{\ '}
\vspace*{0.2cm} \\
\updownarrow & & \updownarrow
\vspace*{0.2cm} \\
\overline{\theta}_{\ \dot{\alpha}} \ ; \ \overline{{\cal{A}}} & \longleftrightarrow & 
\overline{\theta}^{\ \dot{\alpha}} \ =  
\ \left ( \ \varepsilon^{\ '} \ \right )^{\ \dot{\alpha} \dot{\beta}} 
\ \overline{\vartheta}_{\ \dot{\beta}} \ ; \ \overline{{\cal{A}}}^{\ '}
\end{array}
\vspace*{0.4cm} \\
{\cal{A}}^{\ '} \ = \ \varepsilon \ {\cal{A}} \ \varepsilon^{\ -1} 
\hspace*{0.2cm} ; \hspace*{0.2cm}
\overline{{\cal{A}}}^{\ '} \ = \ \varepsilon \ \overline{{\cal{A}}} \ \varepsilon^{\ -1}
\color{red} \ = \ \widetilde{{\cal{A}}}
\end{array}
\end{equation}

\color{blue}
\noindent
The right chiral spinors $\vartheta_{\ \alpha} \ , \ \vartheta^{\ \alpha}$
thus transform under the \color{magenta} equivalent pair 
\color{blue} ${\cal{A}} \ , \ {\cal{A}}^{\ '}$ , whereas left chiral ones under
the associated equivalent pair $\overline{{\cal{A}}} \ , \ \overline{{\cal{A}}}^{\ '}$ .
But right and left chiral spinors transform inequivalently.   

\noindent
In the above 'quadrangular' characterization the Grassmann variables just represent
any spinors, i.e. only the representations under \color{magenta} $SL \ (2,C)$
\color{blue} are relevant.

}

\newpage

\color{blue}

{\large \bf
\begin{center}

\color{magenta} 4a Insertion on the $SL \ (2,C)$ matrices ${\cal{A}} \ , \cdots$

\end{center}
\hspace*{0.1cm}

\noindent
Connection to Lorentz transformations :

\noindent
four vectors $v^{\ \mu} \ , \cdots$ are represented in chiral spinor coordinates as

\vspace*{-0.5cm}
\begin{equation}
\label{eq:19}
\begin{array}{l}
v \ \rightarrow \ v_{\ \alpha \dot{\beta}} \ = \ v^{\ \mu} \ \sigma_{\ \mu \ \alpha \dot{\beta}} 
\ = \ \widetilde{v}
\vspace*{0.2cm} \\ 
\widetilde{v} \ =
\ \left (
\ \begin{array}{ll}
v^{\ 0} \ + \ v^{\ 3} & v^{\ 1} \ - \ i \ v^{\ 2}
\vspace*{0.2cm} \\
v^{\ 1} \ + \ i \ v^{\ 2} & v^{\ 0} \ - \ v^{\ 3} 
\end{array}
\ \right ) 
\vspace*{0.2cm} \\
\widetilde{v} \ = \ v^{\ 0} \ \sigma_{\ 0} \ + \ \vec{v} \ \vec{\sigma}
\hspace*{0.3cm} ; \hspace*{0.3cm} 
v^{\ 2} \ = \ Det \ \widetilde{v}
\end{array}
\end{equation}

\noindent
In eq. (\ref{eq:19}) $\sigma_{\ \mu} \ = \ \left ( \ \sigma_{\ 0} \ , \ \vec{\sigma} \ \right )$
denote the 2 by 2 unit matrix ($\sigma_{\ 0}$) and the three Pauli matrices 
($\sigma_{\ k}$) respectively. 

\noindent
Thus the transformation

\vspace*{-0.5cm}
\begin{equation}
\label{eq:20}
\begin{array}{l}
\widetilde{v} \ \rightarrow \ \widetilde{w} \ = 
\ {\cal{A}} \ \widetilde{v} \ {\cal{A}}^{\ \dagger}
\vspace*{0.2cm} \\
\rightarrow \ w \ = \ \Lambda \ v \ \mbox{or} \ w^{\ \mu} \ = \ \Lambda^{\ \mu}_{\ \ \nu} \ v^{\ \nu}
\vspace*{0.2cm} \\
Det \ \Lambda \ = \ 1 
\hspace*{0.3cm} ; \hspace*{0.3cm}
\Lambda^{\ 0}_{\ \ 0} \ \geq \ 1
\end{array}
\end{equation}

\noindent
$\Lambda$ induces a (real) Lorentz transformation. The latter preserves the time ordering for
causal v \\
( $v^{\ 2} \ \geq \ 0$ ) and has 4 by 4 determinant 1. 

}

\newpage

\color{blue}

{\large \bf

\noindent
In fact the association ${\cal{A}} \ \rightarrow \ \Lambda \ ( \ {\cal{A}} \ )$
is 2 to 1 

\vspace*{-0.5cm}
\begin{equation}
\label{eq:21}
\begin{array}{l}
\Lambda \ ( \ {\cal{A}} \ ) \ = \ \Lambda \ ( \ - {\cal{A}} \ )
\ \rightarrow \ SO \ (1,3) \ \simeq \ SL \ (2,C) \ / \ Z_{\ 2}
\end{array}
\end{equation}

\begin{center}

\color{red}
maximal compact subgroups of $SO \ (1,3)$ and $SL \ (2,C)$ :

\end{center}

\color{blue}

\noindent
The maximal compact subgroups are $SO3$ and $SU2$ respectively
( $SO3 \ \simeq \ SU2 \ / \ Z_{\ 2}$ ) .

\noindent
Assuming the first ($SO3$) known, we look at rotation matrices, i.e.
Lorentz transformations, which do not change the time-component and the association in 
eq. (\ref{eq:20})

\vspace*{-0.5cm}
\begin{equation}
\label{eq:22}
\begin{array}{l}
\Lambda \ = \ R \ \rightarrow \ w^{\ 0} \ = \ R^{\ 0}_{\ \ \mu} \ v^{\ \mu} \ = \ v^{\ 0}
\hspace*{0.2cm} \forall \ v \hspace*{0.2cm} \rightarrow
\vspace*{0.2cm} \\
2 \ w^{\ 0} \ = \ \ tr \ {\cal{A}} \ \widetilde{v} \ {\cal{A}}^{\ \dagger} \ =
\ tr \ {\cal{A}}^{\ \dagger} \ {\cal{A}} \ \widetilde{v} \ = \ tr \ \widetilde{v}
\hspace*{0.2cm} \forall \ v
\vspace*{0.2cm} \\
\color{red}
\rightarrow \ {\cal{A}}^{\ \dagger} \ {\cal{A}} \ = \ \P
\hspace*{0.2cm} \mbox{i.e.} \ {\cal{A}} \ = \ U \ ;
\hspace*{0.2cm} ; \ \mbox{qed} 
\end{array}
\end{equation} \footnote{\hspace*{0.1cm} Show the content of eq. (\ref{eq:22}) .}

\color{blue}
\noindent
Let us construct ${\cal{A}}_{\ 3}$ associated with special Lorentz transformations along the 3-axis 
and $U_{\ 3}$ associated with rotations around the 3-axis to conclude this section. 

}

\newpage

\color{blue}

{\large \bf

\begin{center}
\color{red}

boost in the 3 direction 

\end{center}

\color{blue}

\noindent
The $SO \ (1,3)$ transformation is, using lightcone coordinates

\vspace*{-0.5cm}
\begin{equation}
\label{eq:23}
\begin{array}{l}
v^{\ \pm} \ = \ v^{\ 0} \ \pm \ v^{\ 3} \ :
\ w^{\ \pm} \ = \ e^{\ \pm \ \chi} \ v^{\ \pm} 
\vspace*{0.2cm} \\
v_{\ \perp} \ = \ \left ( \ v^{\ 1} \ , \ v^{\ 2} \ \right ) \ :
\ w_{\ \perp} \ = \ v_{\ \perp} \ \rightarrow
\vspace*{0.2cm} \\
\widetilde{w} \ =
\ \left (
\ \begin{array}{ll}
e^{\ \chi} \ v^{\ +} & v^{\ 1} \ - i \ v^{\ 2}
\vspace*{0.2cm} \\
v^{\ 1} \ + i \ v^{\ 2} & e^{\ - \chi} \ v^{\ -}
\end{array}
\ \right )
\vspace*{0.2cm} \\
\color{red} \rightarrow
\ {\cal{A}}_{\ 3} \ ( \ \chi \ ) \ = \ \pm
\ \left (
\ \begin{array}{ll}
e^{\ \chi \ / \ 2} & 0
\vspace*{0.2cm} \\
0 & e^{\ - \chi \ / \ 2}
\end{array}
\ \right )
\end{array}
\end{equation}

\begin{center}
\color{red}

rotation around the 3 direction (right-screw convention)

\end{center}

\color{blue}

\noindent
The rotation around the 3 axis yields

\vspace*{-0.5cm}
\begin{equation}
\label{eq:24}
\begin{array}{l}
w^{\ 1} \ \pm \ i \ w^{\ 2} \ = \ e^{\ \pm \ i \ \varphi} 
\ \left ( \ v^{\ 1} \ \pm \ i \ v^{\ 2} \ \right )
\hspace*{0.2cm} ; \hspace*{0.2cm}
w^{ \pm} \ = \ v^{\ \pm}
\vspace*{0.2cm} \\
 \widetilde{w} \ =
\ \left (
 \begin{array}{cc}
v^{\ +} & e^{\ - \ i \ \varphi} \ ( \ v^{\ 1} \ - i \ v^{\ 2} \ )
\vspace*{0.2cm} \\
e^{\ + \ i \ \varphi} \ ( \ v^{\ 1} \ + i \ v^{\ 2} \ ) & v^{\ -}
\end{array}
 \right )
\vspace*{0.2cm} \\
\color{red} \rightarrow
\ {\cal{U}}_{\ 3} \ ( \ \varphi \ ) \ = \ \pm
\ \left (
\ \begin{array}{ll}
e^{\ - \ i \ \varphi \ / \ 2} & 0
\vspace*{0.2cm} \\
0 & e^{\ + \ i \ \varphi \ / \ 2}
\end{array}
\ \right )
\end{array}
\end{equation}

}

\newpage

\color{blue}

{\large \bf

\begin{center}
\color{red}

4b) Base susy algebra

\end{center}

\color{blue}

\noindent
We consider the fermionic operators $Q_{\ \alpha} \ , \ Q^{\ *}_{\ \dot{\beta}}$ \\
in conjunction with Grassmann variables 
$\eta_{\ \alpha} \ , \ \overline{\eta}_{\ \dot{\beta}} \ ; \ \vartheta_{\ \alpha} 
\ , \ \overline{\vartheta}_{\ \dot{\beta}} \ ; \ \cdots$ .
The operators $Q_{\ \alpha} \ , \ Q^{\ *}_{\ \dot{\beta}}$ shall obey the
anticommutation algebra

\vspace*{-0.5cm}
\begin{equation}
\label{eq:25}
\begin{array}{l}
\left \lbrace \ Q_{\ \alpha} \ , \ Q^{\ *}_{\ \dot{\beta}} \ \right \rbrace
\ = \ P^{\ \mu} \ \sigma_{\ \mu \ \alpha \dot{\beta}}
\vspace*{0.2cm} \\
\left \lbrace \ Q_{\ \alpha} \ , \ Q_{\ \beta} \ \right \rbrace \ = \ 0
\hspace*{0.2cm} ; \hspace*{0.2cm}
\left \lbrace \ Q^{\ *}_{\ \dot{\alpha}} \ , \ Q^{\ *}_{\ \dot{\beta}} \ \right \rbrace
\ = \ 0
\end{array}
\end{equation}

\noindent
In eq. (\ref{eq:25}) $P^{\ \mu}$ denote the components of the (self-adjoint) energy momentum
four vector.

\noindent
The nontrivial anticommutation relation in eq. (\ref{eq:25}) can be 'bosonized'
by means of Grassmann variables $\eta \ , \ \varepsilon$ to become a commutation
relation

\vspace*{-0.5cm}
\begin{equation}
\label{eq:26}
\begin{array}{l}
\left \lbrack \ \eta \ Q \ , \ \overline{Q} \ \overline{\vartheta} \ \right \rbrack
\ = \ P^{\ \mu} \ v_{\ \mu} \ ( \ \eta \ , \ \overline{\vartheta} \ )
\vspace*{0.2cm} \\
\eta \ Q \ = \ \eta^{\ \alpha} \ Q_{\ \alpha}
\hspace*{0.2cm} ; \hspace*{0.2cm}
\overline{Q} \ \overline{\vartheta} \ = 
\ Q^{\ *}_{\ \dot{\beta}} \ \overline{\varepsilon}^{\ \dot{\beta}}
\vspace*{0.2cm} \\
v_{\ \mu} \ ( \ \eta \ , \ \overline{\vartheta} \ ) \  = 
\ \eta \ \sigma_{\ \mu} \ \overline{\vartheta} \ =
\ \eta^{\ \alpha} \ \sigma_{\ \mu \ \alpha \dot{\beta}} \ \overline{\vartheta}^{\ \dot{\beta}}
\end{array}
\end{equation}

\noindent
So we consider the action of the 'unitary' aperators

\vspace*{-0.5cm}
\begin{equation}
\label{eq:27}
\begin{array}{l}
U \ ( \ \eta \ ) \ =
\ \exp \ \left \lbrack \ \frac{1}{i} \ \left ( \ \eta \ Q \ + 
\ \overline{Q} \ \overline{\eta} \ \right ) \ \right \rbrack
\end{array}
\end{equation}

}

\newpage

\color{blue}

{\large \bf

\noindent
on the substrate formed by

\vspace*{-0.5cm}
\begin{equation}
\label{eq:28}
\begin{array}{l}
F \ ( \ \vartheta \ , \ \overline{\vartheta} \ , \ x \ ) \ =
\ \exp \ \left \lbrack \ \frac{1}{i} \ \left ( \ \vartheta \ Q \ + 
\ \overline{Q} \ \overline{\vartheta} \ \right ) \ \right \rbrack
\ e^{\ i \ x^{\ \mu} \ P_{\ \mu}}
\end{array}
\end{equation}

\noindent
by left-multiplication

\vspace*{-0.5cm}
\begin{equation}
\label{eq:29}
\begin{array}{l}
F_{\ U} \ ( \ \vartheta \ , \ \overline{\vartheta} \ , \ x \ ) \ = 
\ U \ ( \ \eta \ ) \ F \ ( \ \vartheta \ , \ \overline{\vartheta} \ , \ x \ )
\vspace*{0.2cm} \\
\hspace*{0.4cm} = \ F \ ( \ \vartheta \ + \ \eta \ , \ \overline{\vartheta} \ + 
\ \overline{\eta} \ , \ x \ + \ y \ )
\vspace*{0.2cm} \\
y^{\ \mu} \ = \ \frac{i}{2} \ \left ( \ \eta \ \sigma^{\ \mu} \ \overline{\vartheta}
\ - \ \vartheta \ \sigma^{\ \mu} \ \overline{\eta} \ \right )
\end{array}
\end{equation}

\noindent
From eq. (\ref{eq:29}) we read off the infinitesimal action from

\vspace*{-0.5cm}
\begin{equation}
\label{eq:30}
\begin{array}{l}
\left \lbrack \ \delta_{\ U} \ ( \ \eta \ , \ \overline{\eta} \ ) \ \right \rbrack
F \ ( \ \vartheta \ , \ \overline{\vartheta} \ , \ x \ )
\vspace*{0.2cm} \\
\hspace*{0.3cm} \sim
\ F_{\ U} \ ( \ \vartheta \ , \ \overline{\vartheta} \ , \ x \ ) \ -
\ F \ ( \ \vartheta \ , \ \overline{\vartheta} \ , \ x \ )
\vspace*{0.2cm} \\
\hspace*{0.3cm} = 
\ \eta^{\ \alpha} \ q_{\ \alpha} \ F \ + \ \overline{\eta}^{\ \dot{\alpha}}
\ \overline{q}_{\ \dot{\alpha}} \ F
\vspace*{0.2cm} \\
\color{red}
q_{\ \alpha} \ = \ ( \ \partial_{\ \vartheta} \ )_{\ \alpha}
\ + \ \frac{i}{2} \ \overline{\vartheta}^{\ \dot{\beta}} \ ( \ \partial_{\ x} \ )_{\ \alpha \dot{\beta}} 
\vspace*{0.2cm} \\
\color{red}
\overline{q}_{\ \dot{\beta}} \ = \ ( \ \partial_{\ \overline{\vartheta}} \ )_{\ \dot{\beta}}
\ + \ \frac{i}{2} \ \vartheta^{\ \alpha} \ ( \ \partial_{\ x} \ )_{\ \alpha \dot{\beta}} 
\vspace*{0.2cm} \\
\color{red}
( \ \partial_{\ x} \ )_{\ \alpha \dot{\beta}} \ =
\ ( \ \partial_{\ x} \ )^{\ \mu} \ \sigma_{\ \mu \ \alpha \dot{\beta}}
\vspace*{0.2cm} \\
\color{red}
( \ \partial_{\ \vartheta} \ )_{\ \alpha} \ = \ \partial \ / \ \partial \ \vartheta^{\ \alpha}
\hspace*{0.2cm} ; \hspace*{0.2cm}
( \ \partial_{\ \overline{\vartheta}} \ )_{\ \dot{\beta}} \ =
\ \partial \ / \ \partial \ \overline{\vartheta}^{\ \dot{\beta}}
\end{array}
\end{equation}

}

\newpage

\color{blue}

{\large \bf

\noindent
The pair $\color{red} q_{\ \alpha} \ , \ \overline{q}_{\ \dot{\beta}}$ 
\color{blue} forms by construction
a representation of the algebra generated by $Q_{\ \alpha} \ , \ Q^{\ *}_{\ \dot{\beta}}$
(\ref{eq:25}) where the derivatives 
$i \ ( \ \partial_{\ x} \ )_{\ \mu} \ \leftrightarrow \ P_{\ \mu}$ play the same role relative
to the energy-momentum operator. 

\vspace*{-0.5cm}
\begin{equation}
\label{eq:31}
\begin{array}{l}
\left \lbrace \ q_{\ \alpha} \ , \ \overline{q}_{\ \dot{\beta}} \ \right \rbrace
\ = \ i \ ( \ \partial_{\ x} \ )_{\ \mu} \ \sigma^{\ \mu}_{\ \alpha \dot{\beta}}
\vspace*{0.2cm} \\
\left \lbrace \ q_{\ \alpha} \ , \ q_{\ \beta} \ \right \rbrace \ = \ 0
\hspace*{0.2cm} ; \hspace*{0.2cm}
\left \lbrace \ \overline{q}_{\ \dot{\alpha}} \ , \ \overline{q}_{\ \dot{\beta}} \ \right \rbrace
\ = \ 0
\end{array}
\end{equation}

\noindent
It may be useful at this point to complete the susy algebra eqs. (\ref{eq:25}) to the
full super-Poincare algebra

\vspace*{-0.5cm}
\begin{equation}
\label{eq:32}
\begin{array}{l}
\left \lbrace \ Q_{\ \alpha} \ , \ Q^{\ *}_{\ \dot{\beta}} \ \right \rbrace
\ = \ P^{\ \mu} \ \sigma_{\ \mu \ \alpha \dot{\beta}}
\vspace*{0.2cm} \\
\left \lbrace \ Q_{\ \alpha} \ , \ Q_{\ \beta} \ \right \rbrace \ = \ 0
\hspace*{0.2cm} ; \hspace*{0.2cm}
\left \lbrace \ Q^{\ *}_{\ \dot{\alpha}} \ , \ Q^{\ *}_{\ \dot{\beta}} \ \right \rbrace
\ = \ 0
\vspace*{0.2cm} \\
\left \lbrack \ P^{\ \mu} \ , 
\ \begin{array}{l} 
Q_{\ \alpha}
\vspace*{0.2cm} \\
Q^{\ *}_{\ \dot{\beta}}
\end{array}
\ \right \rbrack \ = \ 0
\vspace*{0.2cm} \\
U^{\ -1} \ ( \ {\cal{A}} \ ) \ Q_{\ \alpha} \ U \ ( \ {\cal{A}} \ )
\ = \ {\cal{A}}_{\ \alpha \beta} \ Q_{\ \beta}
\vspace*{0.2cm} \\
U^{\ -1} \ ( \ {\cal{A}} \ ) \ Q^{\ *}_{\ \dot{\alpha}} \ U \ ( \ {\cal{A}} \ )
\ = \ \overline{{\cal{A}}}_{\ \dot{\alpha} \dot{\beta}} \ Q^{\ *}_{\ \dot{\beta}}
\end{array}
\end{equation}

\noindent
In eq. (\ref{eq:32}) only the extension to fermionic charges of the super-Poincare algebra
is displayed.

}

\newpage

\color{blue}

{\large \bf

\begin{center}
\color{red}

4c) Base susy differential representation 

\end{center}

\color{blue}

\noindent
The base differentials representing ( N=1 ) susy are 
$\color{red} \left ( \ q_{\ \alpha} \ , \ \overline{q}_{\ \dot{\beta}} \ \right )$ 
\color{blue} , defined in eqs. (\ref{eq:30}) and (\ref{eq:31}) .

\noindent
Since the product 
$U \ ( \ \eta \ ) \ F \ ( \ \vartheta \ , \ \overline{\vartheta} \ , \ x \ )$ , defined in 
eq. (\ref{eq:29}) can equally well be interpreted as right-multiplication,
commuting with the left one there exist associated differentials 
$\color{magenta} \left ( \ D_{\ \alpha} \ , \ \overline{D}_{\ \dot{\beta}} \ \right )$ 
\color{blue} , 
anticommuting with the pair $\color{red} \left ( \ q_{\ \alpha} \ , \ \overline{q}_{\ \dot{\beta}}
\ \right )$ \color{blue} .

\vspace*{-0.5cm}
\begin{equation}
\label{eq:33}
\begin{array}{l}
\color{red}
q_{\ \alpha} \ = \ ( \ \partial_{\ \vartheta} \ )_{\ \alpha}
\ + \ \frac{i}{2} \ \overline{\vartheta}^{\ \dot{\beta}} \ ( \ \partial_{\ x} \ )_{\ \alpha \dot{\beta}} 
\vspace*{0.2cm} \\
\color{red}
\overline{q}_{\ \dot{\beta}} \ = \ ( \ \partial_{\ \overline{\vartheta}} \ )_{\ \dot{\beta}}
\ + \ \frac{i}{2} \ \vartheta^{\ \alpha} \ ( \ \partial_{\ x} \ )_{\ \alpha \dot{\beta}}
\vspace*{0.2cm} \\
\color{red} \rightarrow  
\color{magenta}
\hspace*{0.2cm} D_{\ \alpha} \ = \ ( \ \partial_{\ \vartheta} \ )_{\ \alpha}
\ - \ \frac{i}{2} \ \overline{\vartheta}^{\ \dot{\beta}} \ ( \ \partial_{\ x} \ )_{\ \alpha \dot{\beta}}
\vspace*{0.2cm} \\
\color{magenta}
\hspace*{0.7cm}
\overline{D}_{\ \dot{\beta}} \ = \ - \ ( \ \partial_{\ \overline{\vartheta}} \ )_{\ \dot{\beta}}
\ + \ \frac{i}{2} \ \vartheta^{\ \alpha} \ ( \ \partial_{\ x} \ )_{\ \alpha \dot{\beta}}
\end{array}
\end{equation}

\color{blue}
\noindent
with the anticommutation relations

\vspace*{-0.5cm}
\begin{equation}
\label{eq:34}
\begin{array}{l}
\left \lbrace \ q_{\ \alpha} \ , \ \overline{q}_{\ \dot{\beta}}
\ \right \rbrace \ = \ i \ \partial_{\ \alpha \dot{\beta}} \ =
\ \left \lbrace \ D_{\ \alpha} \ , \ \overline{D}_{\ \dot{\beta}} \ \right \rbrace
\vspace*{0.2cm} \\
\left \lbrace \ q_{\ \alpha} \ , 
\ \begin{array}{l}
q_{\ \beta}
\vspace*{0.2cm} \\
D_{\ \beta}
\vspace*{0.2cm} \\
\overline{D}_{\ \dot{\beta}}                        
\end{array} \ \right \rbrace \ = \ 0
\hspace*{0.2cm} = \hspace*{0.1cm}
\left \lbrace \ \overline{q}_{\ \dot{\alpha}} \ , 
\ \begin{array}{l}
\overline{q}_{\ \dot{\beta}}
\vspace*{0.2cm} \\
D_{\ \beta}
\vspace*{0.2cm} \\
\overline{D}_{\ \dot{\beta}}                        
\end{array} \ \right \rbrace
\vspace*{0.2cm} \\
\color{red}
\mbox{and} \hspace*{0.2cm}
\left ( \ q_{\ \alpha} \ , \ \overline{q}_{\ \dot{\beta}} \ \right )
\ \leftrightarrow
\color{magenta}
\ \left ( \ D_{\ \alpha} \ , \ \overline{D}_{\ \dot{\beta}} \ \right )
\end{array}
\end{equation}

}

\newpage

\color{blue}

{\large \bf

\noindent
While \color{red} 
$q_{\ A} \ = \ \left ( \ q_{\ \alpha} \ , \ \overline{q}_{\ \dot{\beta}} \ \right ) \ ; \ A \ = 
\ 1\cdots4$ \\
\color{blue} define infinitesimal susy transformations, 
\color{magenta} $D_{\ A} \ = \ \left ( \ D_{\ \alpha} \ , \ \overline{D}_{\ \dot{\beta}} \ \right ) 
\ ; \ A \ = \ 1\cdots4$ \color{blue} can be used to set constraints on the substrate of fields on which
\color{red} $q_{\ A}$ \color{blue} act.

\noindent
\begin{center}
\color{red}
4d) Explicit construction of chiral (scalar) superfield
\end{center}

\noindent
In section 3a) (eq. \ref{eq:10}) a chiral superfield $\Phi \ ( \ x \ , \ \vartheta_{\ A} \ )$
is introduced, which shall be constructed here {\bf explicitely} .

\noindent
To this end we recall the base quantities $\vartheta_{\ A} \ = 
\ \left ( \ \vartheta_{\ \alpha} \ , \ \overline{\vartheta}_{\ \dot{\beta}} \ \right )$ and polynomial
invariants 

\vspace*{-0.5cm}
\begin{equation}
\label{eq:35}
\begin{array}{l}
\mbox{base :} \ \theta_{\ \alpha} \ , \ \overline{\theta}_{\ \dot{\beta}} \ \rightarrow
\vspace*{0.2cm} \\
\theta^{\ \alpha} \ = \ \varepsilon^{\ ' \ \alpha \beta} \ \vartheta_{\ \beta}
\hspace*{0.2cm} , \hspace*{0.2cm}
\overline{\vartheta}^{\ \dot{\beta}} \ = 
\ \varepsilon^{\ ' \ \dot{\beta} \dot{\gamma}} \ \overline{\vartheta}_{\ \dot{\gamma}}
\end{array}
\end{equation}

\noindent
\color{red} the Grassmann derivatives and raising (lowering) of indices

\color{blue}
\noindent
Derivatives with respect to Grassmann variables shall be denoted by shorthand

\vspace*{-0.5cm}
\begin{equation}
\label{eq:36}
\begin{array}{l}
\partial_{\ \alpha} \ = \ \partial \ / \ \vartheta^{\ \alpha}
\hspace*{0.2cm} , \hspace*{0.2cm}
\partial^{\ \alpha} \ = \ \partial \ / \ \vartheta_{\ \alpha}
\vspace*{0.2cm} \\
\overline{\partial}_{\ \dot{\beta}} \ = \ \partial \ / \ \overline{\vartheta}^{\ \dot{\beta}}
\hspace*{0.2cm} , \hspace*{0.2cm}
\overline{\partial}^{\ \dot{\beta}} \ = \ \partial \ / \ \overline{\vartheta}_{\ \dot{\beta}}
\end{array}
\end{equation}

\noindent
Applying the chain rule we find

}

\newpage

\color{blue}

{\large \bf

\vspace*{-0.5cm}
\begin{equation}
\label{eq:37}
\begin{array}{l}
\partial^{\ \alpha} \ = \ \left ( 
\ ( \ \partial \ / \ \vartheta_{\ \alpha} \ ) \ \vartheta^{\ \beta} \ \right )    
\ \partial_{\ \beta}
\vspace*{0.2cm} \\ 
\hspace*{0.5cm}  =
\ \left ( 
\ ( \ \partial \ / \ \vartheta_{\ \alpha} \ ) \ \varepsilon^{\ ' \ \beta \ \gamma}
\ \vartheta_{\ \gamma} \ \right ) \ \partial_{\ \beta} 
\vspace*{0.2cm} \\
\hspace*{0.5cm} = \ \color{red} - 
\ \color{blue} \varepsilon^{\ ' \ \alpha \ \beta} \ \partial_{\ \beta}
\ \color{red} \rightarrow
\vspace*{0.2cm} \\
\color{blue}
\partial^{\ \alpha} \ = \ - \ \varepsilon^{\ ' \ \alpha \ \beta} \ \partial_{\ \beta}
\hspace*{0.2cm} , \hspace*{0.2cm}
\partial_{\ \alpha} \ = \ - \ \varepsilon_{\ \alpha \ \beta} \ \partial^{\ \beta}
\vspace*{0.2cm} \\
\overline{\partial}^{\ \dot{\alpha}} \ = \ - \ \varepsilon^{\ ' \ \dot{\alpha} \ \dot{\beta}} 
\ \partial_{\ \dot{\beta}}
\hspace*{0.2cm} , \hspace*{0.2cm}
\overline{\partial}_{\ \dot{\alpha}} \ = \ - \ \varepsilon_{\ \dot{\alpha} \ \dot{\beta}} 
\ \overline{\partial}^{\ \dot{\beta}}
\end{array}
\end{equation}
\footnote{\hspace*{0.1cm} \color{red} Verify that the -- sign in eq. (\ref{eq:37}) opposite
to the base convention of lowering and raising spinor components is correct.}

\color{blue}
\noindent
It follows that the raising and lowering of derivative operator components
\color{red}
$\left ( \ q_{\ \alpha} \ , \ \overline{q}_{\ \dot{\beta}} \ \right )$
and \color{magenta} $\left ( \ D_{\ \alpha} \ , \ \overline{D}_{\ \dot{\beta}} \ \right )$ 
\color{blue} is to be performed with the characteristic -- sign relative to the
base components $\left ( \ \vartheta_{\ \alpha} \ , \ \overline{\vartheta}_{\ \dot{\beta}}
\ \right )$ .

\noindent
For clarity I list all components for the so defined base pair
\color{magenta} $\left ( \ D_{\ \alpha} \ , \ \overline{D}_{\ \dot{\beta}} \ \right )$
\color{blue}

}

\newpage

\color{blue}

{\large \bf

\vspace*{-0.5cm}
\begin{equation}
\label{eq:38}
\begin{array}{l}
D_{\ \alpha} \ = \ \partial_{\ \alpha} \ - \ \frac{i}{2} 
\ \overline{\vartheta}^{\ \dot{\beta}} \ \sigma^{\ \mu}_{\ \alpha \ \dot{\beta}} 
\ \partial_{\ \mu}
\vspace*{0.2cm} \\
\overline{D}_{\ \dot{\beta}} \ = \ - \ \overline{\partial}_{\ \dot{\beta}}
\ + \ \frac{i}{2} 
\ \vartheta^{\ \alpha} \ \sigma^{\ \mu}_{\ \alpha \ \dot{\beta}}
\vspace*{0.2cm} \\
D^{\ \alpha} \ = \ \partial^{\ \alpha} \ - \ \frac{i}{2} 
\ \overline{\vartheta}_{\ \dot{\beta}} \ \sigma^{\ \mu \ \dot{\beta} \ \alpha}
\ \partial_{\ \mu}
\vspace*{0.2cm} \\
\overline{D}^{\ \dot{\beta}} \ = \ - \ \overline{\partial}^{\ \dot{\beta}}
\ + \ \frac{i}{2} 
\ \vartheta_{\ \alpha} \ \sigma^{\ \mu \ \dot{\beta} \ \alpha}
\ \partial_{\ \mu}
\vspace*{0.2cm} \\
D^{\ \alpha} \ = \ - \ \varepsilon^{\ ' \ \alpha \ \beta} \ D_{\ \beta}
\hspace*{0.2cm} , \hspace*{0.2cm}
\overline{D}^{\ \dot{\beta}} \ = \ - \ \varepsilon^{\ ' \ \dot{\beta} \ \dot{\alpha}} 
\ \overline{D}_{\ \dot{\alpha}}
\vspace*{0.2cm} \\
\color{red}
\left \lbrace \ D_{\ \alpha} \ , \ \overline{D}_{\ \dot{\beta}} \ \right \rbrace
\ = \ i \ \partial_{\ \alpha \ \dot{\beta}}
\hspace*{0.2cm} , \hspace*{0.2cm}
\left \lbrace \ D^{\ \alpha} \ , \ \overline{D}^{\ \dot{\beta}} \ \right \rbrace
\ = \ i \ \partial^{\ \dot{\beta} \ \alpha}
\vspace*{0.2cm} \\
\color{magenta}
\partial^{\ \dot{\beta} \ \alpha} \ = \ \varepsilon^{\ ' \ \dot{\beta} \ \dot{\delta}}
\ \varepsilon^{\ ' \ \alpha \ \gamma} \ \partial_{\ \gamma \ \dot{\delta}} 
\end{array}
\end{equation}
\color{blue} \footnote{\hspace*{0.1cm} Verify the relations in eq. (\ref{eq:38}) and
display the quantities 
$( \ \partial_{\ \alpha \ \dot{\beta}} \ , \ \partial^{\ \dot{\beta} \ \alpha} \ )$
by \\ \hspace*{0.8cm} Pauli matrices and the contravariant four vector \\
\hspace*{0.8cm} $\nabla^{\ \mu} \ = \ ( \nabla^{\ 0} \ , \ \vec{\nabla} \ ) \ = 
\ ( \ \partial_{\ t} \ , \ - \ \vec{\partial}_{\ \vec{x}} \ )$ .}

\noindent
\color{red} Grassmannian Lorentz invariants

\color{blue}
\noindent
Quadratic invariants are formed from 
$\left ( \ \vartheta_{\ \alpha} \ , \ \overline{\vartheta}_{\ \dot{\beta}} \ \right )$

}

\newpage

\color{blue}

{\large \bf

\vspace*{-0.5cm}
\begin{equation}
\label{eq:39}
\begin{array}{l}
\color{red} \vartheta_{\ 1} \ = \ \vartheta^{\ 2}
\hspace*{0.2cm} , \hspace*{0.2cm}
\color{red} \vartheta_{\ 2} \ = \ - \ \vartheta^{\ 1}
\vspace*{0.2cm} \\
\color{blue}
\theta^{\ 2} \ = \ \frac{1}{2} \ \vartheta^{\ \alpha} \ \vartheta_{\ \alpha} \ =
\color{red} \ \vartheta_{\ 1} \ \vartheta_{\ 2}
\hspace*{0.2cm} \color{magenta} \rightarrow
\vspace*{0.2cm} \\
\color{blue}
\overline{\theta}^{\ 2} \ = \ \frac{1}{2} \ \overline{\vartheta}_{\ \dot{\beta}}
\ \overline{\vartheta}^{\ \dot{\beta}} \ =
\color{red} \ \overline{\vartheta}_{\ \dot{2}}
\ \overline{\vartheta}_{\ \dot{1}}
\vspace*{0.2cm} \\
\color{blue}
\mbox{and} \ \left ( \ \partial \ \right )^{\ 2} \ = 
\ \frac{1}{2} \ \partial^{\ \alpha} \ \partial_{\ \alpha}
\hspace*{0.2cm} , \hspace*{0.2cm}
\left ( \ \overline{\partial} \ \right )^{\ 2} \ =
\ \frac{1}{2} \ \overline{\partial}_{\ \dot{\beta}} \ \overline{\partial}^{\ \dot{\beta}}
\vspace*{0.2cm} \\
\left ( \ \partial \ \right )^{\ 2} \ = 
\ \color{red} \ ( \ \partial \  / \ \partial \ \vartheta_{\ 2} \ )
\ ( \ \partial \  / \ \partial \ \vartheta_{\ 1} \ )
\vspace*{0.2cm} \\
\color{blue}
\left ( \ \overline{\partial} \ \right )^{\ 2} \ = 
\ \color{red} \ ( \ \partial \  / \ \partial \ \overline{\vartheta}_{\ \dot{1}} \ )
\ ( \ \partial \  / \ \partial \ \overline{\vartheta}_{\ \dot{2}} \ )
\ \color{magenta} \rightarrow
\vspace*{0.2cm} \\
\color{blue}
\left ( \ \partial \ \right )^{\ 2} \ \theta^{\ 2} \ = \ 1
\hspace*{0.2cm} , \hspace*{0.2cm}
\left ( \ \overline{\partial} \ \right )^{\ 2} \ \overline{\theta}^{\ 2} \ = \ 1
\vspace*{0.2cm} \\
\hline
\end{array}
\end{equation}

\noindent
From the relations derived in this section (4a-4d eq. (\ref{eq:39}))
we construct the right-chiral superfield 
$\Phi \ ( \ x \ , \ \color{red} \vartheta \ , \ \overline{\vartheta} \ \color{blue} )$

\color{blue}

\vspace*{-0.7cm}
\begin{equation}
\label{eq:40}
\begin{array}{l}
\mbox{constraint :} \ \overline{D}_{\ \dot{\alpha}} \ \Phi \ = \ 0 \ \color{red} \rightarrow
\vspace*{0.2cm} \\
\Phi \ = \ \left \lbrace \ \begin{array}{l}
\theta^{\ 2} \ H \ ( \ x^{\ -} \ )
\vspace*{0.2cm} \\
+ \ \vartheta^{\ \alpha} \ \eta_{\ \alpha} \ ( \ x^{\ -} \ )
\vspace*{0.2cm} \\
\ + \ \varphi \ ( \ x^{\ -} \ ) 
\end{array}
\right .
\vspace*{0.2cm} \\
\color{magenta}
\left ( \ x^{\ -} \ \right )^{\ \mu} \ = \ x^{\ \mu} \ - 
\ \frac{i}{2} \ \vartheta^{\ \alpha} \ \sigma^{\ \mu}_{\ \alpha \dot{\beta}} 
\ \overline{\vartheta}^{\ \dot{\beta}}
\hspace*{0.2cm} ; \hspace*{0.2cm}
\overline{D}_{\ \dot{\alpha}} \ x^{\ -} \ = \ 0
\end{array}
\end{equation}

\color{blue}

\noindent
From eq. (\ref{eq:40}) the forms given in eqs. (\ref{eq:12}-\ref{eq:14}) follows.

}

\newpage

\color{blue}

{\large \bf

\noindent
The highest component $\left . \ \right |_{\ \theta^{\ 2} \ \overline{\theta}^{\ 2}}$
of a general  superfield can be projected using the Grassman integration rules
and differentials

\vspace*{-0.7cm}
\begin{equation}
\label{eq:41}
\begin{array}{l}
d^{\ 4} \ \vartheta \ = \ d \ \overline{\vartheta}_{\ \dot{1}}
\ d \ \overline{\vartheta}_{\ \dot{2}} \ d \ \vartheta_{\ 2} \ d \ \vartheta_{\ 1}
\vspace*{0.2cm} \\
{\displaystyle{\int}} \ d \ \vartheta_{\ \alpha} \ \vartheta_{\ \beta} \ = \ \delta_{\ \alpha \beta}
\hspace*{0.2cm} ; \hspace*{0.2cm}
{\displaystyle{\int}} \ d \ \overline{\vartheta}_{\ \dot{\alpha}} 
\ \overline{\vartheta}_{\ \dot{\beta}} \ = \ \delta_{\ \dot{\alpha} \dot{\beta}}
\end{array}
\end{equation}

\noindent
The 'component form' of the relations in eq. (\ref{eq:41}) is not explicitely
covariant, but could be formally rendered such just writing the index of the differentials
as upper index (not 'raising' it) .

\noindent
We note the identities

\vspace*{-0.7cm}
\begin{equation}
\label{eq:42}
\begin{array}{l}
\vartheta_{\ \alpha} \ \vartheta_{\ \beta} \ = \ \varepsilon_{\ \alpha \beta}
\ \theta^{\ 2}
\hspace*{0.2cm} ; \hspace*{0.2cm}
\overline{\vartheta}_{\ \dot{\beta}} \ \overline{\vartheta}_{\ \dot{\alpha}}
\ = \ \varepsilon_{\ \dot{\alpha} \dot{\beta}} \ \overline{\theta}^{\ 2}
\end{array}
\end{equation}

\noindent
\begin{center}
\color{red}
4e) Chiral invariants from chiral (scalar) superfields (superpotentials)
\end{center}

\color{blue}

\noindent
Lets consider an analytic function of a 'dummy' complex variable z , for which
then the right-chiral superfield $\Phi$ is substituted

\vspace*{-0.7cm}
\begin{equation}
\label{eq:43}
\begin{array}{l}
W \ ( \ z \ ) \ \sim \ \sum_{\ n = 0}^{\ \infty} \ w_{\ n} \ z^{\ n}
\hspace*{0.2cm} ; \hspace*{0.2cm} \color{red} \rightarrow
\vspace*{0.2cm} \\
\color{blue}
W \ = \ W \ ( \ \Phi \ )
\hspace*{0.2cm} ; \hspace*{0.2cm}
\color{red} \overline{D}_{\ \dot{\alpha}} \ \Phi \ = 0
\hspace*{0.2cm} \rightarrow \hspace*{0.2cm}
\overline{D}_{\ \dot{\alpha}} \ W \ = 0
\end{array}
\end{equation}

\color{blue}

\noindent
By the substitution in eq. (\ref{eq:43}) W becomes a (composite) chiral superfield.

}

\newpage

\color{blue}

{\large \bf

\noindent
The highest chiral component $\left . \ W \ \right |_{\ \theta^{\ 2}}$
represents a susy invariant (modulo a total divergence) 

\vspace*{-0.5cm}
\begin{equation}
\label{eq:44}
\begin{array}{l}
\left .  W \ \right |_{\ \theta^{\ 2}} \ = \ {\displaystyle{\int}} \ d^{\ 2} \ \vartheta \ W
\hspace*{0.2cm} \rightarrow \hspace*{0.2cm}
d^{\ 2} \ \vartheta \ = \ d \ \vartheta_{\ 2} \ d \ \vartheta_{\ 1}
\vspace*{0.2cm} \\
\left .  W \ \right |_{\ \theta^{\ 2}} \ = \ W_{\ 1} \ ( \ \varphi \ ) \ H
\ - \ \frac{1}{2} \ W_{\ 2} \ ( \ \varphi \ ) \ \eta^{\ \alpha} \ \eta_{\ \alpha}
\vspace*{0.2cm} \\
W_{\ n} \ ( \ z \ ) \ = \ \left ( \ \partial_{\ z} \ \right )^{\ n} \ W \ ( \ z \ )
\end{array}
\end{equation}

\noindent
By construction $\left .  W \ \right |_{\ \theta^{\ 2}}$ does not contain space-time derivatives
of the components of $\Phi$ .

\noindent
\begin{center}
\color{red}
The elementary Wess Zumino model
\end{center}

\color{blue}

\noindent
Considering -- only in this subsection -- $\Phi$ to represent an elementary superfield
the most general (perturbatively) renormalizable Lagrangean density takes the form

\vspace*{-0.5cm}
\begin{equation}
\label{eq:45}
\begin{array}{l}
{\cal{L}} \ = \ \left . \overline{\Phi} \ \Phi \ \right |_{\ \theta^{\ 2} \ \overline{\theta}^{\ 2}}
\ + \ \left ( \ \left .  W \ \right |_{\ \theta^{\ 2}} \ + \ h.c. \ \right )
\vspace*{0.2cm} \\
W \ = \ W^{\ (3)} \ = \ b_{\ 1} \ z + \ \frac{1}{2} \ b_{\ 2} \ z^{\ 2} \ +
\ \frac{1}{3 \ !} \ b_{\ 3} \ z^{\ 3}  
\vspace*{0.2cm} \\
W_{\ 1} \ = \ b_{\ 1} \ + \ b_{\ 2} \ z \ + \ \frac{1}{2} \ b_{\ 3} \ z^{\ 2}
\hspace*{0.2cm} ; \hspace*{0.2cm} b_{\ 1} \ \color{red} \ \rightarrow \ 0
\vspace*{0.2cm} \\
W_{\ 2} \ = \ b_{\ 2} \ + \ b_{\ 3} \ z
\end{array}
\end{equation}

\noindent
We decompose ${\cal{L}} \ = \ {\cal{L}}_{\ kin} \ + \ {\cal{L}}_{\ pot}$ 
into kinetic and potential parts

}

\newpage

\color{blue}

{\large \bf

\noindent
and obtain

\vspace*{-0.5cm}
\begin{equation}
\label{eq:46}
\begin{array}{l}
{\cal{L}}_{\ pot} \ = \ H^{\ *} \ H \ + 
 \left ( \ W_{\ 1} \ H \ - \ \frac{1}{2} \ W_{\ 2} \ \eta^{\ \alpha} \ \eta_{\ \alpha} 
\ + \ h.c.  \right )
\vspace*{0.2cm} \\
{\cal{L}}_{\ pot} \ = \ - \ V  
\end{array}
\end{equation}

\noindent
The auxiliary field H is at this stage determined by the extremum condition

\vspace*{-0.5cm}
\begin{equation}
\label{eq:47}
\begin{array}{l}
\delta \ V \ / \ \delta \ H^{\ *} \ = \ \delta \ V \ / \ \delta \ H \ = \ 0
\hspace*{0.2cm} \color{red} \rightarrow
\vspace*{0.2cm} \\
\color{blue}
H \ = \ - \ W_{\ 1}^{\ *} \hspace*{0.2cm} \color{red} \rightarrow
\vspace*{0.2cm} \\
\color{blue}
V \ = \ W_{\ 1}^{\ *} \ W_{\ 1} \ + \ \left ( \ \frac{1}{2} \ W_{\ 2} 
\ \eta^{\ \alpha} \ \eta_{\ \alpha} \ + \ h.c. \ \right )
\end{array}
\end{equation}

\noindent
The extremum yields the maximum of V with respect to $( \ H \ , \ H^{\ *} \ )$ .

\noindent
Now we look for the minimum of V with respect to $\varphi \ = \ z \ + \ \Delta \ \varphi$ ,
i.e. implying eventually a nontrivial vacuum expected value of the scalar field $\varphi$

\vspace*{-0.5cm}
\begin{equation}
\label{eq:48}
\begin{array}{l}
V_{\ , \ z} \ = \ W_{\ 1}^{\ *} \ W_{\ 2} 
\hspace*{0.2cm} ; \hspace*{0.2cm}
V_{\ , \ \overline{z}} \ = \ W_{\ 1} \ W_{\ 2}^{\ *}
\end{array}
\end{equation}

\noindent
From the extremum conditions in eq. (\ref{eq:48}) we find two solutions

\vspace*{-0.5cm}
\begin{equation}
\label{eq:49}
\begin{array}{l}
a) \ : \ W_{\ 1} \ ( \ z \ ) \ = \ 0 
\hspace*{0.2cm} ; \hspace*{0.2cm}
b) \ : \ W_{\ 2} ( \ z \ ) \ = 0
\end{array}
\end{equation}

\noindent
which we inspect in turn.

}

\newpage

\color{blue}

{\large \bf

\noindent
We do not set $b_{\ 1} \ = 0$ here. It follows

\vspace*{-0.5cm}
\begin{equation}
\label{eq:50}
\begin{array}{l}
a) \ : \ W_{\ 1} \ ( \ z \ ) \ = \ 0 \ = \ \frac{1}{2} \ b_{\ 3}
\ \left ( \ z \ + \ r_{\ 2} \ \right )^{\ 2} \ + \ r_{\ 1}
\vspace*{0.2cm} \\
b) \ : \ W_{\ 2} \ ( \ z \ ) \ = \ 0 \ = \ b_{\ 3} \ \left ( \ z \ + \ r_{\ 2} \ \right )
\vspace*{0.2cm} \\
\color{red} r_{\ 2} \ = \ b_{\ 2} \ / \ b_{\ 3}
\hspace*{0.2cm} ; \hspace*{0.2cm}
r_{\ 1} \ = \ b_{\ 1} \ - \ \frac{1}{2} \ b_{\ 2}^{\ 2} \ / \ b_{\ 3}
\end{array}
\end{equation}

\color{blue}

\noindent
It follows from eq. (\ref{eq:50}) that for case b) we have

\vspace*{-0.5cm}
\begin{equation}
\label{eq:51}
\begin{array}{l}
W_{\ 2} \ = \ 0 \ \rightarrow \ r_{\ 1} \ = \ 0
\hspace*{0.2cm} \mbox{for a minimum of V}
\vspace*{0.2cm} \\
\rightarrow \ z \ = \ - \ r_{\ 2}
\hspace*{0.2cm} \rightarrow \hspace*{0.2cm}
\color{red} W_{\ 1} \ ( \ \Delta \ \varphi \ ) \ = \ \frac{1}{2} \ b_{\ 3} 
\ \left ( \ \Delta \ \varphi \ \right )^{\ 2}
\vspace*{0.2cm} \\
\color{blue} \ V \ = \ \frac{1}{4} \ | \ b_{\ 3} \ |^{\ 2} 
\ \left | \ \Delta \ \varphi \ \right |^{\ 4}
\end{array}
\end{equation}

\noindent
Thus case b) and $r_{\ 1} \ = \ 0$ correspond to a {\it unique} minimum of V and
the physical masses of $\Delta \ \varphi$ and $\eta_{\ \alpha}$ are both 0 .

\noindent
For $r_{\ 1} \ = \ \frac{1}{2} \ b_{\ 3} \ \varrho^{\ 2} \ \neq \ 0$ and case a) we have two minima
and thus define two secondary vaiables $\Delta_{\ \pm} \ \varphi$ :

\vspace*{-0.5cm}
\begin{equation}
\label{eq:52}
\begin{array}{l}
\color{red} W_{\ 1} \ ( \ \Delta \ \varphi \ ) \ = \ \frac{1}{2} \ b_{\ 3}
\ \left ( \ \Delta \ \varphi \ - \ \varrho \ \right ) 
\ \left ( \ \Delta \ \varphi \ + \ \varrho \ \right )  
\vspace*{0.2cm} \\
\color{blue}
\Delta_{\ \pm} \ \varphi \ = \ \Delta \ \varphi \ \mp \ \varrho
\end{array}
\end{equation}

\noindent
Substituting (either/or) $\Delta_{\ \pm}$ , $W_{\ 1}$ becomes

}

\newpage

\color{blue}

{\large \bf

\vspace*{-0.5cm}
\begin{equation}
\label{eq:53}
\begin{array}{l}
\color{red} W_{\ 1} \ = \ \frac{1}{2} \ b_{\ 3} \ \Delta_{\ \pm} \ \varphi
\ \left ( \ \Delta_{\ \pm} \ \varphi  \ \pm \ 2 \ \varrho \ \right )
\vspace*{0.2cm} \\
\color{blue}
V \ = \ \left | \ b_{\ 3} \ \varrho \ \right |^{\ 2} \ \left | \ \Delta_{\ \pm} \ \varphi \ \right |^{\ 2}
\ + \ \frac{1}{4} \ | \ b_{\ 3} \ |^{\ 2} 
\ \left | \ \Delta_{\ \pm} \ \varphi \ \right |^{\ 4}
\vspace*{0.2cm} \\
\hspace*{0.8cm} \pm \ \frac{1}{2} \ \left | \ \ b_{\ 3} \ \right |^{\ 2} 
\ \left | \ \Delta_{\ \pm} \ \varphi \ \right |^{\ 2}
\ \left ( \ \varrho \ ( \ \Delta_{\ \pm} \ \varphi \ )^{\ *} \ + \ h.c. \ \right )  
\end{array}
\end{equation}

\noindent
The derivations (eqs. \ref{eq:50} - \ref{eq:53}) were done under the assumption $b_{\ 3} \ \neq \ 0$ .

\noindent
The limit $b_{\ 3} \ \rightarrow \ 0$ can readily be performed. It corresponds
to a free theory of a complex scalar $\varphi$ and a Majorana spinor $\eta_{\ \alpha}$
with a common mass.

\noindent
However the case $b_{\ 3} \ \neq \ 0$ always allows to shift $\varphi \ \rightarrow 
\ \Delta_{\ (\pm)} \ \varphi$
such that the originally present constant $b_{\ 1} \ \rightarrow \ 0$ . 
It does describe genuine interactions and still leads in the above semiclassical shift-
implementation around any of the two minima of V -- to equal masses 

\vspace*{-0.5cm}
\begin{equation}
\label{eq:54}
\begin{array}{l}
m \ = \ | \ b_{\ 3} \ \varrho \ |
\end{array}
\end{equation}

\noindent
for each of the two scalar 
and fermionic one particle states.

\noindent
Yet this semiclassical shift is to be proven correct because
of the existence of an instanton solution \\
( for $\varrho \ \neq \ 0$ and also $b_{\ 3} \ \neq \ 0$ ) , interpolating
between the two minima of V.

}

\newpage

\color{blue}

{\large \bf

\noindent
\begin{center}
\color{red}
4f) The chiral chain $\Phi \ \rightarrow \ \Phi^{\ '}$ or 
$\Phi_{\ n} \ \rightarrow \ \Phi_{\ n + 1}$
\end{center}

\color{blue}

\noindent
Let the chain start (\cite{shore}) with a chiral superfield $\Phi \ \equiv \ \Phi_{\ 0}$
as given in eq. (\ref{eq:40})

\vspace*{-0.5cm}
\begin{equation}
\label{eq:55}
\begin{array}{l}
\mbox{constraint :} \ \overline{D}_{\ \dot{\alpha}} \ \Phi \ = \ 0 \ \color{red} \rightarrow
\vspace*{0.2cm} \\
\Phi \ = \ \left \lbrace \ \begin{array}{l}
\theta^{\ 2} \ H \ ( \ x^{\ -} \ )
\vspace*{0.2cm} \\
+ \ \vartheta^{\ \alpha} \ \eta_{\ \alpha} \ ( \ x^{\ -} \ )
\vspace*{0.2cm} \\
\ + \ \varphi \ ( \ x^{\ -} \ ) 
\end{array} 
\right .
\vspace*{0.2cm} \\
\color{magenta}
\left ( \ x^{\ -} \ \right )^{\ \mu} \ = \ x^{\ \mu} \ - 
\ \frac{i}{2} \ \vartheta^{\ \alpha} \ \sigma^{\ \mu}_{\ \alpha \dot{\beta}} 
\ \overline{\vartheta}^{\ \dot{\beta}}
\hspace*{0.2cm} ; \hspace*{0.2cm}
\overline{D}_{\ \dot{\alpha}} \ x^{\ -} \ = \ 0
\end{array}
\end{equation}

\color{blue}

\noindent
Then, defining $\overline{D}^{\ 2} \ = \ \frac{1}{2} \ \overline{D}_{\ \dot{\alpha}} 
\ \overline{D}^{\ \dot{\alpha}}$ , the next chiral superfield along the chain is

\vspace*{-0.5cm}
\begin{equation}
\label{eq:56}
\begin{array}{l}
\left \lbrace 
\ \begin{array}{l}
\Phi_{\ 0}
\vspace*{0.2cm} \\ 
\Phi_{\ n}
\end{array}
\ \right \rbrace
\ \rightarrow
\ \left \lbrace 
\ \begin{array}{l}
\Phi_{\ 1}
\vspace*{0.2cm} \\ 
\Phi_{\ n + 1}
\end{array}
\ \right \rbrace 
\ = \ \overline{D}^{\ 2}
\ \left \lbrace 
\ \begin{array}{l}
\overline{\Phi}_{\ 0}
\vspace*{0.2cm} \\
\overline{\Phi}_{\ n} 
\end{array}
\ \right \rbrace
\vspace*{0.2cm} \\
\color{red}
\mbox{with : }
\ \Phi_{\ 1} \ = \ \left \lbrace \ \begin{array}{l}
\theta^{\ 2} \ \left ( \ - \ \framebox[0.3cm][b]{} \ \varphi^{\ *} \ \right )
\vspace*{0.2cm} \\
+ \ \vartheta^{\ \alpha} \ \sigma^{\ \mu \ \dot{\delta}}_{\ \alpha} 
\ i \ \partial_{\ \mu} \ \eta^{\ *}_{\ \dot{\delta}}
\vspace*{0.2cm} \\
\ + \ H^{\ *} 
\end{array} 
\right \rbrace
\ ( \ x^{\ -} \ )
\vspace*{0.2cm} \\
\color{blue}
\Phi_{\ n + 2} \ = \ - \ \framebox[0.3cm][b]{} \ \Phi_{\ n}
\hspace*{0.2cm} ; \hspace*{0.2cm}
\overline{D}_{\ \dot{\alpha}} \ \Phi_{\ n} \ = \ 0
\end{array}
\end{equation}

}

\newpage

\color{blue}

{\large \bf

\noindent
For details of susy transformations only the first two members of the chain are relevant.

\noindent
\begin{center}
\color{red}
5) The N = 1 super-Yang-Mills structure
\end{center}

\color{blue}

\noindent
First we discuss the the susy gauge connection fields, with respect
to an arbitrary ( hermitian ) representation ${\cal{D}}$ of the Lie algebra pertaining
to a {\it simple and compact} gauge group G

\vspace*{-0.5cm}
\begin{equation}
\label{eq:57}
\begin{array}{l}
{\cal{D}} 
\hspace*{0.2cm} : \hspace*{0.2cm}
T^{\ F} \ \mbox{with : } 
\left \lbrack \ T^{\ A} \ , \ T^{\ B} \ \right \rbrack \ =
\ i \ f_{\ ABC} \ T^{\ C} 
\vspace*{0.2cm} \\
F \ , \ A \ , \ B \ , \ C \ = \ 1 \ , \ \cdots \ , \ dim \ ( \ G \ )
\end{array}
\end{equation}

\noindent
In eq. (\ref{eq:57}) $f_{\ ABC}$ denote the (totally antisymmetric) structure constants 
of (Lie-) G, the normalization
of which is in general arbitrary, but usually corresponds to some implicit convention.
\footnote{\hspace*{0.1cm} E.g. for $G \ = \ SU2$ : $f_{\ ABC} \ = \ \varepsilon_{\ ABC}$ ,
 $dim \ ( \ SU2 \ ) \ = \ 3$ .}  
$dim \ ( \ G \ )$ stands for the (real) dimension of G.

\noindent
\begin{center}
\color{red}
5a) The ${\cal{D}}$ valued gauge connection from a hermitian vectorfield in
the Wess-Zumino gauge
\end{center}

\color{blue}

\noindent
We begin with a hermitian ${\cal{D}}$ valued vector field, which is amputated in the
Wess-Zumino gauge to the form

\vspace*{-0.5cm}
\begin{equation}
\label{eq:58}
\begin{array}{l}
{\cal{D}}
\hspace*{0.2cm} : \hspace*{0.2cm}
V \ = \ V^{\ A} \ T^{\ A}
\end{array}
\end{equation}

}

\newpage

\color{blue}

{\large \bf

\noindent
$V \ = \ V \ ( \ x \ )$ in eq. (\ref{eq:58}) takes the form

\vspace*{-0.5cm}
\begin{equation}
\label{eq:59}
\begin{array}{l}
V \ = 
\ \left \lbrace
 \begin{array}{lcr}
 & \vartheta^{\ 2} \ \overline{\vartheta}^{\ 2} \ D &
\vspace*{0.2cm} \\
\vartheta^{\ 2} \ \overline{\vartheta}_{\ \dot{\beta}}  \ ( \ \lambda^{\ *} \ )^{\ \dot{\beta}} 
 & + & \overline{\vartheta}^{\ 2} \ \vartheta^{\ \alpha} \ \lambda_{\ \alpha}
\vspace*{0.2cm} \\
 & \vartheta^{\ \alpha} \ \overline{\vartheta}^{\ \dot{\beta}} \ v_{\ \alpha \dot{\beta}} &
\end{array}
 \right \rbrace
\vspace*{0.2cm} \\
\color{red}
D \ = \ D^{\ A} \ T^{\ A} 
\hspace*{0.2cm} ; \hspace*{0.2cm}
\lambda_{\ \alpha} \ = \ \lambda_{\ \alpha}^{\ A} \ T^{\ A}  
\vspace*{0.2cm} \\
\color{red}
v_{\ \alpha \dot{\beta}} \ = \ v_{\ \mu}^{\ A} \ T^{\ A} \ \sigma^{\ \mu}_{\ \alpha \dot{\beta}}
\end{array}
\end{equation}

\noindent
In eq. (\ref{eq:59}) $D^{\ A} \ , \ v_{\ \mu}^{\ A}$ denote hermitian fields, whereas
the four component spinors

\vspace*{-0.5cm}
\begin{equation}
\label{eq:60}
\begin{array}{l}
\psi_{\ \kappa}^{\ A} \ = 
\ \left (
\ \begin{array}{l}
\lambda_{\ \alpha}
\vspace*{0.2cm} \\
\lambda^{\ * \ \dot{\beta}}
\end{array}
\ \right )^{\ A}
\end{array}
\end{equation} 

form (for $A \ = \ 1 \ , \ \cdots \ , \ dim \ ( \ G \ )$) dim G Majorana spinors 
in the chiral representation, satisfying the hermiticity constraint 

\vspace*{-0.5cm}
\begin{equation}
\label{eq:61}
\begin{array}{l}
\left (
\ \psi \ = \ C \ \gamma_{\ 0} \ \psi^{\ *}
\ \right )^{\ A}
\vspace*{0.2cm} \\
C \ = \ \left (
\ \begin{array}{ll}
\varepsilon & 0
\vspace*{0.2cm} \\
0 &  \varepsilon^{\ '}
\end{array}
\ \right )
\hspace*{0.2cm} ; \hspace*{0.2cm}
\gamma_{\ \mu} \ = \ \left (
\ \begin{array}{ll}
0 & \sigma_{\ \mu}
\vspace*{0.2cm} \\
\widetilde{\sigma}_{\ \mu} & 0
\end{array}
\ \right )
\end{array}
\end{equation}

\noindent
The $2 \ \times \ 2$ matrices $\sigma_{\ \mu} \ , \ \widetilde{\sigma}_{\ \mu}$ are 

}

\newpage

\color{blue}

{\large \bf

\vspace*{-0.5cm}
\begin{equation}
\label{eq:62}
\begin{array}{l}
\sigma_{\ \mu} \ = \ \left (
\ \sigma_{\ 0} \ , \ \ \ \vec{\sigma} \ \right )
\ \rightarrow \ \left ( \ \sigma_{\ \mu} \ \right )_{\ \alpha \dot{\beta}}
\vspace*{0.2cm} \\
\widetilde{\sigma}_{\ \mu} \ = \ \left (
\ \sigma_{\ 0} \ , \ - \ \vec{\sigma} \ \right )
\ \rightarrow \ \left ( \ \sigma_{\ \mu} \ \right )^{\ \dot{\beta} \alpha}
\end{array}
\end{equation}

\noindent
The amputated structure of V implies that only V and $V^{\ 2}$ are not vanishing.
The latter quantity only has a highest component

\vspace*{-0.5cm}
\begin{equation}
\label{eq:63}
\begin{array}{l}
\frac{1}{2} \ V^{\ 2} \ = 
\ \vartheta^{\ 2} \ \overline{\vartheta}^{\ 2} \ v^{\ 2}
\vspace*{0.2cm} \\
v^{\ 2} \ = \ v^{\ \varrho \ A} \ v_{\ \varrho}^{\ B} 
\ \frac{1}{2} \ \left \lbrace \ T^{\ A} \ , \ T^{\ B} \ \right \rbrace
\end{array}
\end{equation}

\noindent
\begin{center}
\color{red}
gauge connection
\end{center}

\color{blue}

\noindent
We make the Ansatz

\vspace*{-0.5cm}
\begin{equation}
\label{eq:64}
\begin{array}{l}
W_{\ \alpha} \ = \ e^{\ -V} \ D_{\ \alpha} \ e^{\ V}
\hspace*{0.2cm} ; \hspace*{0.2cm}
D_{\ \alpha} \ = \ \partial_{\ \alpha} \ - \ \frac{i}{2} 
\ \overline{\vartheta}^{\ \dot{\beta}} \ \partial_{\ \alpha \dot{\beta}}  
\end{array}
\end{equation}

\noindent
In order to verify susy gauge invariance properties, we choose a set of
chiral superfields transforming -- component by component -- according
to the ${\cal{D}}$ - representation of the local gauge group G

\vspace*{-0.5cm}
\begin{equation}
\label{eq:65}
\begin{array}{l}
{\cal{D}} 
\hspace*{0.2cm} : \hspace*{0.2cm}
\Phi \ \rightarrow \ \Phi^{\ a}
\vspace*{0.2cm} \\
\left ( \ \Phi^{\ \Omega} \ \right )^{\ a} \ ( \ x \ ) \ = 
\ \Omega^{\ a}_{\ \ b} \ ( \ x \ ) \ \Phi^{\ b}
\vspace*{0.2cm} \\
\left ( \ \Omega \ = \ \exp \ \frac{1}{i} \ \omega^{\ A} \ ( \ x \ ) 
\ T^{\ A} \ \right )^{\ a}_{\ \ b} 
\vspace*{0.2cm} \\
\color{red}
\mbox{in short : } \Phi^{\ \Omega} \ = \ \Omega \ \Phi
\end{array}
\end{equation}

}

\newpage

\color{blue}

{\large \bf

\noindent
In order to preserve chirality and susy covariance we promote
each component $\omega^{\ A}$ to a chiral superfield 

\vspace*{-0.5cm}
\begin{equation}
\label{eq:66}
\begin{array}{l}
\mbox{constraint :} \ \overline{D}_{\ \dot{\alpha}} \ \omega \ = \ 0 
\hspace*{0.2cm} , \hspace*{0.2cm}
D_{\ \alpha} \ \overline{\omega} \ = \ 0
\ \color{red} \rightarrow
\vspace*{0.2cm} \\
\omega \ = \ \omega^{\ A} \ T^{\ A}
\ = \ \left \lbrace \ \begin{array}{l}
\theta^{\ 2} \ H \ ( \ x^{\ -} \ )
\vspace*{0.2cm} \\
+ \ \vartheta^{\ \alpha} \ \eta_{\ \alpha} \ ( \ x^{\ -} \ )
\vspace*{0.2cm} \\
\ + \ \varphi \ ( \ x^{\ -} \ ) 
\end{array} 
\right .
\vspace*{0.2cm} \\
\overline{\omega} \ = \ \overline{\omega}^{\ A} \ T^{\ A}
\vspace*{0.2cm} \\
\overline{\omega}^{\ A} \ =
\ \left \lbrace \ \begin{array}{l}
\overline{\theta}^{\ 2} \ H^{\ * \ A} \ ( \ x^{\ +} \ )
\vspace*{0.2cm} \\
+ \ \overline{\vartheta}_{\ \dot{\alpha}} \ ( \ \eta^{\ *} \ )^{\ \dot{\alpha} \ A} \ ( \ x^{\ +} \ )
\vspace*{0.2cm} \\
\ + \ \varphi^{\ * \ A} \ ( \ x^{\ +} \ ) 
\end{array} 
\right . 
\vspace*{0.2cm} \\
\color{magenta}
\left ( \ x^{\ +} \ \right )^{\ \mu} \ = \ x^{\ \mu} \ + 
\ \frac{i}{2} \ \vartheta^{\ \alpha} \ \sigma^{\ \mu}_{\ \alpha \dot{\beta}} 
\ \overline{\vartheta}^{\ \dot{\beta}}
\hspace*{0.2cm} ; \hspace*{0.2cm}
D_{\ \alpha} \ x^{\ +} \ = \ 0
\end{array}
\end{equation}

\color{blue}

\noindent
Since $( \ \omega_{\ (0)} \ = \ \varphi \ )^{\ A}$ becomes a complex scalar field, the
gauge group is hereby complexified, i.e. extended to complex (infinitesimal) angles.

\vspace*{-0.5cm}
\begin{equation}
\label{eq:67}
\begin{array}{l}
\color{red} \rightarrow
\color{blue}
\ \begin{array}{lll} 
\Omega \ = \ \exp \ \frac{1}{i} \ \omega & ; & \overline{D}_{\ \dot{\alpha}} \ \Omega \ = \ 0
\vspace*{0.2cm} \\
\overline{\Omega} \ = \ \exp \ i \ \overline{\omega} & ; & D_{\ \alpha} \ \overline{\Omega} \ = \ 0
\end{array}
\end{array}
\end{equation} 

}

\newpage

\color{blue}

{\large \bf

\noindent
\begin{center}
\color{red}
$e^{\ V}$ as gauge compensator field
\end{center}

\color{blue}

\noindent
Now we generalize the expressions for the kinetic hermitian superfield in eq. (\ref{eq:45})

\vspace*{-0.5cm}
\begin{equation}
\label{eq:68}
\begin{array}{l}
{\cal{L}}_{\ kin} \ = \ \overline{\Phi} \ e^{\ V} \ \Phi \ =
\ \overline{\Phi}_{\ a} \ \left ( \ e^{\ V} \ \right )^{\ a}_{\ \ b} \ \Phi^{\ b}
\vspace*{0.2cm} \\
{\cal{L}}_{\ kin} \ = \ {\cal{L}}_{\ kin} \ ( \ \overline{\Phi} \ , \ V \ , \ \Phi \ )
\vspace*{0.2cm} \\
\mbox{susy gauge invariance : } \color{red} \rightarrow
\vspace*{0.2cm} \\
\color{blue}
{\cal{L}}_{\ kin} \ ( \ \overline{\Phi^{\ \Omega}} \ , \ V^{\ \Omega} \ , \ \Phi^{\ \Omega} \ )
\ = \ {\cal{L}}_{\ kin} \ ( \ \overline{\Phi} \ , \ V \ , \ \Phi \ )
\vspace*{0.2cm} \\
\Phi^{\ \Omega} \ = \ \Omega \ \Phi
\hspace*{0.2cm} , \hspace*{0.2cm} 
\overline{\Phi^{\ \Omega}} \ = \ \overline{\Phi} \ \overline{\Omega}
\end{array}
\end{equation}

\noindent
It follows for $\exp \ V^{\ \Omega}$

\vspace*{-0.5cm}
\begin{equation}
\label{eq:69}
\begin{array}{l}
\exp \ V^{\ \Omega} \ = \ \exp \ - \ \overline{\Omega} \ \exp \ V \ \exp \ - \ \Omega
\vspace*{0.2cm} \\
\exp \ - \ V^{\ \Omega} \ = \ \exp \ \Omega \ \exp \ - \ V \ \exp \ \overline{\Omega}
\end{array}
\end{equation}

\noindent
It is easier to implement eq. (\ref{eq:69}) for $\exp \ V^{\ \Omega}$ than
for $V^{\ \Omega}$ itself.

\noindent
It follows that the gauge connection (eq. \ref{eq:64}) transforms
covariantly

\vspace*{-0.5cm}
\begin{equation}
\label{eq:70}
\begin{array}{l}
W_{\ \alpha}^{\ \Omega} \ = \ \exp \ \Omega \ W_{\ \alpha} \ \exp \ - \ \Omega
\end{array}
\end{equation}

\noindent
Similarly we have for $\overline{W}_{\ \dot{\alpha}} \ = \ - \ e^{\ V} \ \overline{D}_{\ \dot{\alpha}}
\ e^{\ - \ V}$

}

\newpage

\color{blue}

{\large \bf

\vspace*{-0.5cm}
\begin{equation}
\label{eq:71}
\begin{array}{l}
\overline{W}_{\ \dot{\alpha}} \ = 
\ - \ \exp \ V \ \overline{D}_{\ \dot{\alpha}} \ \exp \ - \ V
\vspace*{0.2cm} \\
\overline{W}_{\ \dot{\alpha}}^{\ \Omega} \ = \ \exp \ - \ \overline{\Omega} 
\ \overline{W}_{\ \dot{\alpha}} \ \exp \ \overline{\Omega}
\hspace*{0.3cm} \color{red} \leftrightarrow
\vspace*{0.2cm} \\
W_{\ \alpha}^{\ \Omega} \ = \ \exp \ \Omega \ W_{\ \alpha} \ \exp \ - \ \Omega
\end{array}
\end{equation}

\noindent
Because $\Omega \ , \ \overline{\Omega}$ are right- and left-chiral superfields
respectively, acting with $\overline{D}_{\ \dot{\gamma}}$ on $W_{\ \alpha}$
and with $D_{\ \gamma}$ on $\overline{W}_{\ \dot{\alpha}}$ does not change the
gauge transformation properties in eq. (\ref{eq:71}) .

\noindent
\begin{center}
\color{red}
chiral projection of the gauge connection
\end{center}

\color{blue}

\noindent
Despite interesting properties of the non-chiral fields
$W_{\ \alpha} \ , \ \overline{W}_{\ \dot{\alpha}}$ we proceed here directly
to their chiral projections

\vspace*{-0.5cm}
\begin{equation}
\label{eq:72}
\begin{array}{l}
w_{\ \alpha} \ = \ \overline{D}^{\ 2} \ W_{\ \alpha}
\hspace*{0.2cm} \color{red} \leftrightarrow \hspace*{0.2cm}
\color{magenta} \overline{w}_{\ \dot{\alpha}} \ = \ D^{\ 2} \ \overline{W}_{\ \dot{\alpha}}
\ \equiv \ \overline{\left ( \ w_{\ \alpha} \ \right )}
\vspace*{0.2cm} \\
\color{blue}
\overline{D}^{\ 2} \ = \frac{1}{2} \ \overline{D}_{\ \dot{\gamma}} \ \overline{D}^{\ \dot{\gamma}}
\hspace*{0.2cm} ; \hspace*{0.2cm}
\overline{D}^{\ \dot{\gamma}} \ = \ - \ \left ( \ \varepsilon^{\ '} \ \right )^{\ \dot{\gamma} \dot{\delta}}
\ \overline{D}_{\ \dot{\delta}}  
\vspace*{0.2cm} \\
\mbox{with : } 
\ \overline{D}_{\ \dot{\gamma}} \ w_{\ \alpha} \ = \ 0
\hspace*{0.2cm} \color{red} \leftrightarrow \hspace*{0.2cm}
\color{magenta} D_{\ \gamma} \ \overline{w}_{\ \dot{\alpha}} \ = \ 0
\end{array}
\end{equation}

\color{blue}
\noindent
Pro memoria lets recall section 4c) and definitions in eq. (\ref{eq:33})

\vspace*{-0.5cm}
\begin{equation}
\label{eq:73}
\begin{array}{l}
\color{magenta}
D_{\ \alpha} \ = \ ( \ \partial_{\ \vartheta} \ )_{\ \alpha}
\ - \ \frac{i}{2} \ \overline{\vartheta}^{\ \dot{\beta}} \ ( \ \partial_{\ x} \ )_{\ \alpha \dot{\beta}}
\vspace*{0.2cm} \\
\color{magenta}
\hspace*{0.7cm}
\overline{D}_{\ \dot{\beta}} \ = \ - \ ( \ \partial_{\ \overline{\vartheta}} \ )_{\ \dot{\beta}}
\ + \ \frac{i}{2} \ \vartheta^{\ \alpha} \ ( \ \partial_{\ x} \ )_{\ \alpha \dot{\beta}}
\end{array}
\end{equation}

}

\newpage

\color{blue}

{\large \bf

\noindent
The chiral fields $w_{\ \alpha}$ ( and $\rightarrow \ \overline{w}_{\ \dot{\alpha}}$ )
become 

\vspace*{-0.5cm}
\begin{equation}
\label{eq:74}
\begin{array}{l}
w_{\ \alpha} \ = \ w_{\ \alpha}^{\ A} \ T^{\ A}
\vspace*{0.2cm} \\
 = \ \left \lbrace  \begin{array}{l}
\vartheta^{\ 2} \ \sigma^{\ \mu}_{\ \alpha \dot{\beta}}
\ \left ( \ i \ \partial_{\ \mu} \ \lambda^{\ * \ \dot{\beta}}
\ - \ \left \lbrack \ v_{\ \mu} \ , \ \lambda^{\ * \ \dot{\beta}} \ \right \rbrack
\ \right )
\vspace*{0.2cm} \\
\ + \ \vartheta_{\ \alpha} \ D \ + \ \vartheta^{\ \gamma}
\ \frac{1}{2} \ \left ( \ \sigma^{\ \mu \nu} \ \right )_{\ \left \lbrace \ \gamma \alpha
\ \right \rbrace} \ {\cal{F}}_{\ \mu \nu} 
\vspace*{0.2cm} \\
\ + \ \lambda_{\ \alpha}
\end{array}
\right \rbrace  ( x^{\ -} )
\vspace*{0.2cm} \\
{\cal{W}}_{\ \mu} \ = \ i \ v_{\ \mu}
\hspace*{0.2cm} ; \hspace*{0.2cm}
{\cal{W}}_{\ \mu} \ = \ i \ v_{\ \mu}^{\ A} \ T^{\ A}
\vspace*{0.2cm} \\
D_{\ \mu} \ = \ \partial_{\ \mu} \ + \ \left \lbrack 
\ {\cal{W}}_{\ \mu} \ , \right .
\vspace*{0.2cm} \\
i \ \partial_{\ \mu} \ \lambda^{\ * \ \dot{\beta}}
\ - \ \left \lbrack \ v_{\ \mu} \ , \ \lambda^{\ * \ \dot{\beta}} \ \right \rbrack
\ = \ i \ D_{\ \mu} \ \lambda^{\ * \ \dot{\beta}}
\vspace*{0.2cm} \\
{\cal{F}}_{\ \mu \nu} \ = \ \partial_{\ \mu} \ {\cal{W}}_{\ \nu}
\ - \  \partial_{\ \nu} \ {\cal{W}}_{\ \mu} \ + 
\ \left \lbrack \ {\cal{W}}_{\ \mu} \ , \ {\cal{W}}_{\ \nu} \ \right \rbrack
\end{array}
\end{equation}

\noindent
In eq. (\ref{eq:74}) we recognize in $D_{\ \mu}$ the 
restricted (i.e. non-susy) gauge covariant differential,
in ${\cal{W}}_{\ \mu}$ the correspoding gauge connection
and in ${\cal{F}}_{\ \mu \nu}$ the associated field strength tensor.
\footnote{\hspace*{0.1cm} \color{red} Which are the (anti-) chiral chains associated with
$w_{\ \alpha} \ , \ \overline{w}_{\ \dot{\alpha}}$ ? ( compare with section 4f) ) .}

\noindent
The ( SL2C- ) spin one matrices
$\left ( \ \sigma^{\ \mu \nu} \ \right )_{\ \left \lbrace \ \gamma \alpha \ \right \rbrace}$ in 
eq. (\ref{eq:74}) are described in the next section ( 5b) ) .  

}

\newpage

\color{blue}

{\large \bf

\begin{center}
\color{red}
5b) The ( SL2C- ) spin one matrices
$\left ( \ \sigma^{\ \mu \nu} \ \right )_{\ \left \lbrace \ \gamma \alpha \ \right \rbrace}$
\end{center}

\color{blue}

\noindent
It is useful to consider the mixed spinor indices

\vspace*{-0.5cm}
\begin{equation}
\label{eq:75}
\begin{array}{l}
\left ( \ \sigma^{\ \mu \nu} \ \right )_{\ \gamma}^{\ \hspace*{0.3cm} \delta}
\ = \ \left ( \ \varepsilon^{\ '} \ \right )^{\ \delta \alpha}
\ \left ( \ \sigma^{\ \mu \nu} \ \right )_{\ \left \lbrace \ \gamma \alpha \ \right \rbrace}
\vspace*{0.2cm} \\
\left ( \ \sigma^{\ \mu} \ \right )_{\ \gamma \dot{\beta}}
\ \left ( \ \sigma^{\ \nu} \ \right )^{\ \dot{\beta} \delta}
\ = \ \eta^{\ \mu \nu} \ \delta_{\ \gamma}^{\ \hspace*{0.3cm} \delta}
\ + \ \left ( \ \sigma^{\ \mu \nu} \ \right )_{\ \gamma}^{\ \hspace*{0.3cm} \delta}
\vspace*{0.2cm} \\
\eta^{\ \mu \nu} \ = \ diag \ ( \ 1 \ , \ - 1 \ , \ - 1 \ , \ - 1 \ )
\vspace*{0.2cm} \\
\left (
\ \sigma^{\ \mu \nu} \ = 
\ \left \lbrace \ \begin{array}{ll}
\sigma_{\ r} & \mbox{for  } ^{\mu} \ = 0 \ , \ ^{\nu} \ = \ r
\vspace*{0.2cm} \\
\frac{1}{i} \ \varepsilon_{\ s t r} \ \sigma_{\ r} & \mbox{for  } ^{\mu} \ = s \ , \ ^{\nu} \ = \ t
\end{array} \right \rbrace
\ \right )_{\ \gamma}^{\ \hspace*{0.3cm} \delta}
\end{array}
\end{equation}

\noindent
Lets consider the dual to $\sigma^{\ \mu \nu} $, dropping the mixed spinor indices for
simplicity

\vspace*{-0.5cm}
\begin{equation}
\label{eq:76}
\begin{array}{l}
\widetilde{\sigma}_{\ \mu \nu} \ = \ \frac{1}{2} \ \varepsilon_{\ \mu \nu \varrho \tau}
\ \sigma^{\ \varrho \tau}
\vspace*{0.2cm} \\
\widetilde{\sigma}^{\ \mu \nu} \ =
\ \left \lbrace \ \begin{array}{ll}
i \ \sigma_{\ r} & \mbox{for  } ^{\mu} \ = 0 \ , \ ^{\nu} \ = \ r
\vspace*{0.2cm} \\
\varepsilon_{\ s t r} \ \sigma_{\ r} & \mbox{for  } ^{\mu} \ = s \ , \ ^{\nu} \ = \ t
\end{array} \right \rbrace                                    
\vspace*{0.2cm} \\
\color{red} \rightarrow
\hspace*{0.3cm} \color{blue}
\widetilde{\sigma}^{\ \mu \nu} \ = \ i \ \sigma^{\ \mu \nu}
\end{array}
\end{equation}

\noindent
$\left ( \ \sigma \ , \ \widetilde{\sigma} \ \right ) \ \rightarrow 
\left ( \ \sigma_{\ R} \ , \ \widetilde{\sigma}_{\ R} \ \right )$
defined in eq. (\ref{eq:75}) bear a (here suppressed)
right chiral suffix (R) . 

}

\newpage

\color{blue}

{\large \bf

\noindent
The corresponding left chiral matrices are

\vspace*{-0.5cm}
\begin{equation}
\label{eq:77}
\begin{array}{l}
\left ( \ \sigma^{\ \mu \nu}_{\ L} \ \right )^{\ \dot{\gamma}}_{\ \hspace*{0.3cm} \dot{\delta}}
\ = \ \left ( \ \varepsilon^{\ '} \ \right )^{\ \dot{\gamma} \dot{\alpha}}
\ \left ( \ \sigma^{\ \mu \nu}_{\ L} \ \right )_{\ \left \lbrace \ \dot{\alpha} \dot{\delta} 
\ \right \rbrace}
\vspace*{0.2cm} \\
\left ( \ \sigma^{\ \mu} \ \right )^{\ \dot{\gamma} \alpha}
\ \left ( \ \sigma^{\ \nu} \ \right )_{\ \alpha \dot{\delta}}
\ = \ \eta^{\ \mu \nu} \ \delta^{\ \dot{\gamma}}_{\ \hspace*{0.3cm} \dot{\delta}}
\ + \ \left ( \ \sigma^{\ \mu \nu}_{\ L} \ \right )^{\ \dot{\gamma}}_{\ \hspace*{0.3cm} \dot{\delta}}
\vspace*{0.2cm} \\
\left (
\ \sigma^{\ \mu \nu}_{\ L} \ = 
\ \left \lbrace \ \begin{array}{ll}
- \ \sigma_{\ r} & \mbox{for  } ^{\mu} \ = 0 \ , \ ^{\nu} \ = \ r
\vspace*{0.2cm} \\
\frac{1}{i} \ \varepsilon_{\ s t r} \ \sigma_{\ r} & \mbox{for  } ^{\mu} \ = s \ , \ ^{\nu} \ = \ t
\end{array} \right \rbrace
\ \right )^{\ \dot{\gamma}}_{\ \hspace*{0.3cm} \dot{\delta}}
\end{array}
\end{equation} 

\noindent
Considering the left chiral dual, analogous to the right chiral one in eq. (\ref{eq:76})
we have, dropping spinor indices

\vspace*{-0.5cm}
\begin{equation}
\label{eq:78}
\begin{array}{l}
\widetilde{\sigma}_{\ \mu \nu \ L} \ = \ \frac{1}{2} \ \varepsilon_{\ \mu \nu \varrho \tau}
\ \sigma^{\ \varrho \tau}_{\ L}
\vspace*{0.2cm} \\
\widetilde{\sigma}^{\ \mu \nu}_{\ L} \ =
\ \left \lbrace \ \begin{array}{ll}
i \ \sigma_{\ r} & \mbox{for  } ^{\mu} \ = 0 \ , \ ^{\nu} \ = \ r
\vspace*{0.2cm} \\
- \ \varepsilon_{\ s t r} \ \sigma_{\ r} & \mbox{for  } ^{\mu} \ = s \ , \ ^{\nu} \ = \ t
\end{array} \right \rbrace                                    
\vspace*{0.2cm} \\
\color{red} \rightarrow
\hspace*{0.3cm} \color{blue}
\widetilde{\sigma}^{\ \mu \nu \ L} \ = \ - \ i \ \sigma^{\ \mu \nu \ L}
\end{array}
\end{equation}

\noindent
Projecting a photon $F_{\ \mu \nu}$ onto $\left ( \ \sigma^{\ \mu \nu} \ \right )_{\ R \ (L)}$
we have \footnote{\hspace*{0.1cm} \color{red} Show that $F \ . \ \sigma_{\ R \ (L)}$
actually projects onto right (R) - and left (L) - circular photon states respectively .}

\vspace*{-0.5cm}
\begin{equation}
\label{eq:79}
\begin{array}{l}
\frac{1}{2} \ F_{\ \mu \nu} 
\ \left ( \ \sigma^{\ \mu \nu} \ \right )_{\ R \ (L)} \ =
\ \left \lbrace \ \begin{array}{ll}
\left ( \ \frac{1}{i} \ \vec{B} \ - \ \vec{E} \ \right ) \ \vec{\sigma}
 & \mbox{for R}
\vspace*{0.2cm} \\
\left ( \ \frac{1}{i} \ \vec{B} \ + \ \vec{E} \ \right ) \ \vec{\sigma}
 & \mbox{for L}
\end{array}
\right .
\end{array}
\end{equation}

}

\newpage

\color{blue}

{\large \bf

\color{red}
\begin{center}
5c) The chiral field strengths multiplets  
$w_{\ \alpha} \ ,  \ \overline{w}_{\ \dot{\alpha}}$ \\
and the chiral Lagrangean multiplet 
${\cal{L}} \ = \ {\cal{N}} \ tr \ w^{\ \alpha} \ w_{\ \alpha} \ , \ {\cal{N}}^{\ -1} \ = 
\ 4 \ C \ ( \ {\cal{D}} \ ) \ g^{\ 2}$ 
\end{center}

\color{blue}

\noindent
We repeat the form of $w_{\ \alpha}$ given in eq. (\ref{eq:74})

\vspace*{-0.5cm}
\begin{equation}
\label{eq:80}
\begin{array}{l}
w_{\ \alpha} \ = \ w_{\ \alpha}^{\ A} \ T^{\ A}
\vspace*{0.2cm} \\
 = \ \left \lbrace  \begin{array}{l}
\vartheta^{\ 2} \ \sigma^{\ \mu}_{\ \alpha \dot{\beta}}
\ i \ D_{\ \mu} \ \lambda^{\ * \ \dot{\beta}}
\vspace*{0.2cm} \\
\ + \ \vartheta_{\ \alpha} \ D \ + \ \vartheta^{\ \gamma}
\ \frac{1}{2} \ \left ( \ \sigma^{\ \mu \nu} \ \right )_{\ \left \lbrace \ \gamma \alpha
\ \right \rbrace} \ {\cal{F}}_{\ \mu \nu} 
\vspace*{0.2cm} \\
\ + \ \lambda_{\ \alpha}
\end{array}
\right \rbrace  ( x^{\ -} )
\vspace*{0.2cm} \\
{\cal{W}}_{\ \mu} \ = \ i \ v_{\ \mu}
\hspace*{0.2cm} ; \hspace*{0.2cm}
{\cal{W}}_{\ \mu} \ = \ i \ v_{\ \mu}^{\ A} \ T^{\ A}
\vspace*{0.2cm} \\
D_{\ \mu} \ = \ \partial_{\ \mu} \ + \ \left \lbrack 
\ {\cal{W}}_{\ \mu} \ , \right .
\vspace*{0.2cm} \\
i \ \partial_{\ \mu} \ \lambda^{\ * \ \dot{\beta}}
\ - \ \left \lbrack \ v_{\ \mu} \ , \ \lambda^{\ * \ \dot{\beta}} \ \right \rbrack
\ = \ i \ D_{\ \mu} \ \lambda^{\ * \ \dot{\beta}}
\vspace*{0.2cm} \\
{\cal{F}}_{\ \mu \nu} \ = \ \partial_{\ \mu} \ {\cal{W}}_{\ \nu}
\ - \  \partial_{\ \nu} \ {\cal{W}}_{\ \mu} \ + 
\ \left \lbrack \ {\cal{W}}_{\ \mu} \ , \ {\cal{W}}_{\ \nu} \ \right \rbrack
\end{array}
\end{equation}

\begin{center}
\color{red}
the adjoint representation
\end{center}

\color{blue}

\noindent
Let the structure constants be defined through the (general irreducible) representation
formed by the matrices $T^{\ A}$

}

\newpage

\color{blue}

{\large \bf

\vspace*{-0.5cm}
\begin{equation}
\label{eq:81}
\begin{array}{l}
\left \lbrack \ T^{\ A} \ , \ T^{\ B} \ \right \rbrack
\ = \ i \ f_{\ A B C} \ T^{\ C}
\hspace*{0.3cm} \color{red} \rightarrow
\vspace*{0.2cm} \\
\color{blue}
\left ( \ {\cal{F}}^{\ A} \ \right )_{\ B C} \ = i \ f_{\ B A C}
\hspace*{0.2cm} : \hspace*{0.1cm} 
\vspace*{0.2cm} \\
\left \lbrack \ {\cal{F}}^{\ A} \ , \ {\cal{F}}^{\ B} \ \right \rbrack
\ = \ i \ f_{\ A B C} \ {\cal{F}}^{\ C}
\hspace*{0.2cm} : \hspace*{0.2cm}
{\cal{D}} \ = \ {\cal{F}}
\end{array}
\end{equation} 

\noindent
The matrices ${\cal{F}}^{\ A}$ form the adjoint representation of The Lie algebra
of G.

\noindent
The last relation in eq. (\ref{eq:81}) is equivalent to the Jacobi identity for
any triple commutator, e.g.

\vspace*{-0.5cm}
\begin{equation}
\label{eq:82}
\begin{array}{l}
\left \lbrack \ T^{\ A} \ , \ \left \lbrack \ T^{\ B} \ , \ T^{\ C} \ \right \rbrack
\ \right \rbrack \ \equiv \ {\cal{T}}^{\ A B C}
\hspace*{0.2cm} \color{red} \rightarrow
\vspace*{0.2cm} \\
\color{blue}
{\cal{T}}^{\ A B C} \ + \ {\cal{T}}^{\ C A B } \ + \ {\cal{T}}^{\ B C A} \ = \ 0 
\end{array}
\end{equation}

\noindent
For any irreducible representation ${\cal{D}}$ it follows that in an appropriate basis
for the matrices $T^{\ A} \ , \ {\cal{D}}$ we have

\vspace*{-0.5cm}
\begin{equation}
\label{eq:83}
\begin{array}{l}
tr \ T^{\ A} \ T^{\ B} \ = \ C \ ( \ {\cal{D}} \ ) \ \delta_{\ A B}
\end{array}
\end{equation} 

\noindent
The normalization constant $C \ ( \ {\cal{D}} \ )$ is related to the second
Casimir invariant pertinent to the representation ${\cal{D}}$

}

\newpage

\color{blue}

{\large \bf

\vspace*{-0.5cm}
\begin{equation}
\label{eq:84}
\begin{array}{l}
\sum_{\ A} \ \left ( \ T^{\ A} \ T^{\ A} \ \right )_{b c} \ = \ C_{\ 2} \ ( \ {\cal{D}} \ ) 
\ \delta_{\ b c}
\vspace*{0.2cm} \\
b \ , \ c \ = \ 1 \ , \ \cdots \ , \ dim \ {\cal{D}}
\hspace*{0.2cm} \color{red} \rightarrow
\vspace*{0.2cm} \\
\color{blue}
C \ ( \ {\cal{D}} \ ) \ = \ C_{\ 2} \ ( \ {\cal{D}} \ ) \ dim \ {\cal{D}}
\ / \ dim \ G
\hspace*{0.2cm} \color{red} \rightarrow
\vspace*{0.2cm} \\
\color{blue}
C \ ( \ {\cal{F}} \ ) \ = \ C_{\ 2} \ ( \ {\cal{F}} \ )
\end{array}
\end{equation}

The absolute normalization of the structure constants $f_{\ A B C}$ is subject to
{\it convention} which is illustrated in the section 5d) below,
while ratios 

\vspace*{-0.5cm}
\begin{equation}
\label{eq:85}
\begin{array}{l}
C_{\ 2} \ ( \ {\cal{D}} \ ) \ / \ C_{\ 2} \ ( \ {\cal{F}} \ )
\end{array}
\end{equation}

\noindent
are structural invariants.

%\documentstyle[,12pt]{article}

%\begin{document}

%\thispagestyle{empty}

\color{blue}
\begin{center}
{\bf \large \color{red} 5d)  Some numbers for the simple compact Lie groups}
\end{center}
\vspace*{0.5cm}

\color{blue}

\begin{center}
{\bf \small
P. Minkowski, 3. March, 1994} 
\end{center}
\vspace*{0.5cm}

{\bf \small

\noindent
Below I list the following characteristic numbers for the simple 
compact Lie groups  

\begin{displaymath}
\begin{array}{c}
\mbox{group :} \ \mbox{G}
\hspace*{0.4cm} , \hspace*{0.4cm}  
\mbox{dimension :} \ d
\hspace*{0.4cm} , \hspace*{0.4cm}  
\mbox{rank :} \ r 
\vspace*{0.3cm} \\
\mbox{dimension / rank :} \ \delta \ = \ d \ / \ r
\hspace*{0.1cm} , \hspace*{0.1cm}  
\mbox{value of Casimir operator :} \ C_{\ 2}
\vspace*{0.3cm} \\
%\hspace*{0.4cm} , \hspace*{0.4cm}  
\mbox{characteristic number :} \ \nu \ = \delta \ - \ C_{\ 2}
\end{array}
\end{displaymath}

\noindent
The simple compact Lie groups are denoted according to 
mathematical conventions 

\begin{equation}
\label{geq:1}
\begin{array}{c}
\mbox{ classical groups}
\vspace*{0.4cm} \\
\mbox{A}_{\ n}
\ =
\ \mbox{SU n+1}
\hspace*{0.4cm} , \hspace*{0.4cm}  
\mbox{B}_{\ n}
\ =
\ \mbox{SO 2n+1}
\hspace*{0.4cm} , \hspace*{0.4cm}  
n \ \ge \ 3
\vspace*{0.4cm} \\
\mbox{C}_{\ n}
\ =
\ \mbox{Sp n}
\hspace*{0.4cm} , \hspace*{0.4cm}  
n \ \ge \ 2
\hspace*{0.4cm} ; \hspace*{0.4cm}  
\mbox{D}_{\ n}
\ =
\ \mbox{SO 2n}
\hspace*{0.4cm} , \hspace*{0.4cm}  
n \ \ge \ 4
\vspace*{0.4cm} \\
\mbox{ exceptional groups}
\vspace*{0.4cm} \\
\mbox{G}_{\ 2}
\hspace*{0.4cm} , \hspace*{0.4cm}  
\mbox{F}_{\ 4}
\hspace*{0.4cm} , \hspace*{0.4cm}  
\mbox{E}_{\ 6}
\hspace*{0.4cm} , \hspace*{0.4cm}  
\mbox{E}_{\ 7}
\hspace*{0.4cm} , \hspace*{0.4cm}  
\mbox{E}_{\ 8}
\hspace*{0.8cm} \longrightarrow 
\end{array}
\end{equation}
\vspace*{0.3cm} 

}

\newpage

\color{blue}

{\small \bf

\begin{equation}
\label{geq:2}
\begin{array}{|@{\hspace*{0.4cm}}c@{\hspace*{0.4cm}}| 
@{\hspace*{0.4cm}}c@{\hspace*{0.4cm}} 
@{\hspace*{0.4cm}}c@{\hspace*{0.4cm}} 
@{\hspace*{0.4cm}}c@{\hspace*{0.4cm}} 
@{\hspace*{0.4cm}}c@{\hspace*{0.4cm}} 
@{\hspace*{0.2cm}}r@{\hspace*{0.7cm}}|} 
\hline
 & & & & & \\
 \mbox{G} & d & r & \delta & C_{\ 2} & \nu \\
 & & & & & \\
 \hline
 & & & & & \\
\mbox{A}_{\ n} & 
n \ ( \ n \ + \ 2 \ ) & 
n &
n \ + \ 2 &
n \ + \ 1 &
1 \\
 & & & & & \\
\mbox{B}_{\ n} &
n \ ( \ 2n \ + \ 1 \ ) & 
n &
2n \ + \ 1 &
2n \ - \ 1 &
2 \\
 & & & & & \\
\mbox{C}_{\ n} &
n \ ( \ 2n \ + \ 1 \ ) & 
n &
2n \ + \ 1 &
2n \ + \ 2 &
- 1 \\
 & & & & & \\
\mbox{D}_{\ n} &
n \ ( \ 2n \ - \ 1 \ ) & 
n &
2n \ - \ 1 &
2n \ - \ 2 &
 1 \\
 & & & & & \\
\mbox{G}_{\ 2} &
14 & 
2 &
7 &
4 &
 3 \\
 & & & & & \\
\mbox{F}_{\ 4} &
52 & 
4 &
13 &
9 &
 4 \\
 & & & & & \\
\mbox{E}_{\ 6} &
78 & 
6 &
13 &
12 &
 1 \\
 & & & & & \\
\mbox{E}_{\ 7} &
133 & 
7 &
19 &
18 &
 1 \\
 & & & & & \\
\mbox{E}_{\ 8} &
248 & 
8 &
31 &
30 &
 1 \\
 & & & & & \\
 \hline
%\hspace*{0.4cm} , \hspace*{0.4cm}  
%\vspace*{0.6cm} \\
\end{array}
\end{equation}
\vspace*{0.3cm} 

\noindent
We give the entriesd in eq. ( \ref{geq:2} ) for the ranks of the
exceptional groups : 2 , 4 , 6 , 7 , 8 :

}

\newpage

\color{blue}

{\small \bf

\begin{equation}
\label{geq:3}
\begin{array}{|@{\hspace*{0.4cm}}c@{\hspace*{0.4cm}}| 
@{\hspace*{0.4cm}}c@{\hspace*{0.4cm}} 
@{\hspace*{0.4cm}}c@{\hspace*{0.4cm}} 
@{\hspace*{0.4cm}}c@{\hspace*{0.4cm}} 
@{\hspace*{0.4cm}}c@{\hspace*{0.4cm}} 
@{\hspace*{0.2cm}}r@{\hspace*{0.7cm}}|} 
\hline
 & & & & & \\
 \mbox{G} & d & r & \delta & C_{\ 2} & \nu \\
 & & & & & \\
 \hline
 & & & & & \\
\mbox{A}_{\ 2} & 
%\mbox{A}_{\ n} & 
8 & 
%n \ ( \ n \ + \ 2 \ ) & 
2 &
%n &
4 &
%n \ + \ 2 &
3 &
%n \ + \ 1 &
1 \\
% & & & & & \\
%\mbox{B}_{\ n} &
%n \ ( \ 2n \ + \ 1 \ ) & 
%n &
%2n \ + \ 1 &
%2n \ - \ 1 &
%2 \\
 & & & & & \\
\mbox{C}_{\ 2} &
%\mbox{C}_{\ n} &
10 & 
%n \ ( \ 2n \ + \ 1 \ ) & 
2 &
%n &
5 &
%2n \ + \ 1 &
6 &
%2n \ + \ 2 &
- 1 \\
% & & & & & \\
%\mbox{D}_{\ n} &
%n \ ( \ 2n \ - \ 1 \ ) & 
%n &
%2n \ - \ 1 &
%2n \ - \ 2 &
% 1 \\
 & & & & & \\
\mbox{G}_{\ 2} &
14 & 
2 &
7 &
4 &
 3 \\
% & & & & & \\
%\mbox{F}_{\ 4} &
%52 & 
%4 &
%13 &
%9 &
% 4 \\
% & & & & & \\
%\mbox{E}_{\ 6} &
%78 & 
%6 &
%13 &
%12 &
% 1 \\
% & & & & & \\
%\mbox{E}_{\ 7} &
%133 & 
%7 &
%19 &
%18 &
% 1 \\
% & & & & & \\
%\mbox{E}_{\ 8} &
%248 & 
%8 &
%31 &
%30 &
% 1 \\
 & & & & & \\
 \hline
%\hspace*{0.4cm} , \hspace*{0.4cm}  
%\vspace*{0.6cm} \\
\end{array}
\end{equation}
\vspace*{0.3cm} 

\begin{equation}
\label{geq:4}
\begin{array}{|@{\hspace*{0.4cm}}c@{\hspace*{0.4cm}}| 
@{\hspace*{0.4cm}}c@{\hspace*{0.4cm}} 
@{\hspace*{0.4cm}}c@{\hspace*{0.4cm}} 
@{\hspace*{0.4cm}}c@{\hspace*{0.4cm}} 
@{\hspace*{0.4cm}}c@{\hspace*{0.4cm}} 
@{\hspace*{0.2cm}}r@{\hspace*{0.7cm}}|} 
\hline
 & & & & & \\
 \mbox{G} & d & r & \delta & C_{\ 2} & \nu \\
 & & & & & \\
 \hline
 & & & & & \\
\mbox{A}_{\ 4} & 
%\mbox{A}_{\ n} & 
24 & 
%n \ ( \ n \ + \ 2 \ ) & 
4 &
%n &
6 &
%n \ + \ 2 &
5 &
%n \ + \ 1 &
1 \\
 & & & & & \\
\mbox{B}_{\ 4} &
%\mbox{B}_{\ n} &
36 & 
%n \ ( \ 2n \ + \ 1 \ ) & 
4 &
%n &
9 &
%2n \ + \ 1 &
7 &
%2n \ - \ 1 &
2 \\
 & & & & & \\
\mbox{C}_{\ 4} &
%\mbox{C}_{\ n} &
36 & 
%n \ ( \ 2n \ + \ 1 \ ) & 
4 &
%n &
9 &
%2n \ + \ 1 &
10 &
%2n \ + \ 2 &
- 1 \\
 & & & & & \\
\mbox{D}_{\ 4} &
%\mbox{D}_{\ n} &
28 & 
%n \ ( \ 2n \ - \ 1 \ ) & 
4 &
%n &
7 &
%2n \ - \ 1 &
6 &
%2n \ - \ 2 &
 1 \\
% & & & & & \\
%\mbox{G}_{\ 2} &
%14 & 
%2 &
%7 &
%4 &
% 3 \\
 & & & & & \\
\mbox{F}_{\ 4} &
52 & 
4 &
13 &
9 &
 4 \\
% & & & & & \\
%\mbox{E}_{\ 6} &
%78 & 
%6 &
%13 &
%12 &
% 1 \\
% & & & & & \\
%\mbox{E}_{\ 7} &
%133 & 
%7 &
%19 &
%18 &
% 1 \\
% & & & & & \\
%\mbox{E}_{\ 8} &
%248 & 
%8 &
%31 &
%30 &
% 1 \\
 & & & & & \\
 \hline
%\hspace*{0.4cm} , \hspace*{0.4cm}  
%\vspace*{0.6cm} \\
\end{array}
\end{equation}
\vspace*{0.3cm} 

}

\newpage

\color{blue}

{\small \bf

\begin{equation}
\label{geq:5}
\begin{array}{|@{\hspace*{0.4cm}}c@{\hspace*{0.4cm}}| 
@{\hspace*{0.4cm}}c@{\hspace*{0.4cm}} 
@{\hspace*{0.4cm}}c@{\hspace*{0.4cm}} 
@{\hspace*{0.4cm}}c@{\hspace*{0.4cm}} 
@{\hspace*{0.4cm}}c@{\hspace*{0.4cm}} 
@{\hspace*{0.2cm}}r@{\hspace*{0.7cm}}|} 
\hline
 & & & & & \\
 \mbox{G} & d & r & \delta & C_{\ 2} & \nu \\
 & & & & & \\
 \hline
 & & & & & \\
\mbox{A}_{\ 6} & 
%\mbox{A}_{\ n} & 
48 & 
%n \ ( \ n \ + \ 2 \ ) & 
6 &
%n &
8 &
%n \ + \ 2 &
7 &
%n \ + \ 1 &
1 \\
 & & & & & \\
\mbox{B}_{\ 6} &
%\mbox{B}_{\ n} &
78 & 
%n \ ( \ 2n \ + \ 1 \ ) & 
6 &
%n &
13 &
%2n \ + \ 1 &
11 &
%2n \ - \ 1 &
2 \\
 & & & & & \\
\mbox{C}_{\ 6} &
%\mbox{C}_{\ n} &
78 & 
%n \ ( \ 2n \ + \ 1 \ ) & 
6 &
%n &
13 &
%2n \ + \ 1 &
14 &
%2n \ + \ 2 &
- 1 \\
 & & & & & \\
\mbox{D}_{\ 6} &
%\mbox{D}_{\ n} &
66 & 
%n \ ( \ 2n \ - \ 1 \ ) & 
6 &
%n &
11 &
%2n \ - \ 1 &
10 &
%2n \ - \ 2 &
 1 \\
% & & & & & \\
%\mbox{G}_{\ 2} &
%14 & 
%2 &
%7 &
%4 &
% 3 \\
% & & & & & \\
%\mbox{F}_{\ 4} &
%52 & 
%4 &
%13 &
%9 &
% 4 \\
 & & & & & \\
\mbox{E}_{\ 6} &
78 & 
6 &
13 &
12 &
 1 \\
% & & & & & \\
%\mbox{E}_{\ 7} &
%133 & 
%7 &
%19 &
%18 &
% 1 \\
% & & & & & \\
%\mbox{E}_{\ 8} &
%248 & 
%8 &
%31 &
%30 &
% 1 \\
 & & & & & \\
 \hline
%\hspace*{0.4cm} , \hspace*{0.4cm}  
%\vspace*{0.6cm} \\
\end{array}
\end{equation}
\vspace*{0.3cm} 

}

\newpage

\color{blue}

{\small \bf

\begin{equation}
\label{geq:6}
\begin{array}{|@{\hspace*{0.4cm}}c@{\hspace*{0.4cm}}| 
@{\hspace*{0.4cm}}c@{\hspace*{0.4cm}} 
@{\hspace*{0.4cm}}c@{\hspace*{0.4cm}} 
@{\hspace*{0.4cm}}c@{\hspace*{0.4cm}} 
@{\hspace*{0.4cm}}c@{\hspace*{0.4cm}} 
@{\hspace*{0.2cm}}r@{\hspace*{0.7cm}}|} 
\hline
 & & & & & \\
 \mbox{G} & d & r & \delta & C_{\ 2} & \nu \\
 & & & & & \\
 \hline
 & & & & & \\
\mbox{A}_{\ 7} & 
%\mbox{A}_{\ n} & 
63 & 
%n \ ( \ n \ + \ 2 \ ) & 
7 &
%n &
9 &
%n \ + \ 2 &
8 &
%n \ + \ 1 &
1 \\
 & & & & & \\
\mbox{B}_{\ 7} &
%\mbox{B}_{\ n} &
105 & 
%n \ ( \ 2n \ + \ 1 \ ) & 
7 &
%n &
15 &
%2n \ + \ 1 &
13 &
%2n \ - \ 1 &
2 \\
 & & & & & \\
\mbox{C}_{\ 7} &
%\mbox{C}_{\ n} &
105 & 
%n \ ( \ 2n \ + \ 1 \ ) & 
7 &
%n &
15 &
%2n \ + \ 1 &
16 &
%2n \ + \ 2 &
- 1 \\
 & & & & & \\
\mbox{D}_{\ 7} &
%\mbox{D}_{\ n} &
91 & 
%n \ ( \ 2n \ - \ 1 \ ) & 
7 &
%n &
13 &
%2n \ - \ 1 &
12 &
%2n \ - \ 2 &
 1 \\
% & & & & & \\
%\mbox{G}_{\ 2} &
%14 & 
%2 &
%7 &
%4 &
% 3 \\
% & & & & & \\
%\mbox{F}_{\ 4} &
%52 & 
%4 &
%13 &
%9 &
% 4 \\
% & & & & & \\
%\mbox{E}_{\ 6} &
%78 & 
%6 &
%13 &
%12 &
% 1 \\
 & & & & & \\
\mbox{E}_{\ 7} &
133 & 
7 &
19 &
18 &
 1 \\
% & & & & & \\
%\mbox{E}_{\ 8} &
%248 & 
%8 &
%31 &
%30 &
% 1 \\
 & & & & & \\
 \hline
%\hspace*{0.4cm} , \hspace*{0.4cm}  
%\vspace*{0.6cm} \\
\end{array}
\end{equation}
\vspace*{0.3cm} 

}

\newpage

\color{blue}

{\small \bf

\begin{equation}
\label{geq:7}
\begin{array}{|@{\hspace*{0.4cm}}c@{\hspace*{0.4cm}}| 
@{\hspace*{0.4cm}}c@{\hspace*{0.4cm}} 
@{\hspace*{0.4cm}}c@{\hspace*{0.4cm}} 
@{\hspace*{0.4cm}}c@{\hspace*{0.4cm}} 
@{\hspace*{0.4cm}}c@{\hspace*{0.4cm}} 
@{\hspace*{0.2cm}}r@{\hspace*{0.7cm}}|} 
\hline
 & & & & & \\
 \mbox{G} & d & r & \delta & C_{\ 2} & \nu \\
 & & & & & \\
 \hline
 & & & & & \\
\mbox{A}_{\ 8} & 
%\mbox{A}_{\ n} & 
80 & 
%n \ ( \ n \ + \ 2 \ ) & 
8 &
%n &
10 &
%n \ + \ 2 &
9 &
%n \ + \ 1 &
1 \\
 & & & & & \\
\mbox{B}_{\ 8} &
%\mbox{B}_{\ n} &
136 & 
%n \ ( \ 2n \ + \ 1 \ ) & 
8 &
%n &
17 &
%2n \ + \ 1 &
15 &
%2n \ - \ 1 &
2 \\
 & & & & & \\
\mbox{C}_{\ 8} &
%\mbox{C}_{\ n} &
136 & 
%n \ ( \ 2n \ + \ 1 \ ) & 
8 &
%n &
17 &
%2n \ + \ 1 &
18 &
%2n \ + \ 2 &
- 1 \\
 & & & & & \\
\mbox{D}_{\ 8} &
%\mbox{D}_{\ n} &
120 & 
%n \ ( \ 2n \ - \ 1 \ ) & 
8 &
%n &
15 &
%2n \ - \ 1 &
14 &
%2n \ - \ 2 &
 1 \\
% & & & & & \\
%\mbox{G}_{\ 2} &
%14 & 
%2 &
%7 &
%4 &
% 3 \\
% & & & & & \\
%\mbox{F}_{\ 4} &
%52 & 
%4 &
%13 &
%9 &
% 4 \\
% & & & & & \\
%\mbox{E}_{\ 6} &
%78 & 
%6 &
%13 &
%12 &
% 1 \\
% & & & & & \\
%\mbox{E}_{\ 7} &
%133 & 
%7 &
%19 &
%18 &
% 1 \\
 & & & & & \\
\mbox{E}_{\ 8} &
248 & 
8 &
31 &
30 &
 1 \\
 & & & & & \\
 \hline
%\hspace*{0.4cm} , \hspace*{0.4cm}  
%\vspace*{0.6cm} \\
\end{array}
\end{equation}
\vspace*{0.3cm} 

\noindent
We note, that \'{E}lie Cartan \cite{Cartan}
constructed the large exceptional groups first
from his alphabetical ordering A , B , C , D , E , F , G .

}

%\end{document}

}

\newpage

\color{blue}

{\large \bf

\color{red}
\begin{center}
5c) continued,
$w_{\ \alpha} \ ,  \ \overline{w}_{\ \dot{\alpha}}$ \\
and the chiral Lagrangean multiplet
${\cal{L}} \ = \ {\cal{N}} \ tr \ w^{\ \alpha} \ w_{\ \alpha} \ , \ {\cal{N}}^{\ -1} \ = 
\ 4 \ C \ ( \ {\cal{D}} \ ) \ g^{\ 2}$
\end{center}

\color{blue}
\noindent
We consider the $\vartheta^{\ 2}$ component of the chiral bilinear

\vspace*{-0.5cm}
\begin{equation}
\label{eq:93}
\begin{array}{l}
\left ( \ {\cal{N}}^{\ -1} \ {\cal{L}} \ = \ tr \ w^{\ \alpha} \ w_{\ \alpha} 
\ \right )_{\ \vartheta^{\ 2}} \ =
\vspace*{0.2cm} \\
2 \ C \ ( \ {\cal{D}} \ ) \ \vartheta^{\ 2} \ \times
\vspace*{0.2cm} \\
\times \ \left \lbrack \ \begin{array}{c}
\lambda^{\ * \ B}_{\ \dot{\beta}}
\ \left \lbrace \ \begin{array}{c}
i \ \stackrel{\leftharpoondown}{\partial}_{\ \mu} \ \delta_{\ B A}
\vspace*{0.2cm} \\
+ \ i \ {\cal{W}}_{\ \mu}^{\ C} \ \left ( \ {\cal{F}}^{\ C} \ \right )_{\ B A}
\end{array}                     
\right \rbrace
\ \sigma^{\ \mu \ \dot{\beta} \alpha} \ \lambda^{\ A}_{\ \alpha}
\vspace*{0.2cm} \\
+ \ D^{\ A} \ D^{\ A}
\vspace*{0.2cm} \\
- \ \left ( \ \vec{B}^{\ A} \ - i \ \vec{E}^{\ A} \ \right )
\ \left ( \ \vec{B}^{\ A} \ - i \ \vec{E}^{\ A} \ \right )
\end{array}
\ \right \rbrack
\end{array}
\end{equation}

defining the hermition field strengths $F_{\ \mu \nu} \ \rightarrow \ F_{\ \mu \nu}^{\ A}$
and associated electric $E^{\ A \ k} \ \rightarrow \ \vec{E}^{\ A}$ and magnetic 
$B^{\ A \ k} \ \rightarrow \ \vec{B}^{\ A}$ ( $^{k} \ = \ 1,2,3$ ) field components

\vspace*{-0.5cm}
\begin{equation}
\label{eq:94}
\begin{array}{l}
i \ {\cal{F}}_{\ \mu \nu} \ = \ \partial_{\ \nu} \ v_{\ \mu}
\ - \  \partial_{\ \mu} \ v_{\ \nu} \ + 
\ i \ \left \lbrack \ v_{\ \nu} \ , \ v_{\ \mu} \ \right \rbrack
\ = \ F_{\ \mu \nu} 
\vspace*{0.2cm} \\
F_{\ \mu \nu} \ = \ F_{\ \mu \nu}^{\ A} \ T^{\ A}
\vspace*{0.2cm} \\
F_{\ \mu \nu}^{\ A} \ = \ \partial_{\ \nu} \ v_{\ \mu}^{\ A}
\ - \  \partial_{\ \mu} \ v_{\ \nu}^{\ A} \ -
\ f_{\ A B C} \ v_{\ \nu}^{\ B} \ v_{\ \mu}^{\ C} 
\vspace*{0.2cm} \\
E^{\ A \ k} \ = \ F^{\ A \ 0 k}
\hspace*{0.2cm} , \hspace*{0.2cm}
B^{\ A \ k} \ = \ \frac{1}{2} \ \varepsilon_{\ k m n} \ F^{\ A \ m n}
\end{array}
\end{equation}

}

\newpage

\color{blue}

{\large \bf

\noindent
The symbol $X \ \stackrel{\leftharpoondown}{\partial}_{\ \mu} $ in eq. (\ref{eq:93})
denote the derivative acting with a minus sign to the left

\vspace*{-0.5cm}
\begin{equation}
\label{eq:95}
\begin{array}{l}
X \ \stackrel{\leftharpoondown}{\partial}_{\ \mu} \ = \ - \ \partial_{\ \mu} \ X
\end{array}
\end{equation}

\noindent
We rescale the fermion fields $\lambda_{\ \alpha}^{\ A} \ = \ \sqrt{2} \ \Lambda_{\ \alpha}^{\ A}$
and cast eq. (\ref{eq:93}) to the form

\vspace*{-0.5cm}
\begin{equation}
\label{eq:96}
\begin{array}{l}
\left ( \ {\cal{N}}^{\ -1} \ {\cal{L}} \ = \ tr \ w^{\ \alpha} \ w_{\ \alpha} 
\ \right )_{\ \vartheta^{\ 2}} \ =
\vspace*{0.2cm} \\
4 \ C \ ( \ {\cal{D}} \ ) \ \vartheta^{\ 2} \ \times
\vspace*{0.2cm} \\
\times \ \left \lbrack \ \begin{array}{c}
\Lambda^{\ * \ B}_{\ \dot{\beta}}
\ \left \lbrace \ \begin{array}{c}
\frac{i}{2} \ \stackrel{\leftharpoondown \hspace*{-0.3cm} \rightharpoonup}{\partial}_{\ \mu} 
\ \delta_{\ B A}
\vspace*{0.2cm} \\
+ \ i \ {\cal{W}}_{\ \mu}^{\ C} \ \left ( \ {\cal{F}}^{\ C} \ \right )_{\ B A}
\end{array}                     
\right \rbrace
\ \sigma^{\ \mu \ \dot{\beta} \alpha} \ \Lambda^{\ A}_{\ \alpha}
\vspace*{0.2cm} \\
\hspace*{0.5cm} \color{red}
- \ \frac{i}{2} \ \partial_{\ \mu}
\ \left ( \ \Lambda^{\ * \ A}_{\ \dot{\beta}} \ \sigma^{\ \mu \ \dot{\beta} \alpha}
\ \Lambda^{\ A}_{\ \alpha} \ \right )
\vspace*{0.2cm} \\
\color{blue}
+ \ \frac{1}{2} \ D^{\ A} \ D^{\ A}
\vspace*{0.2cm} \\
+ \ \frac{1}{2} \ \left ( \ \vec{E}^{\ A} \ \vec{E}^{\ A} \ - \ \vec{B}^{\ A} \ \vec{B}^{\ A} \ \right )
\vspace*{0.2cm} \\
\hspace*{0.5cm} \color{red} + \ i \ \left ( \ \vec{E}^{\ A} \ \vec{B}^{\ A} \ \right )
\end{array}
\color{blue}
\ \right \rbrack
\end{array}
\end{equation}
 
\color{blue}

\noindent
The {\color{red} red} entries to ${\cal{L}}$ in eq. (\ref{eq:96}) are antihermitian ,
whereas the remaining ones are hermitian.

\noindent
We rewrite the Lagrangean multiplet component in eq. (\ref{eq:96}) in terms of
the field strength tensor and the overall chiral fermion current
$j^{\ \mu}_{\ \Lambda} \ = \ \Lambda^{\ * \ A}_{\ \dot{\beta}} \ \sigma^{\ \mu \ \dot{\beta} \alpha}
\ \Lambda^{\ A}_{\ \alpha}$ 

}

\newpage

\color{blue}

{\large \bf

\vspace*{-0.5cm}
\begin{equation}
\label{eq:97}
\begin{array}{l}
\Phi \ =
\ \left ( \ 4 \ C \ ( \ {\cal{D}} \ ) \ \right )^{\ -1} \ tr \ w^{\ \alpha} \ w_{\ \alpha} 
\ =
\vspace*{0.4cm} \\
\left \lbrace \begin{array}{l}
\vartheta^{\ 2}
 \left \lbrack \ \begin{array}{c}
\Lambda^{\ * \ B}_{\ \dot{\beta}}
\ \left \lbrace \ \begin{array}{c}
\frac{i}{2} \ \stackrel{\leftharpoondown \hspace*{-0.3cm} \rightharpoonup}{\partial}_{\ \mu} 
\ \delta_{\ B A}
\vspace*{0.2cm} \\
- \ v_{\ \mu}^{\ C} \ \left ( \ {\cal{F}}^{\ C} \ \right )_{\ B A}
\end{array}                     
\right \rbrace
\ \sigma^{\ \mu \ \dot{\beta} \alpha} \ \Lambda^{\ A}_{\ \alpha}
\vspace*{0.2cm} \\
\hspace*{0.5cm} \color{red}
- \ \frac{i}{2} \ \partial_{\ \mu} \ j^{\ \mu}_{\ \Lambda} 
\vspace*{0.2cm} \\
\color{blue}
+ \ \frac{1}{2} \ D^{\ A} \ D^{\ A}
\vspace*{0.2cm} \\
- \ \frac{1}{4} \ F^{\ \mu \nu \ A} \ \left ( \ F_{\ \mu \nu}^{\ A} 
\ \color{red} \ - \ i \ \widetilde{F}_{\ \mu \nu}^{\ A} \ \color{blue} \right )
\end{array}
\color{blue}
 \right \rbrack
\vspace*{0.2cm} \\
+ \ \vartheta^{\ \alpha} \ \frac{1}{\sqrt{2}}
\left \lbrack \ D^{\ A} \ \Lambda^{\ A}_{\ \alpha} \ +
\ {\cal{F}}_{\ \alpha}^{\ A \ \gamma} 
 \Lambda^{\ A}_{\ \gamma}
\ \right \rbrack
\vspace*{0.2cm} \\
+ \ \frac{1}{2} \ \Lambda^{\ A \ \alpha} \ \Lambda^{\ A}_{\ \alpha}
\end{array} \right .
\vspace*{0.4cm} \\ 
\hline \vspace*{-0.2cm} \\
\color{red}
j^{\ \mu}_{\ \Lambda} \ = \ \Lambda^{\ * \ A}_{\ \dot{\beta}} \ \sigma^{\ \mu \ \dot{\beta} \alpha}
\ \Lambda^{\ A}_{\ \alpha}
\hspace*{0.2cm} \color{blue} ; \hspace*{0.2cm} \color{blue}
{\cal{F}}_{\ \alpha}^{\ A \ \gamma} \ = \ \frac{1}{2} \ {\cal{F}}_{\ \mu \nu}
\ \left ( \ \sigma^{\ \mu \nu} \ \right )_{\ \alpha}^{\hspace*{0.3cm} \gamma}
\vspace*{0.4cm} \\
\color{blue}
X \ \stackrel{\leftharpoondown \hspace*{-0.3cm} \rightharpoonup}{\partial}_{\ \mu} \ Y
\ = \ - \ ( \ \partial_{\ \mu} \ X \ ) \ Y \ + \ X \ ( \ \partial_{\ \mu} \ Y \ ) 
\end{array}
\end{equation}

\color{blue}

\noindent
The $\vartheta^{\ 2}$ component of $\Phi$ in eq. (\ref{eq:97}) restricted
to its hermition part and multiplied wit $1 \ / g^{\ 2}$ serves as starting point
Lagrangean density to the N=1 super-Yang-Mills structure.

}

\newpage

\color{blue}

{\large \bf

\color{red}
\begin{center}
6) The minimal source extension of N=1 super-Yang-Mills structure
 \cite{LeibMink}
\end{center}

\color{blue}

\noindent
We consider an external chiral multiplet , called the (space-time dependent -)
coupling constant multiplet, denoted by $J \ ( \ y \ )$

\vspace*{-0.5cm}
\begin{equation}
\label{eq:98}
\begin{array}{l}
\Phi \ =
\hspace*{0.2cm}
\left \lbrace \begin{array}{l}
\vartheta^{\ 2} \ H
\vspace*{0.2cm} \\
+ \ \vartheta^{\ \alpha} \ \eta_{\ \alpha}
\vspace*{0.2cm} \\
+ \ \varphi  
\end{array}
\right \rbrace \ ( \ x^{\ -} \ ) \ \color{red} \leftrightarrow
\vspace*{0.3cm} \\
J \ =
\hspace*{0.2cm}
\left \lbrace \begin{array}{l}
\vartheta^{\ 2} \ ( \ - \ m \ ) 
\vspace*{0.2cm} \\
+ \ \vartheta^{\ \alpha} \ \psi_{\ \alpha}
\vspace*{0.2cm} \\
+ \ j 
\end{array}
\right \rbrace \ ( \ x^{\ -} \ )
\vspace*{0.4cm} \\
j \ ( \ y \ ) \ = \ \frac{1}{2} 
\ \left \lbrack \ ( \ 1 \ / \ g^{\ 2} \ ( \ y \ ) \ ) \ - \ \frac{i}{8 \pi^{ 2}} \ \Theta \ ( \ y \ )
\ \right \rbrack
\end{array}
\end{equation}

\color{blue}
The external sources contained in $J \ ( \ y \ )$ are thus the four complex functions

\vspace*{-0.5cm}
\begin{equation}
\label{eq:99}
\begin{array}{l}
J \ ( \ y \ ) \ \rightarrow 
\color{red} 
\ \left ( \ m \ , \ \psi_{\ \alpha} \ , \ j \ \right ) \ ( \ y \ )
\end{array}
\end{equation}

\noindent
The Lagrangean density, denoted $L$, is defined by the source dependent products

\vspace*{-0.5cm}
\begin{equation}
\label{eq:100}
\begin{array}{l}
L \ ( \ J \ ) \ = \ {\displaystyle{\int}} \ d^{\ 2} \ \vartheta 
\ \left ( \ \Phi \ J \ \right ) \ + \ h.c.
\end{array}
\end{equation}

}

\newpage

\color{blue}

{\large \bf

\noindent
The chiral multiplets shall be denoted

\vspace*{-0.5cm}
\begin{equation}
\label{eq:101}
\begin{array}{l}
{\cal{L}} \ ( \ J \ ) \ =  
\ \Phi \ J 
\hspace*{0.2cm} , \hspace*{0.2cm}
\overline{{\cal{L}}} \ ( \ \overline{J} \ ) \ = 
\ \overline{J} \ \overline{\Phi} 
\end{array}
\end{equation}

\noindent
The Lagrangean in the absence of sources involves the 'limiting conditions'

\vspace*{-0.5cm}
\begin{equation}
\label{eq:102}
\begin{array}{l}
\lim_{\ x \ \rightarrow \ \infty}
\ J \ ( \ x \ ) \ = \ J_{\ \infty}
\hspace*{0.2cm} :
\vspace*{0.4cm} \\ 
\lim_{\ x \ \rightarrow \ \infty}
\ \left \lbrace \ \begin{array}{llll} 
m \ ( \ x \ ) & = & m_{\ \infty} & \color{red} \rightarrow \ 0
\vspace*{0.2cm} \\
\color{blue}
\psi_{\ \alpha} \ ( \ x \ ) & = & \psi_{\ \alpha \ \infty} & \color{red} \rightarrow \ 0
\vspace*{0.2cm} \\
\color{blue}
g \ ( \ x \ ) & = & g_{\ \infty} &
\vspace*{0.2cm} \\
\Theta \ ( \ x \ ) & = & \Theta_{\ \infty} &
\vspace*{0.2cm} \\
\end{array} \right . 
\end{array}
\end{equation}

\color{blue}
and the relative sources

\vspace*{-0.5cm}
\begin{equation}
\label{eq:103}
\begin{array}{l}
J_{\ \Delta} \ ( \ x \ ) \ = \ J \ ( \ x \ ) \ - \ J_{\ \infty} \ \color{magenta} \rightarrow \ 0
\end{array}
\end{equation}

\color{blue}

\noindent
whereby the \color{magenta} magenta $\rightarrow \ 0$ \color{blue} limit (eq. \ref{eq:103})
is to be performed \color{magenta} first \color{blue} and the \color{red} red $\rightarrow \ 0$
\color{blue} limit (eq. \ref{eq:102}) \color{red} last\color{blue}.

\noindent
We proceed to calculate the multiplets ${\cal{L}}$ , $\overline{{\cal{L}}}$ :
first the $\vartheta^{\ 2}$ components

}

\newpage

\color{blue}

{\large \bf

\vspace*{-0.5cm}
\begin{equation}
\label{eq:104}
\begin{array}{l}
\left . {\cal{L}} \ ( \ J \ ) \ \right |_{\ \vartheta^{\ 2}} \ =
\ H \ j \ - \ \eta^{\ \alpha} \ \psi_{\ \alpha} \ - \ m \ \varphi
\vspace*{0.4cm} \\
H \ =
\ \left \lbrack \ \begin{array}{c}
\Lambda^{\ * \ B}_{\ \dot{\beta}}
\ \left \lbrace \ \begin{array}{c}
\frac{i}{2} \ \stackrel{\leftharpoondown \hspace*{-0.3cm} \rightharpoonup}{\partial}_{\ \mu} 
\ \delta_{\ B A}
\vspace*{0.2cm} \\
- \ v_{\ \mu}^{\ C} \ \left ( \ {\cal{F}}^{\ C} \ \right )_{\ B A}
\end{array}                     
\right \rbrace
\ \sigma^{\ \mu \ \dot{\beta} \alpha} \ \Lambda^{\ A}_{\ \alpha}
\vspace*{0.2cm} \\
\hspace*{0.5cm} \color{red}
- \ \frac{i}{2} \ \partial_{\ \mu} \ j^{\ \mu}_{\ \Lambda} 
\vspace*{0.2cm} \\
\color{blue}
+ \ \frac{1}{2} \ D^{\ A} \ D^{\ A}
\vspace*{0.2cm} \\
- \ \frac{1}{4} \ F^{\ \mu \nu \ A} \ \left ( \ F_{\ \mu \nu}^{\ A} 
\ \color{red} \ - \ i \ \widetilde{F}_{\ \mu \nu}^{\ A} \ \color{blue} \right )
\end{array}
\color{blue}
 \right \rbrack
\vspace*{0.2cm} \\
j \ = \ \frac{1}{2} \ 
\ \left \lbrack \ ( \ 1 \ / \ g^{\ 2} \ \color{red} - \ \frac{i}{8 \pi^{ 2}} 
\ \Theta
\color{blue}
\ \right \rbrack
\vspace*{0.4cm} \\
\color{red}
j^{\ \mu}_{\ \Lambda} \ = \ \Lambda^{\ * \ A}_{\ \dot{\beta}} \ \sigma^{\ \mu \ \dot{\beta} \alpha}
\ \Lambda^{\ A}_{\ \alpha}
\vspace*{0.4cm} \\
\color{blue}
\eta_{\ \alpha} \ =
\ \frac{1}{\sqrt{2}}
\left \lbrack \ D^{\ A} \ \Lambda^{\ A}_{\ \alpha} \ +
\ {\cal{F}}_{\ \alpha}^{\ A \ \gamma} 
\ \Lambda^{\ A}_{\ \gamma}
\ \right \rbrack
\vspace*{0.4cm} \\
\varphi \ =
\ \frac{1}{2} \ \Lambda^{\ A \ \alpha} \ \Lambda^{\ A}_{\ \alpha}
\end{array}
\end{equation}

\noindent
From eq. (\ref{eq:104}) we infer the Lagrangean density \\
( eq. \ref{eq:100} )

}

\newpage

\color{blue}

{\large \bf

\vspace*{-0.5cm}
\begin{equation}
\label{eq:105}
\begin{array}{l}
L \ ( \ J \ ) \ =
\vspace*{0.2cm} \\
\ \left \lbrack \ \begin{array}{c}
\frac{1}{g^{\ 2}}
\ \Lambda^{\ * \ B}_{\ \dot{\beta}}
\ \left \lbrace \ \begin{array}{c}
\frac{i}{2} \ \stackrel{\leftharpoondown \hspace*{-0.3cm} \rightharpoonup}{\partial}_{\ \mu} 
\ \delta_{\ B A}
\vspace*{0.2cm} \\
- \ v_{\ \mu}^{\ C} \ \left ( \ {\cal{F}}^{\ C} \ \right )_{\ B A}
\end{array}                     
\right \rbrace
\ \sigma^{\ \mu \ \dot{\beta} \alpha} \ \Lambda^{\ A}_{\ \alpha}
\vspace*{0.2cm} \\
- \ \frac{1}{4 \ g^{\ 2}} \ F^{\ \mu \nu \ A} \ F_{\ \mu \nu}^{\ A}
\ + \ \frac{1}{2 \ g^{\ 2}} \ D^{\ A} \ D^{\ A}
\vspace*{0.2cm} \\
\color{red}
+ \ \frac{1}{8 \pi^{ 2}} \ \Theta
\ \left ( \ \frac{1}{4} \ F^{\ \mu \nu \ A} \ \widetilde{F}_{\ \mu \nu}^{\ A}
\ - \ \frac{1}{2} \ \partial_{\ \mu} \ j^{\ \mu}_{\ \Lambda} 
\ \right )
\vspace*{0.2cm} \\
\color{blue}
+ \  \frac{1}{\sqrt{2}}
\left \lbrack \ D^{\ A} \ \Lambda^{\ A}_{\ \alpha} \ +
\ {\cal{F}}_{\ \alpha}^{\ A \ \gamma}
\ \Lambda^{\ A}_{\ \gamma}
\ \right \rbrack \ \psi^{\ \alpha} \ + \ h.c.
\vspace*{0.2cm} \\
- \ \frac{1}{2} \ m \ \Lambda^{\ A \ \alpha} \ \Lambda^{\ A}_{\ \alpha} \ - \ h.c. 
\end{array}
\right \rbrack
\end{array}
\end{equation}

\noindent
Upon retaining the fermionic source $\psi_{\ \alpha}$ the elimination of the auxiliary fields 
$D^{\ A}$ leaves the latter nontrivial 

\vspace*{-0.5cm}
\begin{equation}
\label{eq:106}
\begin{array}{l}
\delta \ L \ ( \ J \ ) \ / \ \delta \ D^{\ A}
\ =
\vspace*{0.2cm} \\
\hspace*{0.5cm} 
= \ \frac{1}{g^{\ 2}} \ D^{\ A} \ + \ \frac{1}{\sqrt{2}}
\ \left ( \ \Lambda^{\ A}_{\ \alpha} \ \psi^{\ \alpha} \ + \ h.c. \ \right )
\hspace*{0.2cm} \color{red} \rightarrow \ 0
\end{array}
\end{equation}

\noindent
The net term induced through $D^{\ A}$ becomes

\vspace*{-0.5cm}
\begin{equation}
\label{eq:107}
\begin{array}{l}
L_{\ D} \ ( \ J \ )
= 
- \ \frac{g^{ 2}}{4}
\ \left \lbrack \ \begin{array}{l}
\left ( \ \Lambda^{\ A}_{\ \alpha} \ \psi^{\ \alpha} \ + 
\ \psi^{\ * \ \dot{\beta}} \ \Lambda^{\ * \ A}_{\ \dot{\beta}} 
\ \right ) \ \times
\vspace*{0.2cm} \\
\ \times \ \left ( \ \Lambda^{\ A}_{\ \gamma} \ \psi^{\ \gamma} \ + 
\ \psi^{\ * \ \dot{\delta}} \ \Lambda^{\ * \ A}_{\ \dot{\delta}} 
\ \right )
\end{array} \right \rbrack
\end{array}
\end{equation}

}

\newpage

\color{blue}

{\large \bf

\noindent
We reorder the source fermion fields $\psi_{\ \alpha} \ , \ \psi^{*}_{\ \dot{\beta}}$
in eq. (\ref{eq:107}) all to the left

\vspace*{-0.5cm}
\begin{equation}
\label{eq:108}
\begin{array}{l}
L_{\ D} \ ( \ J \ )
= 
\vspace*{0.2cm} \\
\hspace*{0.4cm} =
\ - \ \frac{g^{ 2}}{4}
\ \left \lbrack \ \begin{array}{l}
\left ( \ \psi_{\ \alpha} \ \psi^{ \ \gamma} \ \right )
\ \left ( \ \Lambda^{\ A \ \alpha} \ \Lambda^{\ A}_{\ \gamma} \ \right ) \ + \ h.c.  
\vspace*{0.2cm} \\
\ + \ 2 \ \left ( \ \psi^{\ * \ \dot{\delta}} \ \psi^{\ \alpha} \ \right ) 
\ \left ( \ \ \Lambda^{\ * \ A}_{\ \dot{\delta}} \ \Lambda^{\ A}_{\ \alpha} 
\ \right )
\end{array} \right \rbrack
\end{array}
\end{equation}

\noindent
and defining in an equivalent way to $\vartheta^{\ 2} \ , \ \overline{\vartheta}^{\ 2}$ in
eq. (\ref{eq:39}) the Lorentz invariants pertaining to the fermionic sources

\vspace*{-0.5cm}
\begin{equation}
\label{eq:109}
\begin{array}{l}
\psi^{\ 2} \ = \ \frac{1}{2} \ \psi^{\ \alpha} \ \psi_{\ \alpha} \ = \ \psi_{\ 1} \ \psi_{\ 2}
\vspace*{0.2cm} \\
\overline{\psi}^{\ 2} \ = \ \frac{1}{2} \ \psi^{\ *}_{\ \dot{\beta}} \ \psi^{\ * \ \dot{\beta}}
\ = \ \psi^{\ *}_{\ \dot{2}} \ \psi^{\ *}_{\ \dot{1}}
\end{array}
\end{equation}

\noindent
we obtain

\vspace*{-0.5cm}
\begin{equation}
\label{eq:110}
\begin{array}{l}
L_{\ D} \ ( \ J \ )
= 
\vspace*{0.2cm} \\
\hspace*{0.4cm} =
\ - \ \frac{g^{ 2}}{4}
\ \left \lbrack \ \begin{array}{l}
- \ \psi^{\ 2}
\ \Lambda^{\ A \ \alpha} \ \Lambda^{\ A}_{\ \alpha} \ + \ h.c.  
\vspace*{0.2cm} \\
\ + \ 2 \ \left ( \ \psi^{\ * \ \dot{\delta}} \ \psi^{\ \alpha} \ \right ) 
\ \left ( \ \ \Lambda^{\ * \ A}_{\ \dot{\delta}} \ \Lambda^{\ A}_{\ \alpha} 
\ \right )
\end{array} \right \rbrack
\end{array}
\end{equation}

\noindent
The last term on the right hand side of eqs. (\ref{eq:108} and \ref{eq:110})
represents a hermitian current-current coupling. We use the identity

}

\newpage

\color{blue}

{\large \bf

\vspace*{-0.5cm}
\begin{equation}
\label{eq:111}
\begin{array}{l}
\sigma^{\ \mu \ \dot{\delta} \alpha} \ \sigma_{\ \mu \ \beta \dot{\gamma}}
\ = \ 2 \ \delta^{\ \dot{\delta}}_{\ \dot{\gamma}} 
\ \delta^{\ \alpha}_{\ \beta}
\end{array}
\end{equation}

\noindent
whereupon $L_{\ D} \ ( \ J \ )$ takes the form

\vspace*{-0.5cm}
\begin{equation}
\label{eq:112}
\begin{array}{l}
L_{\ D} \ ( \ J \ )
= 
\vspace*{0.2cm} \\
\hspace*{0.3cm} =
\ - \ \frac{g^{ 2}}{4}
\ \left \lbrack \ \begin{array}{c}
- \ \psi^{\ 2}
\ \Lambda^{\ A \ \alpha} \ \Lambda^{\ A}_{\ \alpha}
\ + \ \frac{1}{2} \ j^{\ \mu}_{\ \psi} \ j_{\ \mu \ \Lambda} \ 
\end{array} \right \rbrack
\ + \ h.c.
\vspace*{0.4cm} \\
j^{\ \mu}_{\ \psi} \ = \ \psi^{\ *}_{\ \dot{\delta}} \ \sigma^{\ \mu \ \dot{\delta} \alpha}
\ \psi_{\ \alpha}
\vspace*{0.2cm} \\
j^{\ \mu}_{\ \Lambda} \ = \ \Lambda^{\ * \ A}_{\ \dot{\delta}} \ \sigma^{\ \mu \ \dot{\delta} \alpha}
\ \Lambda^{\ A}_{\ \alpha} 
\end{array}
\end{equation}

\noindent
We rewrite $L \ ( \ J \ )$ in eq. (\ref{eq:105}) exhibiting sources in \color{red} red
\color{blue} .

}

\newpage

\color{blue}

{\large \bf

\vspace*{-0.5cm}
\begin{equation}
\label{eq:113}
\begin{array}{l}
L \ ( \ J \ ) \ = \ L_{\ div} \ + \ L_{\ 1} \ ( \ J \ )
\vspace*{0.2cm} \\
 L_{\ 1} \ ( \ J \ ) \ =
\vspace*{0.2cm} \\
\ \left \lbrack \ \begin{array}{c}
\color{red}
\frac{1}{g^{\ 2}}
\color{blue}
\ \Lambda^{\ * \ B}_{\ \dot{\beta}}
\ \left \lbrace \ \begin{array}{c}
\frac{i}{2} \ \stackrel{\leftharpoondown \hspace*{-0.3cm} \rightharpoonup}{\partial}_{\ \mu} 
\ \delta_{\ B A}
\vspace*{0.2cm} \\
- \ v_{\ \mu}^{\ C} \ \left ( \ {\cal{F}}^{\ C} \ \right )_{\ B A}
\end{array}                     
\right \rbrace
\ \sigma^{\ \mu \ \dot{\beta} \alpha} \ \Lambda^{\ A}_{\ \alpha}
\vspace*{0.2cm} \\
+ \ \color{red} \frac{1}{g^{\ 2}} \ 
\color{blue} \left ( \ - \ \frac{1}{4} \ F^{\ \mu \nu \ A} \ F_{\ \mu \nu}^{\ A}
\ \right )
\vspace*{0.2cm} \\
+ \ \color{red} \frac{1}{8 \pi^{ 2}} \ \Theta
\color{blue}
\ \left ( \ \frac{1}{4} \ F^{\ \mu \nu \ A} \ \widetilde{F}_{\ \mu \nu}^{\ A}
\ \right )
\vspace*{0.2cm} \\
\ + 
\color{red}
\ \left ( \ \frac{1}{16 \ \pi^{\ 2}} \ \partial_{\ \mu} \ \Theta
\ - \ \frac{g^{\ 2}}{4} \ j_{\ \mu \ \psi} \ \right )
\ \color{blue} j^{\ \mu}_{\ \Lambda} 
\vspace*{0.2cm} \\
\color{blue}
- \  \frac{1}{\sqrt{2}}
\ \left \lbrack \ \color{red} \psi^{\ \alpha} \ \color{blue}  {\cal{F}}_{\ \alpha}^{\ A \ \gamma}
\ \Lambda^{\ A}_{\ \gamma}
\ \right \rbrack \ + \ h.c.
\vspace*{0.2cm} \\
- \ \frac{1}{2} \ \color{red} \left ( \ m \ - \ \frac{g^{\ 2}}{4} \ \psi^{\ 2} \ \right ) 
\ \color{blue} \Lambda^{\ A \ \alpha} \ \Lambda^{\ A}_{\ \alpha} 
\ + \ h.c. 
\end{array}
\right \rbrack
\vspace*{0.4cm} \\
\color{blue}
L_{\ div} \ = \ - \ \partial_{\ \mu}
\ \left (
\ \color{red} \frac{1}{16 \ \pi^{\ 2}} \ \Theta
\ \color{blue} j^{\ \mu}_{\ \Lambda}
\ \right )
\end{array}
\end{equation}

}

\newpage

\color{blue}

{\large \bf

\color{red}
\begin{center}
6a) The minimal source extension of the $D^{\ A}$-eliminated Lagrangean multiplet for
N=1 super-Yang-Mills
\end{center}

\color{blue}
It is obviously important to reconstruct the chiral Lagrangean multiplet
${\cal{L}} \ ( \ J \ )$ as defined in eqs. (\ref{eq:98} - \ref{eq:101}) ,
taking into account the elimination of the auxiliary fields $D^{\ A}$ . 

\noindent
We turn to this task next, repeating the chiral multiplet structures pertaining
to dynamical fields $\Phi$ and sources $\color{red} J$
\color{blue}

\vspace*{-0.5cm}
\begin{equation}
\label{eq:114}
\begin{array}{l}
\Phi \ =
\hspace*{0.2cm}
\left \lbrace \begin{array}{l}
\vartheta^{\ 2} \ H
\vspace*{0.2cm} \\
+ \ \vartheta^{\ \alpha} \ \eta_{\ \alpha}
\vspace*{0.2cm} \\
+ \ \varphi  
\end{array}
\right \rbrace \ ( \ x^{\ -} \ ) \ \color{red} \leftrightarrow
\vspace*{0.3cm} \\
\color{red} J \ \color{blue} =
\hspace*{0.2cm}
\left \lbrace \begin{array}{l}
\vartheta^{\ 2} \ \color{red} ( \ - \ m \ ) 
\vspace*{0.2cm} \\
\color{blue}
+ \ \vartheta^{\ \alpha} \ \color{red} \psi_{\ \alpha}
\vspace*{0.2cm} \\
\color{blue}
+ \ \color{red} j 
\end{array}
\color{blue}
\right \rbrace \ ( \ x^{\ -} \ )
\vspace*{0.4cm} \\
\color{red}
j \ ( \ y \ ) \ = \ \frac{1}{2} 
\ \left \lbrack \ ( \ 1 \ / \ g^{\ 2} \ ( \ y \ ) \ ) \ - \ \frac{i}{8 \pi^{ 2}} \ \Theta \ ( \ y \ )
\ \right \rbrack
\vspace*{0.4cm} \\
\color{blue}
\eta_{\ \alpha} \ =
\ \frac{1}{\sqrt{2}}
\left \lbrack \ D^{\ A} \ \Lambda^{\ A}_{\ \alpha} \ +
\ {\cal{F}}_{\ \alpha}^{\ A \ \gamma} 
\ \Lambda^{\ A}_{\ \gamma}
\ \right \rbrack
\end{array}
\end{equation}

}

\newpage

\color{blue}

{\large \bf

\noindent
The fermion field $\eta_{\ \alpha}$ contains the auxiliary fields (eq. \ref{eq:104}) and
becomes using eq. (\ref{eq:106}) expanded  below

\vspace*{-0.5cm}
\begin{equation}
\label{eq:115}
\begin{array}{l}
\delta \ L \ ( \ J \ ) \ / \ \delta \ D^{\ A}
\ =
\vspace*{0.2cm} \\
\hspace*{0.5cm} 
= \ \frac{1}{g^{\ 2}} \ D^{\ A} \ + \ \frac{1}{\sqrt{2}}
\ \left ( \ \Lambda^{\ A}_{\ \alpha} \ \psi^{\ \alpha} \ + \ h.c. \ \right )
\hspace*{0.2cm} \color{red} \rightarrow \ 0
\vspace*{0.2cm} \\
\color{blue}
D^{\ A} \ = \ \color{red} \frac{g^{\ 2}}{\sqrt{2}} \ \color{blue} \left (
\ \color{red} \psi^{\ \alpha} 
\color{blue} \ \Lambda^{\ A}_{\ \alpha} \ + \ \Lambda^{\ * \ A}_{\ \dot{\beta}}
\ \color{red} \psi^{\ * \ \dot{\beta}}
\ \color{blue} \right )
\end{array}
\end{equation}

\noindent
Thus $\eta_{\ \alpha}$ becomes ( eq. \ref{eq:114} )

\vspace*{-0.5cm}
\begin{equation}
\label{eq:116}
\begin{array}{l}
\eta_{\ \alpha} \ =
\ \left \lbrack \ \begin{array}{c}
\ \frac{1}{\sqrt{2}} \ {\cal{F}}_{\ \alpha}^{\ A \ \gamma}
\ \Lambda^{\ A}_{\ \gamma} 
\ +
\vspace*{0.2cm} \\
\ \color{red} \frac{g^{\ 2}}{2} 
\ \color{blue} \left (
\ \color{red} \psi^{\ \gamma} 
\color{blue} \ \Lambda^{\ A}_{\ \gamma} \ + \ \Lambda^{\ * \ A}_{\ \dot{\beta}}
\ \color{red} \psi^{\ * \ \dot{\beta}}
\ \color{blue} \right )
\ \Lambda^{\ A}_{\ \alpha}
\end{array} \ \right \rbrack 
\end{array}
\end{equation}

\noindent
We go step by step to avoid sign errors

\vspace*{-0.5cm}
\begin{equation}
\label{eq:117}
\begin{array}{l}
\eta_{\ \alpha} \ =
\vspace*{0.2cm} \\
\ = \ \left \lbrack \ \begin{array}{c}
\ \frac{1}{\sqrt{2}} \ {\cal{F}}_{\ \alpha}^{\ A \ \gamma}
\ \Lambda^{\ A}_{\ \gamma} 
\vspace*{0.2cm} \\
\ - \ \color{red} \frac{g^{\ 2}}{2} 
\ \color{blue} \left (
\ \color{red} \psi_{\ \gamma} 
\color{blue} \ \Lambda^{\ A \ \gamma}
\ \Lambda^{\ A}_{\ \alpha} 
\ +
\ \color{red} \psi^{\ * \ \dot{\beta}} 
\ \color{blue} \Lambda^{\ * \ A}_{\ \dot{\beta}}
\ \Lambda^{\ A}_{\ \alpha}
\ \right )
\end{array} \ \right \rbrack 
\end{array}
\end{equation}

\noindent
First we use the identity 

\begin{center}
$\Lambda^{\ A \ \gamma} \ \Lambda^{\ A}_{\ \alpha} \ = \ \frac{1}{2} \ \delta^{\ \gamma}_{\ \alpha}
\ \Lambda^{\ A \ \beta} \ \Lambda^{\ A}_{\ \beta}$
\end{center}

}

\newpage

\color{blue}

{\large \bf

\noindent 
and obtain

\vspace*{-0.5cm}
\begin{equation}
\label{eq:118}
\begin{array}{l}
\eta_{\ \alpha} \ =
\ \left \lbrack \ \begin{array}{c}
\ \frac{1}{\sqrt{2}} \ {\cal{F}}_{\ \alpha}^{\ A \ \gamma}
\ \Lambda^{\ A}_{\ \gamma} 
\vspace*{0.2cm} \\
- \ \color{red} \frac{g^{\ 2}}{4} 
\ \color{red} \psi_{\ \alpha} 
\color{blue} \ \Lambda^{\ A \ \delta}
\ \Lambda^{\ A}_{\ \delta} 
\vspace*{0.2cm} \\
-
\ \color{red} \frac{g^{\ 2}}{4} \ \psi^{\ * \ \dot{\gamma}} 
\ \color{blue} \Lambda^{\ * \ A}_{\ \dot{\beta}}
\ \Lambda^{\ A}_{\ \delta} \ \left ( \ 2 \ \delta^{\ \dot{\beta}}_{\ \dot{\gamma}}
\ \delta^{\ \delta}_{\ \alpha} \ \right )
\end{array} \ \right \rbrack 
\end{array}
\end{equation}

\noindent
Next we substitute ( eq. \ref{eq:111} )

\begin{center}
$2 \ \delta^{\ \dot{\beta}}_{\ \dot{\gamma}}
\ \delta^{\ \delta}_{\ \alpha} \ =
\sigma^{\ \mu \ \dot{\beta} \delta} \ \sigma_{\ \mu \ \alpha \dot{\gamma}} $
\end{center}

\noindent
whereby $\eta_{\ \alpha}$ becomes

\vspace*{-0.5cm}
\begin{equation}
\label{eq:119}
\begin{array}{l}
\eta_{\ \alpha} \ =
\ \left \lbrack \ \begin{array}{c}
\ \frac{1}{\sqrt{2}} \ {\cal{F}}_{\ \alpha}^{\ A \ \gamma}
\ \Lambda^{\ A}_{\ \gamma} 
\vspace*{0.2cm} \\
- \ \color{red} \frac{g^{\ 2}}{4} 
\ \color{red} \psi_{\ \alpha} 
\color{blue} \ \Lambda^{\ A \ \delta}
\ \Lambda^{\ A}_{\ \delta} 
\vspace*{0.2cm} \\
-
\ \color{red} \frac{g^{\ 2}}{4} \ \psi^{\ *}_{\ \mu \ \alpha}
\ \color{blue} j^{\ \mu}_{\ \Lambda} 
\end{array} \ \right \rbrack 
\vspace*{0.4cm} \\
\color{red}
\psi^{\ *}_{\ \mu \ \alpha} \ = \ \psi^{\ * \ \dot{\gamma}} 
\ \color{blue} \sigma_{\ \mu \ \alpha \dot{\gamma}}
\vspace*{0.2cm} \\
j^{\ \mu}_{\ \Lambda} \ = \ \Lambda^{\ * \ A}_{\ \dot{\beta}} \ \sigma^{\ \mu \ \dot{\beta} \delta}
\ \Lambda^{\ A}_{\ \delta}
\end{array}
\end{equation}

}

\newpage

\color{blue}

{\large \bf

\noindent
Now we turn to the $\theta^{\ 2}$ component of $\Phi$ ( eq. \ref{eq:104} ) ,
where $D^{\ A}$ are to be substituted ( eq. \ref{eq:115} )

\vspace*{-0.5cm}
\begin{equation}
\label{eq:120}
\begin{array}{l}
H \ =
\ \left \lbrack \ \begin{array}{c}
\Lambda^{\ * \ B}_{\ \dot{\beta}}
\ \left \lbrace \ \begin{array}{c}
\frac{i}{2} \ \stackrel{\leftharpoondown \hspace*{-0.3cm} \rightharpoonup}{\partial}_{\ \mu} 
\ \delta_{\ B A}
\vspace*{0.2cm} \\
- \ v_{\ \mu}^{\ C} \ \left ( \ {\cal{F}}^{\ C} \ \right )_{\ B A}
\end{array}                     
\right \rbrace
\ \sigma^{\ \mu \ \dot{\beta} \alpha} \ \Lambda^{\ A}_{\ \alpha}
\vspace*{0.2cm} \\
\hspace*{0.5cm}
- \ \frac{i}{2} \ \partial_{\ \mu} \ j^{\ \mu}_{\ \Lambda} 
\vspace*{0.2cm} \\
+ \ \frac{1}{2} \ D^{\ A} \ D^{\ A}
\vspace*{0.2cm} \\
- \ \frac{1}{4} \ F^{\ \mu \nu \ A} \ \left ( \ F_{\ \mu \nu}^{\ A}  
\ - \ i \ \widetilde{F}_{\ \mu \nu}^{\ A} \ \color{blue} \right )
\end{array}
 \right \rbrack
\vspace*{0.4cm} \\
D^{\ A} \ = \ \color{red} \frac{g^{\ 2}}{\sqrt{2}} \ \color{blue} \left (
\ \color{red} \psi^{\ \alpha} 
\color{blue} \ \Lambda^{\ A}_{\ \alpha} \ + \ \Lambda^{\ * \ A}_{\ \dot{\beta}}
\ \color{red} \psi^{\ * \ \dot{\beta}}
\ \color{blue} \right )
\end{array}
\end{equation}

\noindent
Thus we proceed to evaluate the quantity $\frac{1}{2} \ D^{\ A} \ D^{\ A}$

\vspace*{-0.5cm}
\begin{equation}
\label{eq:121}
\begin{array}{l}
\frac{1}{2} \ D^{\ A} \ D^{\ A} \ = 
\vspace*{0.2cm} \\
\ = \ \color{red} \frac{g^{\ 4}}{4} \ \color{blue} \left ( \ \begin{array}{l}
\color{red} \psi^{\ \alpha} \ \psi_{\ \gamma} 
\ \color{blue} \Lambda^{\ A}_{\ \alpha} \ \Lambda^{\ A \ \gamma} \ + \ h.c.
\vspace*{0.2cm} \\
+ \ \color{red} \psi^{\ * \ \dot{\beta}} \ \psi^{\ \alpha}
\ \color{blue} 2 \ \Lambda^{\ * \ A}_{\ \dot{\beta}} \ \Lambda^{\ A}_{\ \alpha}
\end{array}
\right )
\end{array}
\end{equation}

\noindent
and using the identity in eq. (\ref{eq:111}) it follows

\vspace*{-0.5cm}
\begin{equation}
\label{eq:122}
\begin{array}{l}
\frac{1}{2} \ D^{\ A} \ D^{\ A} \ = 
\vspace*{0.2cm} \\
\ = \ \color{red} \frac{g^{\ 4}}{4} \ \color{blue} \left ( \ \begin{array}{c}
\color{red} - \ \psi^{\ 2} 
\ \color{blue} \Lambda^{\ A \ \alpha} \ \Lambda_{\ A \ \alpha} \ + \ h.c.
\vspace*{0.2cm} \\
+ \ \color{red} j_{\ \mu \ \psi}
\ \color{blue} j^{\ \mu}_{\ \Lambda}
\end{array}
\right )
\end{array}
\end{equation}

}

\newpage

\color{blue}

{\large \bf

\color{red}
\begin{center}
6b) The minimal source extension  $D^{\ A}$-eliminated \\
$\Phi$ multiplet ( results )
\end{center}

\color{blue}

\noindent
The expressions for 
$\Phi \ = \ \left ( \ H \ , \ \eta_{\ \alpha} \ , \ \varphi \ \right )$ 
in eq. (\ref{eq:114}) become ( eqs. \ref{eq:104} , \ref{eq:119} , \ref{eq:122} )

\vspace*{-0.5cm}
\begin{equation}
\label{eq:123}
\begin{array}{l}
H \ =
\ \left \lbrack \ \begin{array}{c}
\Lambda^{\ * \ B}_{\ \dot{\beta}}
\ \left \lbrace \ \begin{array}{c}
\frac{i}{2} \ \stackrel{\leftharpoondown \hspace*{-0.3cm} \rightharpoonup}{\partial}_{\ \mu} 
\ \delta_{\ B A}
\vspace*{0.2cm} \\
- \ v_{\ \mu}^{\ C} \ \left ( \ {\cal{F}}^{\ C} \ \right )_{\ B A}
\end{array}                     
\right \rbrace
\ \sigma^{\ \mu \ \dot{\beta} \alpha} \ \Lambda^{\ A}_{\ \alpha}
\vspace*{0.2cm} \\
\hspace*{0.5cm}
- \ \frac{\color{magenta} i}{\color{blue} 2} \ \color{blue} \partial_{\ \mu} \ j^{\ \mu}_{\ \Lambda} 
\vspace*{0.2cm} \\
- \ \frac{1}{4} \ F^{\ \mu \nu \ A} \ \left ( \ F_{\ \mu \nu}^{\ A}  
\ - \ \color{magenta} i \ \color{blue} \widetilde{F}_{\ \mu \nu}^{\ A} \ \color{blue} \right )
\vspace*{0.2cm} \\
+ \ \color{red} \frac{g^{\ 4}}{4} \ \color{blue} \left ( \ \begin{array}{c}
\color{red} - \ \psi^{\ 2} 
\ \color{blue} \Lambda^{\ A \ \alpha} \ \Lambda_{\ A \ \alpha} \ + \ h.c.
\vspace*{0.2cm} \\
+ \ \color{red} j_{\ \mu \ \psi}
\ \color{blue} j^{\ \mu}_{\ \Lambda}
\end{array}
\right )
\end{array}
 \right \rbrack
\vspace*{0.7cm} \\
\eta_{\ \alpha} \ =
\ \left \lbrack \ \begin{array}{c}
\ \frac{1}{\sqrt{2}} \ {\cal{F}}_{\ \alpha}^{\ A \ \gamma}
\ \Lambda^{\ A}_{\ \gamma} 
\vspace*{0.2cm} \\
- \ \color{red} \frac{g^{\ 2}}{4} 
\ \color{red} \psi_{\ \alpha} 
\color{blue} \ \Lambda^{\ A \ \delta}
\ \Lambda^{\ A}_{\ \delta} 
\vspace*{0.2cm} \\
-
\ \color{red} \frac{g^{\ 2}}{4} \ \psi^{\ *}_{\ \mu \ \alpha}
\ \color{blue} j^{\ \mu}_{\ \Lambda} 
\end{array} \ \right \rbrack
\vspace*{0.7cm} \\
\varphi \ =
\ \frac{1}{2} \ \Lambda^{\ A \ \alpha} \ \Lambda^{\ A}_{\ \alpha}
\end{array}
\end{equation}

\noindent
The antihermitian parts of $H$ in eq. (\ref{eq:123}) are marked with a color magenta
factor $\color{magenta} i$ \color{blue} .

}

\newpage

{\large \bf

\color{red}
\begin{center}
6c) The minimal source extension  $D^{\ A}$-eliminated \\
Lagrangean multiplet $\color{blue} {\cal{L}} \ ( \ \color{red} J \ \color{blue} )$ 
\end{center}

\color{blue}

\noindent
Having established the $D^{\ A}$-eliminated form of $\Phi$ in eq. (\ref{eq:123}) ,
we go back to the external source multiplet ( eq. \ref{eq:98} ) and $\Phi$ , reproduced below

\vspace*{-0.5cm}
\begin{equation}
\label{eq:124}
\begin{array}{l}
\Phi \ =
\hspace*{0.2cm}
\left \lbrace \begin{array}{l}
\vartheta^{\ 2} \ H
\vspace*{0.2cm} \\
+ \ \vartheta^{\ \alpha} \ \eta_{\ \alpha}
\vspace*{0.2cm} \\
+ \ \varphi  
\end{array}
\right \rbrace \ ( \ x^{\ -} \ ) \ \color{red} \leftrightarrow
\vspace*{0.3cm} \\
\color{red} J \ \color{blue} =
\hspace*{0.2cm}
\left \lbrace \begin{array}{l}
\vartheta^{\ 2} \ ( \ \color{red} - \ m \ \color{blue} ) 
\vspace*{0.2cm} \\
+ \ \vartheta^{\ \alpha} \ \color{red} \psi_{\ \alpha}
\vspace*{0.2cm} \\
\color{blue}
+ \ \color{red} j
\color{blue} 
\end{array}
\right \rbrace \ ( \ x^{\ -} \ )
\vspace*{0.4cm} \\
\color{red}
j \ \color{blue} ( \ y \ ) \ = \ \frac{1}{2} 
\ \left \lbrack \ ( \ \color{red} 1 \ / \ g^{\ 2} \ 
\color{blue} ( \ y \ ) \ ) \ - \ \frac{\color{magenta} i}{\color{blue} 8 \pi^{ 2}} 
\ \color{red} \Theta \ \color{blue} ( \ y \ )
\ \right \rbrack
\end{array}
\end{equation}

\noindent
The next step is to determine the Lagrangean multiplet 
${\cal{L}} \ ( \ \color{red} J \ \color{blue} ) \ = \ \Phi \ \color{red} J$
\color{blue}

\vspace*{-0.5cm}
\begin{equation}
\label{eq:125}
\begin{array}{l}
{\cal{L}} \ ( \ \color{red} J \ \color{blue} ) \ =
\hspace*{0.2cm}
\left \lbrace \begin{array}{l}
\vartheta^{\ 2} \ A
\vspace*{0.2cm} \\
+ \ \vartheta^{\ \alpha} \ B_{\ \alpha}
\vspace*{0.2cm} \\
+ \ C  
\end{array}
\right \rbrace \ ( \ x^{\ -} \ )
\end{array}
\end{equation}

}

\newpage

\color{blue}

{\large \bf

\noindent
The components ${\cal{L}} \ ( \ \color{red} J \ \color{blue} ) \ = 
\ \left ( \ A \ , \ B_{\ \alpha} \ , \ C \ \right )$ are

\vspace*{-0.5cm}
\begin{equation}
\label{eq:126}
\begin{array}{l}
\begin{array}{lll}
A & = & {\color{red} j} \ {\color{blue} H \ -} \ {\color{red} \psi^{\ \alpha}}
\ {\color{blue} \eta_{\ \alpha} \ -} \ {\color{red} m} \ {\color{blue} \varphi} 
\vspace*{0.2cm} \\
{\color{blue}
B_{\ \alpha}} & = & {\color{red} j} \ {\color{blue} \eta_{\ \alpha} \ +}
\ {\color{red} \psi_{\ \alpha}} \ {\color{blue} \varphi}
\vspace*{0.2cm} \\
{\color{blue}
C} & = & {\color{red} j} \ {\color{blue} \varphi}
\end{array}
\vspace*{0.4cm} \\
L \ ( \ \color{red} J \ \color{blue} ) \ = \ A \ + \ A^{\ *}
\vspace*{0.4cm} \\
\color{red} j \ \color{blue} ( \ y \ ) \ = \ \frac{1}{2} 
\ \left \lbrack \ \color{red} \frac{1}{g^{\ 2}} 
\ \color{blue} ( \ y \ ) \ - \ \color{red} \frac{\color{magenta} i}{8 \pi^{ 2}} \ \Theta 
\ \color{blue} ( \ y \ )
\ \right \rbrack
\end{array}
\end{equation}

\noindent
We first give $\color{red} j \ \color{blue} H$ ( eq. \ref{eq:123} )

\vspace*{-0.5cm}
\begin{equation}
\label{eq:127}
\begin{array}{l}
\color{red} j \ \color{blue} H \ = \ A_{\ 1} \ =
\vspace*{0.2cm} \\
 \left \lbrack  \begin{array}{c}
\begin{array}{l}
\color{red} ( \ \frac{1}{2 \ g^{\ 2}} \ - \ \frac{\color{magenta} i}{16 \ \pi^{\ 2}} \ \Theta \ )
\ \times
\vspace*{0.2cm} \\
\hspace*{0.4cm}
\color{blue} 
\times \ 
\Lambda^{\ * \ B}_{\ \dot{\beta}}
\ \left \lbrace \ \begin{array}{c}
\frac{i}{2} \ \stackrel{\leftharpoondown \hspace*{-0.3cm} \rightharpoonup}{\partial}_{\ \mu} 
\ \delta_{\ B A}
\vspace*{0.2cm} \\
- \ v_{\ \mu}^{\ C} \ \left ( \ {\cal{F}}^{\ C} \ \right )_{\ B A}
\end{array}                     
\right \rbrace
\ \sigma^{\ \mu \ \dot{\beta} \alpha} \ \Lambda^{\ A}_{\ \alpha}
\end{array}
\vspace*{0.2cm} \\
- \ \color{red} ( \ \frac{1}{32 \ \pi^{\ 2}} \ \Theta 
\ + \ \frac{\color{magenta} i}{4 \ g^{\ 2}} \ )
\ \color{blue} \partial_{\ \mu} \ j^{\ \mu}_{\ \Lambda} 
\vspace*{0.2cm} \\
- \ \color{red} ( \ \frac{1}{8 \ g^{\ 2}} \ - \ \frac{\color{magenta} i}{64 \ \pi^{\ 2}} \ \Theta \ )
\color{blue} 
\ \ F^{\ \mu \nu \ A} \ \left ( \ F_{\ \mu \nu}^{\ A}  
\ - \ \color{magenta} i \ \color{blue} \widetilde{F}_{\ \mu \nu}^{\ A} \ \color{blue} \right )
\vspace*{0.2cm} \\
 \begin{array}{l} 
\color{blue} + 
\ \color{red} ( \ \frac{g^{\ 2}}{8} \ - \ \frac{\color{magenta} i}{64 \ \pi^{\ 2}} \ g^{\ 4} 
\ \Theta \ ) \ \times
\vspace*{0.2cm} \\
\hspace*{0.7cm}  \color{blue} \times \ \left ( \ \begin{array}{c}
\color{red} - \ \psi^{\ 2} 
\ \color{blue} \Lambda^{\ A \ \alpha} \ \Lambda_{\ A \ \alpha} \ + \ h.c.
\vspace*{0.2cm} \\
+ \ \color{red} j_{\ \mu \ \psi}
\ \color{blue} j^{\ \mu}_{\ \Lambda}
\end{array}
\right )
\end{array}
\end{array}
 \right \rbrack
\end{array}
\end{equation}

}

\newpage

\color{blue}

{\large \bf

\noindent
Next we determine $\color{red} \psi^{\ \alpha} \ \color{blue} \eta_{\ \alpha}$
contributing with a negative sign to $A$ repeating eq. (\ref{eq:123}) below

\vspace*{-0.5cm}
\begin{equation}
\label{eq:128}
\begin{array}{l}
\eta_{\ \alpha} \ =
\ \left \lbrack \ \begin{array}{c}
\ \frac{1}{\sqrt{2}} \ {\cal{F}}_{\ \alpha}^{\ A \ \gamma}
\ \Lambda^{\ A}_{\ \gamma} 
\vspace*{0.2cm} \\
- \ \color{red} \frac{g^{\ 2}}{4} 
\ \color{red} \psi_{\ \alpha} 
\color{blue} \ \Lambda^{\ A \ \delta}
\ \Lambda^{\ A}_{\ \delta} 
\vspace*{0.2cm} \\
-
\ \color{red} \frac{g^{\ 2}}{4} \ \psi^{\ *}_{\ \mu \ \alpha}
\ \color{blue} j^{\ \mu}_{\ \Lambda} 
\end{array} \ \right \rbrack \ \color{red} \rightarrow
\vspace*{0.5cm} \\
\color{red} \psi^{\ \alpha} \ \color{blue} \eta_{\ \alpha} \ = \ - \ A_{\ 2} \ =
\vspace*{0.2cm} \\
\ = \ \left \lbrack \ \begin{array}{c}
\ \frac{1}{\sqrt{2}} \ \color{red} \psi^{\ \alpha} \ \color{blue} {\cal{F}}_{\ \alpha}^{\ A \ \gamma}
\ \Lambda^{\ A}_{\ \gamma} 
\vspace*{0.2cm} \\
- \ \color{red} \frac{g^{\ 2}}{2} 
\ \color{red} \psi^{\ 2} 
\color{blue} \ \Lambda^{\ A \ \delta}
\ \Lambda^{\ A}_{\ \delta} 
\vspace*{0.2cm} \\
+
\ \color{red} \frac{g^{\ 2}}{4} \ j_{\ \mu \ \psi}
\ \color{blue} j^{\ \mu}_{\ \Lambda} 
\end{array} \ \right \rbrack
\end{array}
\end{equation}

Finally (for $A$) we turn to $\color{red} m \ \color{blue} \varphi$ which is
independent of $D^{\ A}$

\vspace*{-0.5cm}
\begin{equation}
\label{eq:129}
\begin{array}{l}
\color{red} m \ \color{blue} \varphi \ = \ - \ A_{\ 3} \ =
\ \frac{1}{2} \ \color{red} m \ \color{blue} \ \Lambda^{\ A \ \alpha} \ \Lambda^{\ A}_{\ \alpha}
\end{array}
\end{equation}

\noindent
Hence we obtain $A \ = \ \sum_{\ k}^{\ 3} \ A_{\ k}$ ( eqs. \ref{eq:127} - \ref{eq:129} )

}

\newpage

\color{blue}

{\large \bf

\vspace*{-0.7cm}
\begin{equation}
\label{eq:130}
\begin{array}{l}
A \ = \ \ \sum_{\ k}^{\ 3} \ A_{\ k} \ =
\vspace*{0.2cm} \\
 \left \lbrack  \begin{array}{c}
\begin{array}{l}
\color{red} ( \ \frac{1}{2 \ g^{\ 2}} \ - \ \frac{\color{magenta} i}{16 \ \pi^{\ 2}} \ \Theta \ )
\ \times
\vspace*{0.2cm} \\
\hspace*{0.4cm}
\color{blue} 
\times \ 
\Lambda^{\ * \ B}_{\ \dot{\beta}}
\ \left \lbrace \ \begin{array}{c}
\frac{i}{2} \ \stackrel{\leftharpoondown \hspace*{-0.3cm} \rightharpoonup}{\partial}_{\ \mu} 
\ \delta_{\ B A}
\vspace*{0.2cm} \\
- \ v_{\ \mu}^{\ C} \ \left ( \ {\cal{F}}^{\ C} \ \right )_{\ B A}
\end{array}                     
\right \rbrace
\ \sigma^{\ \mu \ \dot{\beta} \alpha} \ \Lambda^{\ A}_{\ \alpha}
\end{array}
\vspace*{0.2cm} \\
- \ \color{red} ( \ \frac{1}{32 \ \pi^{\ 2}} \ \Theta 
\ + \ \frac{\color{magenta} i}{4 \ g^{\ 2}} \ )
\ \color{blue} \partial_{\ \mu} \ j^{\ \mu}_{\ \Lambda} 
\vspace*{0.2cm} \\
- \ \color{red} ( \ \frac{1}{8 \ g^{\ 2}} \ - \ \frac{\color{magenta} i}{64 \ \pi^{\ 2}} \ \Theta \ )
\color{blue} 
\ \ F^{\ \mu \nu \ A} \ \left ( \ F_{\ \mu \nu}^{\ A}
\ - \ \color{magenta} i \ \color{blue} \widetilde{F}_{\ \mu \nu}^{\ A} \ \color{blue} \right )
\vspace*{0.2cm} \\
 \begin{array}{l} 
\color{blue} + 
\ \color{red} ( \ \frac{g^{\ 2}}{8} \ - \ \frac{\color{magenta} i}{64 \ \pi^{\ 2}} \ g^{\ 4} 
\ \Theta \ ) \ \times
\vspace*{0.2cm} \\
\hspace*{0.7cm}  \color{blue} \times \ \left ( \ \begin{array}{c}
\color{red} - \ \psi^{\ 2} 
\ \color{blue} \Lambda^{\ A \ \alpha} \ \Lambda_{\ A \ \alpha} \ + \ h.c.
\vspace*{0.2cm} \\
+ \ \color{red} j_{\ \mu \ \psi}
\ \color{blue} j^{\ \mu}_{\ \Lambda}
\end{array}
\right )
\end{array}
\vspace*{0.3cm} \\
+ \ \color{red} \frac{g^{\ 2}}{2} 
\ \color{red} \psi^{\ 2} 
\color{blue} \ \Lambda^{\ A \ \alpha}
\ \Lambda^{\ A}_{\ \alpha} \ -
\ \color{red} \frac{g^{\ 2}}{4} \ j_{\ \mu \ \psi}
\ \color{blue} j^{\ \mu}_{\ \Lambda}
\vspace*{0.3cm} \\
- \ \frac{1}{\sqrt{2}} \ \color{red} \psi^{\ \alpha} \ \color{blue} {\cal{F}}_{\ \alpha}^{\ A \ \gamma}
\ \Lambda^{\ A}_{\ \gamma}
\ - \ \frac{1}{2} \ \color{red} m \ \color{blue} \ \Lambda^{\ A \ \alpha} \ \Lambda^{\ A}_{\ \alpha} 
\end{array}
 \right \rbrack
\vspace*{0.2cm} \\
\mbox{\color{red} result for} \ \color{blue} A
\end{array}
\end{equation}

\noindent
We repeat the structure of ${\cal{L}} \ ( \ \color{red} J \ \color{blue} )$ ( eq. \ref{eq:125} )

\vspace*{-0.5cm}
\begin{equation}
\label{eq:131}
\begin{array}{l}
{\cal{L}} \ ( \ \color{red} J \ \color{blue} ) \ =
\hspace*{0.2cm}
\left \lbrace \begin{array}{l}
\vartheta^{\ 2} \ A
\vspace*{0.2cm} \\
+ \ \vartheta^{\ \alpha} \ B_{\ \alpha}
\vspace*{0.2cm} \\
+ \ C  
\end{array}
\right \rbrace \ ( \ x^{\ -} \ )
\end{array}
\end{equation}

}

\newpage

\color{blue}

{\large \bf

\noindent
Next we turn to the fermionic components $B_{\ \alpha}$ of 
${\cal{L}} \ ( \ \color{red} J \ \color{blue} )$ ( eq. \ref{eq:126} )

\vspace*{-0.5cm}
\begin{equation}
\label{eq:132}
\begin{array}{l}
B_{\ \alpha} \ = \ {\color{red} j} \ {\color{blue} \eta_{\ \alpha} \ +}
\ {\color{red} \psi_{\ \alpha}} \ {\color{blue} \varphi}
\vspace*{0.4cm} \\
\eta_{\ \alpha} \ =
\ \left \lbrack \ \begin{array}{c}
\ \frac{1}{\sqrt{2}} \ {\cal{F}}_{\ \alpha}^{\ A \ \gamma}
\ \Lambda^{\ A}_{\ \gamma} 
\vspace*{0.2cm} \\
- \ \color{red} \frac{g^{\ 2}}{4} 
\ \color{red} \psi_{\ \alpha} 
\color{blue} \ \Lambda^{\ A \ \delta}
\ \Lambda^{\ A}_{\ \delta} 
\vspace*{0.2cm} \\
-
\ \color{red} \frac{g^{\ 2}}{4} \ \psi^{\ *}_{\ \mu \ \alpha}
\ \color{blue} j^{\ \mu}_{\ \Lambda} 
\end{array} \ \right \rbrack
\vspace*{0.4cm} \\
\varphi \ = \  
\ \frac{1}{2} \ \Lambda^{\ A \ \delta} \ \Lambda^{\ A}_{\ \delta}
\end{array}
\end{equation}

\noindent
Thus we find for $B_{\ \alpha}$ and $C \ = \ \color{red} j \ \color{blue} \varphi$

\vspace*{-0.5cm}
\begin{equation}
\label{eq:133}
\begin{array}{l}
B_{\ \alpha} \ =
\ \left \lbrack  \begin{array}{c}
\color{red} ( \ \frac{1}{2 \ g^{\ 2}} \ - \ \frac{\color{magenta} i}{16 \ \pi^{\ 2}} \ \Theta \ )
\ \color{blue} \frac{1}{\sqrt{2}} \ {\cal{F}}_{\ \alpha}^{\ A \ \gamma}
\ \Lambda^{\ A}_{\ \gamma} 
\vspace*{0.2cm} \\
- \ \color{red} ( \ \frac{1}{8} \ - \ \frac{\color{magenta} i \ \color{red} g^{\ 2}}
{16 \ \pi^{\ 2}} \ \Theta \ ) \ \psi_{\ \alpha}
\ \color{blue} \Lambda^{\ A \ \delta} \ \Lambda^{\ A}_{\ \delta} 
\vspace*{0.2cm} \\
- \ \color{red} ( \ \frac{1}{8} \ - \ \frac{\color{magenta} i \ \color{red} g^{\ 2}}
{16 \ \pi^{\ 2}} \ \Theta \ ) \ \psi^{\ *}_{\ \mu \ \alpha}
\ \color{blue} j^{\ \mu}_{\ \Lambda}
\vspace*{0.2cm} \\
+ \ \color{red} \psi_{\ \alpha} 
\ \color{blue} \ \frac{1}{2} \ \Lambda^{\ A \ \delta} \ \Lambda^{\ A}_{\ \delta} 
\end{array}
\ \right \rbrack
\vspace*{0.5cm} \\
C \ = \ \color{red} ( \ \frac{1}{2 \ g^{\ 2}} \ - \ \frac{\color{magenta} i}{16 \ \pi^{\ 2}} \ \Theta \ )
\ \color{blue} \ \frac{1}{2} \ \Lambda^{\ A \ \delta} \ \Lambda^{\ A}_{\ \delta}
\vspace*{0.4cm} \\
\mbox{\color{red} results for} \ \color{blue} B_{\ \alpha} \ , \ C
\end{array}
\end{equation}

}

\newpage

\color{blue}

{\large \bf

\color{red}
\begin{center}
Remark to $D^{\ A}$ elimination \footnote{\color{blue} \hspace*{0.1cm}
Check all results in subsection 6c) .}
\end{center}

\color{blue}

\noindent
It is possible, that the auxiliary fields $D^{\ A}$ reappear through gauge fixing
in a susy covariant way.

\noindent
We can at any point reinsert them in ${\cal{L}} \ = \ ( \ A \ , \ B_{\ \alpha} \ , \ C \ )$
using the quantities ( eq. \ref{eq:120} )

\vspace*{-0.5cm}
\begin{equation}
\label{eq:134}
\begin{array}{l}
\widetilde{D}^{\ A} \ = \ D^{\ A} \ -
\ \color{red} \frac{g^{\ 2}}{\sqrt{2}} \ \color{blue} \left (
\ \color{red} \psi^{\ \alpha} 
\color{blue} \ \Lambda^{\ A}_{\ \alpha} \ + \ \Lambda^{\ * \ A}_{\ \dot{\beta}}
\ \color{red} \psi^{\ * \ \dot{\beta}}
\ \color{blue} \right )
\end{array}
\end{equation}

\noindent
through the substitution

\vspace*{-0.5cm}
\begin{equation}
\label{eq:135}
\begin{array}{l}
{\cal{L}} \ \rightarrow \ {\cal{L}} \ + \ \Delta \ {\cal{L}}
\vspace*{0.3cm} \\
\Delta \ A \ = \ \color{red} j \ \color{blue} \frac{1}{2}
\ \widetilde{D}^{\ A} \ \widetilde{D}^{\ A}
\vspace*{0.3cm} \\
\Delta \ B_{\ \alpha} \ = \ \color{red} j \ \color{blue}
\frac{1}{\sqrt{2}}
\left \lbrack \ \widetilde{D}^{\ A} \ \Lambda^{\ A}_{\ \alpha}
\ \right \rbrack
\vspace*{0.3cm} \\
\Delta \ C \ = \ 0
\end{array}
\end{equation}

}

\newpage

\color{blue}

{\large \bf

\color{red}
\begin{center}
7) The Legendre transform from (minimal) sources to classical fields 
representing the Lagrangean multiplet
\end{center}

\color{blue}

\noindent
We go back to the operator Lagrangean multiplet coupled to the minimal source
extension as shown in eqs. ( \ref{eq:98} - \ref{eq:100} ) reproduced below

\vspace*{-0.5cm}
\begin{equation}
\label{eq:136}
\begin{array}{l}
\underline{\Phi} \ =
\hspace*{0.2cm}
\left \lbrace \begin{array}{l}
\vartheta^{\ 2} \ \underline{H}
\vspace*{0.2cm} \\
+ \ \vartheta^{\ \alpha} \ \underline{\eta}_{\ \alpha}
\vspace*{0.2cm} \\
+ \ \underline{\varphi}  
\end{array}
\right \rbrace \ ( \ x^{\ -} \ ) \ \color{red} \leftrightarrow
\vspace*{0.3cm} \\
\color{red} J \ \color{blue} =
\hspace*{0.2cm}
\left \lbrace \begin{array}{l}
\vartheta^{\ 2} \ ( \ \color{red}  - \ m \ ) 
\vspace*{0.2cm} \\
\color{blue}
+ \ \vartheta^{\ \alpha} \ \color{red} \psi_{\ \alpha}
\vspace*{0.2cm} \\
\color{blue}
+ \ \color{red}  j 
\end{array}
\color{blue}
\right \rbrace \ ( \ x^{\ -} \ )
\vspace*{0.4cm} \\
\color{red} j \ \color{blue} ( \ y \ ) \ = \ \frac{1}{2} 
\ \left \lbrack \ ( \color{red} \ 1 \ / \ g^{\ 2} 
\ \color{blue} ( \ y \ ) \ ) \ - \ \color{red} \frac{i}{8 \pi^{ 2}} \ \Theta 
\ \color{blue} ( \ y \ )
\ \right \rbrack
\end{array}
\end{equation}

\vspace*{-0.5cm}
\begin{equation}
\label{eq:137}
\begin{array}{l}
\underline{L} \ ( \ \color{red} J \ \color{blue} ) \ = \ {\displaystyle{\int}} \ d^{\ 2} \ \vartheta 
\ \left ( \ \underline{\Phi} \ \color{red} J \ \color{blue} \right ) \ + \ h.c.
\vspace*{0.3cm} \\
\underline{{\cal{L}}} \ ( \ \color{red} J \ \color{blue} ) \ = \ \underline{\Phi} \ \color{red} J
\ \color{blue} = \ ( \ \underline{A} \ , \ \underline{B}_{\ \alpha} \ , \ \underline{C} \ ) 
\end{array}
\end{equation}

\noindent
The explicit form of ${\cal{L}}$ is given in eqs. ( \ref{eq:130} and \ref{eq:133} - \ref{eq:135} )  
and the general boundary conditions specified in eq. ( \ref{eq:102} ) .

}

\newpage

\color{blue}

{\large \bf

\noindent
In eqs. ( \ref{eq:136} - \ref{eq:137} ) underlined quantities are operator valued,
distinguished in this section relative to the associated classical fields,
defined below.

\noindent
We do not discuss here gauge fixing and associated ghost action, which
serve to define a gauge invariant measure in path space of the field configurations
associated with the operator valued base variables

\vspace*{-0.5cm}
\begin{equation}
\label{eq:138}
\begin{array}{l}
\left \lbrace \ \underline{v}_{\ \mu}^{\ A} \ , \ \underline{\Lambda}_{\ \alpha}^{\ A} 
\ , \ \underline{\Lambda}_{\ \dot{\alpha}}^{\ * \ A} \ \right \rbrace 
\ \color{red} \longrightarrow
\vspace*{0.2cm} \\
\color{blue} d \ \mu \ \left \lbrace \ v_{\ \mu}^{\ A} \ , \ \Lambda_{\ \alpha}^{\ A} 
\ , \ \Lambda_{\ \dot{\alpha}}^{\ * \ A} \ \right \rbrace
\end{array}
\end{equation}

\noindent
The measure $\mu$ in eq. ( \ref{eq:138} ) , subject to full renormalization,
defines the generating functional

\vspace*{-0.5cm}
\begin{equation}
\label{eq:139}
\begin{array}{l}
Z \ ( \ \color{red} J \ , \ J^{\ *} \ \color{blue} ) \ =
\ {\displaystyle{\int}} \ d \ \mu  \left \lbrace \ v_{\ \mu}^{\ A} \ , \ \Lambda_{\ \alpha}^{\ A} 
\ , \ \Lambda_{\ \dot{\alpha}}^{\ * \ A} \ \right \rbrace
\ \exp \ i \ S
\vspace*{0.3cm} \\
S \ = \ {\displaystyle{\int}} \ d^{\ 4} \ y 
\ \left ( \ {\cal{L}} \ ( \ \color{red} J \ \color{blue} ) \ + 
\ {\cal{L}}^{\ *} \ ( \ \color{red} J^{\ *} \ \color{blue} ) \ \right )  
\vspace*{0.3cm} \\
Z \ ( \ \color{red} J \ , \ J^{\ *} \ \color{blue} ) \ = \ \exp \ i
\ W \ ( \ \color{red} J \ , \ J^{\ *} \ \color{blue} )
\end{array}
\end{equation}

\noindent
W defines the itransition from the operator valued Lagrangean multiplet ( eq. \ref{eq:136} )
to its associated 

}

\newpage

\color{blue}

{\large \bf

\noindent
classical fields

\vspace*{-0.5cm}
\begin{equation}
\label{eq:140}
\begin{array}{l}
\underline{\Phi} \ =
\hspace*{0.2cm}
\left \lbrace \begin{array}{l}
\vartheta^{\ 2} \ \underline{H}
\vspace*{0.2cm} \\
+ \ \vartheta^{\ \alpha} \ \underline{\eta}_{\ \alpha}
\vspace*{0.2cm} \\
+ \ \underline{\varphi}  
\end{array}
\right \rbrace \ ( \ y \ ) \ \rightarrow
\vspace*{0.3cm} \\
\hspace*{1.5cm} \rightarrow \  \Phi \ =
\hspace*{0.2cm}
\left \lbrace \begin{array}{l}
\vartheta^{\ 2} \ H
\vspace*{0.2cm} \\
+ \ \vartheta^{\ \alpha} \ \eta_{\ \alpha}
\vspace*{0.2cm} \\
+ \ \varphi  
\end{array}
\right \rbrace \ ( \ y \ )
\vspace*{0.3cm} \\
\left ( \ \delta \ / \ \delta \ \color{red} J \ \color{blue} ( \ y \ ) \ \right )
\ W \ ( \ \color{red} J \ , \ J^{\ *} \ \color{blue} )
\ = \ \Phi \ ( \ y \ )
\vspace*{0.3cm} \\
\left ( \ \delta \ / \ \delta \ \color{red} J^{\ *} \ \color{blue} ( \ y \ ) \ \right )
\ W \ ( \ \color{red} J \ , \ J^{\ *} \ \color{blue} )
\ = \ \Phi^{\ *} \ ( \ y \ )
\end{array}
\end{equation}

\noindent
The effective potential $\Gamma$ denotes the Legendre transforme of W 

\vspace*{-0.5cm}
\begin{equation}
\label{eq:141}
\begin{array}{l}
\Gamma \ ( \ \Phi \ , \ \Phi^{\ *} \  )
 =  \left \lbrace 
 \begin{array}{c}
{\displaystyle{\int}} \ d^{\ 4} \ y
\ \left \lbrack
 \begin{array}{c}
\Phi \ ( \ y \ ) \ \color{red} J \ \color{blue} ( \ y \ ) 
\vspace*{0.2cm} \\
+ \ \Phi^{\ *} \ ( \ y \ ) \ \color{red} J^{\ *} \ \color{blue} ( \ y \ )
 \end{array}
 \right \rbrack
\vspace*{0.2cm} \\
 - \ W
\end{array}
 \right \rbrace
\end{array}
\end{equation}

\noindent
The arguments of $\Gamma$ ( eq. \ref{eq:141} ) are {\it the} classical fields pertaining
to the operators $\underline{\Phi} \ , \ \underline{\Phi}^{\ *}$ as indicated in 
eq. (\ref{eq:140}) .

}

\newpage

\color{blue}

{\large \bf

\noindent
The former $\Phi \ , \ \Phi^{\ *}$ are to be determined as functionals of the
sources $\color{red} J \ , \ J^{\ *}$ from the defining, generating functional W,
according to eq. (\ref{eq:140}) . The effective potential then determines associated sources,
as multiple Legendre transforms are involutory \cite{LBPM}

\vspace*{-0.5cm}
\begin{equation}
\label{eq:142}
\begin{array}{l}
\left ( \ \delta \ / \ \delta \  \Phi \ ( \ y \ ) \ \right )
\ \Gamma \ ( \ \Phi \ , \ \Phi^{\ *} \ )
\ = \ \color{red} J \ \color{blue} ( \ y \ )
\vspace*{0.2cm} \\
\left ( \ \delta \ / \ \delta \  \Phi^{\ *} \ ( \ y \ ) \ \right )
\ \Gamma \ ( \ \Phi \ , \ \Phi^{\ *} \ )
\ = \ \color{red} J^{\ *} \ \color{blue} ( \ y \ )
\vspace*{0.3cm} \\
W \ ( \ \color{red} J \ , \ J^{\ *} \ \color{blue} )
 =  \left \lbrace 
 \begin{array}{c}
{\displaystyle{\int}} \ d^{\ 4} \ y
 \left \lbrack
 \begin{array}{c}
\Phi \ ( \ y \ ) \ \color{red} J \ \color{blue} ( \ y \ ) 
\vspace*{0.2cm} \\
+ \ \Phi^{\ *} \ ( \ y \ ) \ \color{red} J^{\ *} \ \color{blue} ( \ y \ )
 \end{array}
 \right \rbrack
\vspace*{0.2cm} \\
 - \ \Gamma
\end{array}
 \right \rbrace
\end{array}
\end{equation}

\color{red}
\begin{center}
Vanishing minimal sources, vanishing momenta and effective potential for
the vacuum expected values of $\underline{\Phi} \ , \ \underline{\Phi}^{\ *}$
\end{center}

\color{blue}

\noindent
For vanishing minimal sources $\color{red} J \ , \ J^{\ *} \ \color{blue} = \ 0$
eq. (\ref{eq:142}) requires an extremum (minimum) for the effective potential $\Gamma$

\vspace*{-0.5cm}
\begin{equation}
\label{eq:143}
\begin{array}{l}
\left ( \ \delta \ / \ \delta \  \Phi \ ( \ y \ ) \ \right )
\ \Gamma \ ( \ \Phi \ , \ \Phi^{\ *} \ )
\ = \ 0
\vspace*{0.2cm} \\
\left ( \ \delta \ / \ \delta \  \Phi^{\ *} \ ( \ y \ ) \ \right )
\ \Gamma \ ( \ \Phi \ , \ \Phi^{\ *} \ )
\ = \ 0
\end{array}
\end{equation}

}

\newpage

\color{blue}

{\large \bf

\color{red}
\begin{center}
{\small
Remark on the naive classical limit $\hbar \ = \ 0$}
\end{center}

\color{blue}

\noindent
If there were no renormalization to be done {\it also} in the classical limit, not to
be confused with the classical fields $\Phi \ , \ \Phi^{\ *}$ {\it within} a quantum
field theory, the effective potential reduces to the negative classical action

\vspace*{-0.7cm}
\begin{equation}
\label{eq:144}
\begin{array}{l}
\hbar \ \rightarrow 0 \ :
\vspace*{0.2cm} \\
\ \Gamma \ ( \ \Phi \ , \ \Phi^{\ *} \ ) \ \rightarrow 
\ S_{\ cl} - \ S_{\ cl} \  ( \ \Phi \ , \ \Phi^{\ *} \ ) \ = \ 0
\vspace*{0.2cm} \\ 
S_{\ cl} \ ( \ \Phi \ , \ \Phi^{\ *} \ ) \ = \ {\displaystyle{\int}} \ d^{\ 4} \ y
\ \left . L \ ( \ \Phi \ , \ \Phi^{\ *} \ ) \ \color{red} \right |_{\ \color{red} J \ = \ 0}
\ \color{blue} ( \ y \ )
\end{array}
\end{equation}
\vspace*{-0.5cm} \\

\noindent
The Lagrangean density 
$\left .L \ ( \ \Phi \ , \ \Phi^{\ *} \ ) \ \color{red} \right |_{\ \color{red} J \ = \ 0}$
depends only on the limiting quantities $\color{red} J_{\ \infty}$
and can be constructed in this limit from $A$ in eq. (\ref{eq:130})

\vspace*{-0.8cm}
\begin{equation}
\label{eq:145}
\begin{array}{l}
\left . L \ ( \ \Phi \ , \ \Phi^{\ *} \ ) \ \color{red} \right |_{\ \color{red} J \ = \ 0} 
\ \color{blue} =
\vspace*{0.2cm} \\
 \left \lbrack  \begin{array}{c}
\begin{array}{l}
\color{red} \frac{1}{g^{\ 2}_{\ \infty}}
\ \color{blue} \times
\vspace*{0.2cm} \\
\hspace*{0.4cm}
\color{blue} 
\times \ 
\Lambda^{\ * \ B}_{\ \dot{\beta}}
\ \left \lbrace \ \begin{array}{c}
\frac{i}{2} \ \stackrel{\leftharpoondown \hspace*{-0.3cm} \rightharpoonup}{\partial}_{\ \mu} 
\ \delta_{\ B A}
\vspace*{0.2cm} \\
- \ v_{\ \mu}^{\ C} \ \left ( \ {\cal{F}}^{\ C} \ \right )_{\ B A}
\end{array}                     
\right \rbrace
\ \sigma^{\ \mu \ \dot{\beta} \alpha} \ \Lambda^{\ A}_{\ \alpha}
\end{array}
\vspace*{0.2cm} \\
- \ \color{red} \frac{1}{16 \ \pi^{\ 2}} \ \Theta_{\ \infty} 
\ \color{blue} \partial_{\ \mu} \ j^{\ \mu}_{\ \Lambda} 
\vspace*{0.2cm} \\
- \ \color{red} \frac{1}{4 \ g^{\ 2}_{\ \infty}} 
\color{blue} 
\ F^{\ \mu \nu \ A} \ F_{\ \mu \nu}^{\ A}
\ + \ \color{red} \frac{1}{32 \ \pi^{\ 2}} \ \Theta_{\ \infty}
\ \color{blue} F^{\ \mu \nu \ A} \ \widetilde{F}_{\ \mu \nu}^{\ A}
\vspace*{0.2cm} \\
\ - \ \frac{1}{2} \ \color{red} m_{\ \infty} 
\ \color{blue} \ \Lambda^{\ A \ \alpha} \ \Lambda^{\ A}_{\ \alpha} \ + \ c.c.
\end{array}
 \right \rbrack
\end{array}
\end{equation}

}

\newpage

\color{blue}

{\large \bf

\noindent
Had we chosen sources for the primary fields, denoting them by 

\noindent
$\underline{\chi}_{\ primary} \ = \ \underline{\chi}$ :
$\underline{\Lambda}^{\ A}_{\ \alpha} \ , \ \underline{\Lambda}^{\ * \ A}_{\ \dot{\alpha}}$
and $\underline{v}_{\ \mu}^{\ A}$ , 

\noindent
the corresponding variations in the limit $\hbar \ \rightarrow \ 0$,
under the conditions that the variations  $\delta \ \chi$
vanish at the boundaries of the action integral, the equivalent to eq. (\ref{eq:143}) 
would have reduced to the
Euler-Lagrange field equations

\vspace*{-0.5cm}
\begin{equation}
\label{eq:146}
\begin{array}{l}
\left ( \ \delta \ / \ \delta \  \chi \ ( \ y \ ) \ \right )
\ ( \ - \ S_{\ cl} \ ) \ ( \ \chi \ , \ \chi^{\ *} \ )
\ = \ 0
\vspace*{0.2cm} \\
\left ( \ \delta \ / \ \delta \  \chi^{\ *} \ ( \ y \ ) \ \right )
\ ( \ - \ S_{\ cl} \ ) \ ( \ \chi \ , \ \chi^{\ *} \ )
\ = \ 0 \ \color{red} \rightarrow
\vspace*{0.2cm} \\
\color{blue}
\partial_{\ \mu} \ L_{\ , \ \partial_{\ \mu} \ \chi} \ - \ L_{\ , \ \chi} \ = \ 0
\hspace*{0.2cm} \mbox{and} \hspace*{0.2cm}
\chi \ \rightarrow \ \chi^{\ *}
\end{array}
\end{equation} 

\noindent
This shall illustrate the difference between minimal and primary source extension.
\footnote{\hspace*{0.1cm} \color{red} Prove the above derivation of the classical
Euler-Lagrange equations in the classical limit.}

\color{red}
\begin{center}
{\small
Illustration : the free scalar field and primary source extension}
\end{center}

\color{blue}

\noindent 
We shall calculate for the primary source of a massive, free, complex scalar field $\underline{\chi}$
the associated functionals

}

\newpage

\color{blue}

{\large \bf

\vspace*{-0.5cm}
\begin{equation}
\label{eq:147}
\begin{array}{l}
L \ ( \ \color{red} J \ \color{blue} ) \ \sim 
\vspace*{0.2cm} \\
\ - \ \underline{\chi}^{\ *} \ ( \ m^{\ 2} \ + \ \mbox{\fbox{\rule{0mm}{1.0mm}}} \ )
\ \underline{\chi} \ + 
\ \left \lbrack \ \color{red} J \ \color{blue} ( \ y \ ) \ \underline{\chi} \ + \ h.c. 
\ \right \rbrack \ \color{red} \rightarrow
\vspace*{0.2cm} \\
\color{blue}
\ \underline{\chi}^{\ '} \ = \ \underline{\chi} \ + \ \Delta \ \chi
\vspace*{0.2cm} \\
L \ =
\ \left \lbrack \ \begin{array}{c} 
- \ \underline{\chi}^{\ ' \ *} \ ( \ m^{\ 2} \ + \ \mbox{\fbox{\rule{0mm}{1.0mm}}} \ )
\ \underline{\chi}^{\ '} 
\vspace*{0.2cm} \\
- \ \underline{\chi}^{\ ' \ *} \ ( \ m^{\ 2} \ + \ \mbox{\fbox{\rule{0mm}{1.0mm}}} \ )
\ \Delta \ \chi
\vspace*{0.2cm} \\
- \ \left \lbrace \ ( \ m^{\ 2} \ + \ \mbox{\fbox{\rule{0mm}{1.0mm}}} \ ) \ \Delta \ \chi^{\ *}
\ \right \rbrace \ \underline{\chi}^{\ '}
\vspace*{0.2cm} \\
+ 
\ \color{red} J \ \color{blue} ( \ y \ ) \ \underline{\chi}^{\ '} 
\ + \ \underline{\chi}^{\ ' \ *} \ \color{red} J^{\ *} \ \color{blue} ( \ y \ ) 
\vspace*{0.2cm} \\
+ 
\ \color{red} J \ \color{blue} ( \ y \ ) \ \Delta \ \chi 
\ + \ \Delta \ \chi^{\ *} \ \color{red} J^{\ *} \ \color{blue} ( \ y \ )
\vspace*{0.2cm} \\
- \ \Delta \ \chi^{\ *} \ ( \ m^{\ 2} \ + \ \mbox{\fbox{\rule{0mm}{1.0mm}}} \ )
\ \Delta \ \chi
\end{array}
\ \right \rbrack
\end{array}
\end{equation}

\noindent
We choose $\Delta \ \chi \ , \ \Delta \ \chi^{\ *}$ such as to cancel the source terms 
multiplying the operators $\underline{\chi}^{\ '} \ , \ \underline{\chi}^{\ ' \ *}$ in
eq. (\ref{eq:147})

\vspace*{-0.5cm}
\begin{equation}
\label{eq:148}
\begin{array}{l}
( \ m^{\ 2} \ + \ \mbox{\fbox{\rule{0mm}{1.0mm}}} \ )
\ \Delta \ \chi 
\ = \ \color{red} J^{\ *} \ \color{blue} ( \ y \ )
\hspace*{0.2cm} \mbox{and c.c.} \ \color{red} \rightarrow
\vspace*{0.2cm} \\
\color{blue}
\Delta \ \chi \ = \ ( \ m^{\ 2} \ + \ \mbox{\fbox{\rule{0mm}{1.0mm}}} \ )^{\ -1} 
\ \color{red} J^{\ *} \ \color{blue} ( \ y \ ) 
\hspace*{0.2cm} \mbox{and c.c.}
\vspace*{0.2cm} \\
\mbox{\fbox{\rule{0mm}{1.0mm}}} \ = \ \mbox{\fbox{\rule{0mm}{1.0mm}}}_{\ y}
\end{array}
\end{equation}

}

\newpage

\color{blue}

{\large \bf

\noindent
Treating the Green function $( \ m^{\ 2} \ + \ \mbox{\fbox{\rule{0mm}{1.0mm}}} \ )^{\ -1}$ 
as if it were real and well defined, we obtain for the generating functional $W$

\vspace*{-0.5cm}
\begin{equation}
\label{eq:149}
\begin{array}{l}
W \ ( \ \color{red} J \ , \ \color{red} J^{\ *} \ \color{blue} )
\ =
\vspace*{0.2cm} \\ 
{\displaystyle{\int}} \ d^{\ 4} \ x \ d^{\ 4} \ y
\ \color{red} J \ \color{blue} ( \ x \ ) \ G \ ( \ x \ , \ y \ )
\ \color{red} J^{\ *} \ \color{blue} ( \ y \ )
\vspace*{0.4cm} \\
( \ m^{\ 2} \ + \ \mbox{\fbox{\rule{0mm}{1.0mm}}} \ )^{\ -1}_{\ x} \ . \ =
\ {\displaystyle{\int}} \ d^{\ 4} \ y \ G \ ( \ x \ , \ y \ ) \ .
\end{array}
\end{equation}

\noindent
Eq. (\ref{eq:140}) in this case becomes

\vspace*{-0.5cm}
\begin{equation}
\label{eq:150}
\begin{array}{l}
\left ( \ \delta \ / \ \delta \ \color{red} J \ \color{blue} ( \ y \ ) \ \right )
\ W \ ( \ \color{red} J \ , \ J^{\ *} \ \color{blue} )
\ = \ \chi \ ( \ y \ ) \ ( \ = \ \chi_{\ cl} \ )
\vspace*{0.2cm} \\
\chi \ ( \ y \ ) \ = \ ( \ m^{\ 2} \ + \ \mbox{\fbox{\rule{0mm}{1.0mm}}} \ )^{\ -1}_{\ y}
\ \color{red} J^{\ *} \ \rightarrow 
\vspace*{0.2cm} \\
\color{red} J^{\ *} \ \color{blue} = \ ( \ m^{\ 2} \ + \ \mbox{\fbox{\rule{0mm}{1.0mm}}} \ ) \ \chi
\hspace*{0.2cm} \mbox{and c.c.}
\end{array}
\end{equation}

\noindent
and substituting into the corresponding eq. (\ref{eq:141}) we obtain

\vspace*{-0.5cm}
\begin{equation}
\label{eq:151}
\begin{array}{l}
\Gamma \ ( \ \chi \ , \ \chi^{\ *} \  )
\ =
\vspace*{0.2cm} \\
\ {\displaystyle{\int}} \ d^{\ 4} \ y 
\ \chi^{\ *} \ ( \ m^{\ 2} \ + \ \mbox{\fbox{\rule{0mm}{1.0mm}}} \ )_{\ y} \ \chi
\ = \ - \ S_{\ cl} \ ( \ \chi \ , \ \chi^{\ *} \ ) 
\end{array}
\end{equation} 

\noindent
But the situation changes in its interpretation for a constant nonvanisching source. $\color{red} \rightarrow$

}

\newpage

\color{blue}

{\large \bf

\noindent
For a free field with a source  
$\lim_{\ y \ \rightarrow \ \infty} \ \color{red} J \ \color{blue} ( \ y \ ) \ =
\ \color{red} J_{\ \infty}$ we can deduce the potential energy (density-) term 

\vspace*{-0.5cm}
\begin{equation}
\label{eq:152}
\begin{array}{l}
L \ ( \ \color{red} J \ \color{blue} ) \ = \ L_{\ kin} \ - \ V \ ( \ \color{red} J \ \color{blue} ) \ )
\vspace*{0.2cm} \\
V \ ( \ \color{red} J \ \color{blue} ) \ \sim 
\vspace*{0.2cm} \\
m^{\ 2} \  \underline{\chi}^{\ *} \ \underline{\chi} \ ( \ y \ )
\ - \ \color{red} J_{\ \infty} \ \color{blue} \underline{\chi} \ ( \ y \ )
\ - \ \color{red} J_{\ \infty}^{\ *} \ \color{blue} \underline{\chi}^{\ *} \ ( \ y \ )
\hspace*{0.2cm} \color{red} \rightarrow
\vspace*{0.2cm} \\
\lim_{\ {\cal{V}} \ \rightarrow \ \infty} \ {\cal{V}}^{\ -1} \Gamma_{\ {\cal{V}}} 
\ ( \ \chi \ , \ \chi^{\ *} \ ) \ \rightarrow \ \gamma \ ( \ \chi \ , \ \chi^{\ *} \ ) \ =
\vspace*{0.2cm} \\
m^{\ 2} \ \chi^{\ *} \ \chi \ ( \ y \ )
\ - \ \color{red} J_{\ \infty} \ \color{blue} \chi \ ( \ y \ )
\ - \ \color{red} J_{\ \infty}^{\ *} \ \color{blue} \chi^{\ *} \ ( \ y \ )
\end{array}
\end{equation}

\noindent
In eq. (\ref{eq:152}) ${\cal{V}} \ = \ {\cal{V}}^{\ 4}$ denotes a four dimensional
finitie volume, and the thermodynamic limit corresponds to ${\cal{V}} \ \rightarrow \ \infty$.

\noindent
The sourceless condition now implies in extension (not contradiction) of eqs. 
(\ref{eq:143} and \ref{eq:151})

\vspace*{-0.5cm}
\begin{equation}
\label{eq:153}
\begin{array}{l}
\left ( \ \delta \ / \ \delta \  \chi \ ( \ y \ ) \ \right )
\ \gamma \ ( \ \chi \ , \ \chi^{\ *} \ ) \ = \ 0
\hspace*{0.3cm} \mbox{and c.c.} \ \color{red} \rightarrow
\vspace*{0.2cm} \\
\chi \ ( \ y \ ) \ = \ \chi_{\ \infty} \ = \ \color{red} J_{\ \infty}^{\ *}
\ \color{blue} / \ m^{\ 2}
%\hspace*{0.2cm} , \hspace*{0.2cm}
\vspace*{0.2cm} \\
\chi^{\ *} \ ( \ y \ ) \ = \ \chi_{\ \infty}^{\ *} \ = \ \color{red} J_{\ \infty} 
\ \color{blue} / \ m^{\ 2}  
\end{array}
\end{equation}

\noindent
In the result (eq. \ref{eq:153}) the dangerous infrared limit $m \ \rightarrow \ 0$
becomes apparent.

}

\newpage

\color{blue}

{\large \bf

\noindent
We go back to section 3) (eqs. \ref{eq:12} , \ref{eq:13}) and reproduce the structure 
of the full Lagrangean multiplet ( ${\cal{L}} \ = \ ( \ A \ , \ B_{\ \alpha} \ , \ C \ )$ )
given in \\
eqs. ( \ref{eq:130}, \ref{eq:133} - \ref{eq:135} )
below, in order to derive the infinitesimal susy transformations component by component

\vspace*{-0.5cm}
\begin{equation}
\label{eq:154}
\begin{array}{l}
{\cal{L}}  \ = 
\vspace*{0.3cm} \\
\left \lbrace
\begin{array}{c}
\vartheta^{\ 2} \ \overline{\vartheta}^{\ 2} 
\ \left ( \ - \ \frac{1}{4} \ \mbox{\fbox{\rule{0mm}{1.0mm}}} \ C
\ \right )
\vspace*{0.3cm} \\
\vartheta^{\ 2} \ \overline{\vartheta}_{\ \dot{\delta}}
\ \left ( \ \frac{-i}{2} \ \partial^{\ \dot{\delta} \alpha} \ B_{\ \alpha}
\ \right )
\ + \color{red} \ \overline{\vartheta}^{\ 2} \ \vartheta^{\ \alpha} \ 0
\vspace*{0.3cm} \\
\color{blue}
\vartheta^{\ 2} \ A \ + \  \vartheta^{\ \alpha} 
\ \sigma^{\ \mu}_{ \alpha \dot{\beta}} \ \overline{\vartheta}^{\ \dot{\beta}}
\ \left ( \ \frac{-i}{2} \ \partial_{\ \mu} \ C \ \right )
\ + \ \color{red} \overline{\vartheta}^{\ 2} \ 0
\vspace*{0.3cm} \\
+ \ \color{red} \overline{\vartheta}^{\ \dot{\beta}} \ 0
\color{blue} \ + \ \vartheta^{\ \alpha} \ B_{\ \alpha}
\vspace*{0.3cm} \\
+ \ C
\end{array}
\right \rbrace
\end{array}
\end{equation}

\noindent
Also we recall the infinitesimal susy transformations ( eqs. \ref{eq:30} , \ref{eq:33} 
$\color{red} \rightarrow$ )

\vspace*{-0.5cm}
\begin{equation}
\label{eq:155}
\begin{array}{l}
\color{red}
q_{\ \alpha} \ = \ ( \ \partial_{\ \vartheta} \ )_{\ \alpha}
\ + \ \frac{i}{2} \ \overline{\vartheta}^{\ \dot{\beta}} \ ( \ \partial_{\ x} \ )_{\ \alpha \dot{\beta}} 
\vspace*{0.2cm} \\
\color{red}
\overline{q}_{\ \dot{\beta}} \ = \ ( \ \partial_{\ \overline{\vartheta}} \ )_{\ \dot{\beta}}
\ + \ \frac{i}{2} \ \vartheta^{\ \alpha} \ ( \ \partial_{\ x} \ )_{\ \alpha \dot{\beta}}
\vspace*{0.2cm} \\
\color{magenta}
%\hspace*{0.7cm}
\overline{D}_{\ \dot{\beta}} \ = \ - \ ( \ \partial_{\ \overline{\vartheta}} \ )_{\ \dot{\beta}}
\ + \ \frac{i}{2} \ \vartheta^{\ \alpha} \ ( \ \partial_{\ x} \ )_{\ \alpha \dot{\beta}}
\vspace*{0.2cm} \\
\color{red}
\overline{q}_{\ \dot{\beta}} \ = \ 2 \ ( \ \partial_{\ \overline{\vartheta}} \ )_{\ \dot{\beta}}
\ + \ \overline{D}_{\ \dot{\beta}}
\end{array}
\end{equation}

\color{blue}

}

\newpage

\color{blue}

{\large \bf

\vspace*{-0.5cm}
\begin{equation}
\label{eq:156}
\begin{array}{l}
\left \lbrack \ \delta_{\ U} \ ( \ \eta \ , \ \overline{\eta} \ ) \ \right \rbrack
F \ ( \ \vartheta \ , \ \overline{\vartheta} \ , \ x \ )
\vspace*{0.2cm} \\
\hspace*{0.3cm} \sim
\ F_{\ U} \ ( \ \vartheta \ , \ \overline{\vartheta} \ , \ x \ ) \ -
\ F \ ( \ \vartheta \ , \ \overline{\vartheta} \ , \ x \ )
\vspace*{0.2cm} \\
\hspace*{0.3cm} = 
\ \eta^{\ \alpha} \ q_{\ \alpha} \ F \ + \ \overline{\eta}^{\ \dot{\alpha}}
\ \overline{q}_{\ \dot{\alpha}} \ F
\vspace*{0.2cm} \\
\color{red}
q_{\ \alpha} \ = \ ( \ \partial_{\ \vartheta} \ )_{\ \alpha}
\ + \ \frac{i}{2} \ \overline{\vartheta}^{\ \dot{\beta}} \ ( \ \partial_{\ x} \ )_{\ \alpha \dot{\beta}} 
\vspace*{0.2cm} \\
\color{red}
\overline{q}_{\ \dot{\beta}} \ = \ ( \ \partial_{\ \overline{\vartheta}} \ )_{\ \dot{\beta}}
\ + \ \frac{i}{2} \ \vartheta^{\ \alpha} \ ( \ \partial_{\ x} \ )_{\ \alpha \dot{\beta}} 
\vspace*{0.2cm} \\
\color{red}
( \ \partial_{\ x} \ )_{\ \alpha \dot{\beta}} \ =
\ ( \ \partial_{\ x} \ )^{\ \mu} \ \sigma_{\ \mu \ \alpha \dot{\beta}}
\vspace*{0.2cm} \\
\color{red}
( \ \partial_{\ \vartheta} \ )_{\ \alpha} \ = \ \partial \ / \ \partial \ \vartheta^{\ \alpha}
\hspace*{0.2cm} ; \hspace*{0.2cm}
( \ \partial_{\ \overline{\vartheta}} \ )_{\ \dot{\beta}} \ =
\ \partial \ / \ \partial \ \overline{\vartheta}^{\ \dot{\beta}}
\end{array}
\end{equation}

\noindent
Proceeding step by step we first calculate
$\color{red} ( \ \partial_{\ \vartheta} \ )_{\ \alpha} \ \color{blue} {\cal{L}}$ and
$\color{red} 2 \ ( \ \partial_{\ \overline{\vartheta}} \ )_{\ \dot{\beta}} \ \color{blue} {\cal{L}}
\ \color{blue} = \ \color{red} \overline{q}_{\ \dot{\beta}} \ \color{blue} {\cal{L}}$ .

\vspace*{-0.5cm}
\begin{equation}
\label{eq:157}
\begin{array}{l}
\color{red} ( \ \partial_{\ \vartheta} \ )_{\ \alpha} \ \color{blue} {\cal{L}}  \ = 
\vspace*{0.3cm} \\
\left \lbrace
\begin{array}{c}
\vartheta^{\ 2} \ \overline{\vartheta}^{\ 2} \ \color{red} 0
\vspace*{0.3cm} \\
\vartheta^{\ 2} \ \overline{\vartheta}_{\ \dot{\delta}} \ \color{red} 0
\ \color{blue} + \color{red} \ \overline{\vartheta}^{\ 2} \ \vartheta_{\ \alpha} 
\ \left ( \ - \ \frac{1}{4} \ \mbox{\fbox{\rule{0mm}{1.0mm}}} \ C
\ \right )
\vspace*{0.3cm} \\
\color{blue}
\vartheta^{\ 2} \ \color{red} 0 \ \color{blue} + \  \vartheta_{\ \alpha} 
\ \sigma^{\ \mu \ \dot{\delta} \gamma} \ \overline{\vartheta}_{\ \dot{\delta}}
\ \left ( \ \frac{-i}{2} \ \partial_{\ \mu} \ B_{\ \gamma} \ \right )
\ + \ \color{red} \overline{\vartheta}^{\ 2} \ 0
\vspace*{0.3cm} \\
+ \ \color{red} \sigma^{\ \mu}_{\ \alpha \dot{\delta}} \ \overline{\vartheta}^{\ \dot{\delta}}
\ \left ( \ \frac{-i}{2} \ \partial_{\ \mu} \ C \ \right )
\color{blue} \ + \ \vartheta_{\ \alpha} \ A
\vspace*{0.3cm} \\
+ \ B_{\ \alpha}
\end{array}
\right \rbrace
\end{array}
\end{equation}

}

\newpage

\color{blue}

{\large \bf

\vspace*{-0.5cm}
\begin{equation}
\label{eq:158}
\begin{array}{l}
\color{red} \color{red} \overline{q}_{\ \dot{\beta}} \ \color{blue} {\cal{L}} \ =
\ \color{red} 2 \ ( \ \partial_{\ \overline{\vartheta}} \ )_{\ \dot{\beta}} \ \color{blue} {\cal{L}}  \ = 
\vspace*{0.3cm} \\
\left \lbrace
\begin{array}{c}
\vartheta^{\ 2} \ \overline{\vartheta}^{\ 2} \ \color{red} 0
\vspace*{0.3cm} \\
\vartheta^{\ 2} \ \overline{\vartheta}_{\ \dot{\beta}}
\ \left ( \ \frac{1}{2} \ \mbox{\fbox{\rule{0mm}{1.0mm}}} \ C
\ \right )
\ + \color{red} \ \overline{\vartheta}^{\ 2} \ \vartheta^{\ \gamma} \ 0
\vspace*{0.3cm} \\
\color{blue}
\vartheta^{\ 2} \ \ \left ( \ -i \ \partial_{\ \gamma \dot{\beta}} \ B^{\ \gamma}
\ \right )
\ + \  \vartheta^{\ \gamma} 
\ \sigma^{\ \mu}_{ \gamma \dot{\delta}} \ \overline{\vartheta}^{\ \dot{\delta}}
\ \color{red} 0
\ \color{blue} + \ \color{red} \overline{\vartheta}^{\ 2} \ 0
\vspace*{0.3cm} \\
+ \ \color{red} \overline{\vartheta}^{\ \dot{\beta}} \ 0
\color{blue} \ + \ \vartheta^{\ \gamma} \ \sigma^{\ \mu}_{\ \gamma \dot{\beta}}
\ \left ( \ i \ \partial_{\ \mu} \ C \ \right ) 
\vspace*{0.3cm} \\
+ \ \color{red} 0
\end{array}
\right \rbrace
\end{array}
\end{equation}

\noindent
Next we compare $\color{red}  \overline{q}_{\ \dot{\beta}} \ \color{blue} {\cal{L}}$
with ${\cal{L}}$ ( eq. \ref{eq:154} ) reproduced below

\vspace*{-0.5cm}
\begin{equation}
\label{eq:159}
\begin{array}{l}
{\cal{L}}  \ = 
\vspace*{0.3cm} \\
\left \lbrace
\begin{array}{c}
\vartheta^{\ 2} \ \overline{\vartheta}^{\ 2} 
\ \left ( \ - \ \frac{1}{4} \ \mbox{\fbox{\rule{0mm}{1.0mm}}} \ C
\ \right )
\vspace*{0.3cm} \\
\vartheta^{\ 2} \ \overline{\vartheta}_{\ \dot{\delta}}
\ \left ( \ \frac{-i}{2} \ \partial^{\ \dot{\delta} \alpha} \ B_{\ \alpha}
\ \right )
\ + \color{red} \ \overline{\vartheta}^{\ 2} \ \vartheta^{\ \alpha} \ 0
\vspace*{0.3cm} \\
\color{blue}
\vartheta^{\ 2} \ A \ + \  \vartheta^{\ \alpha} 
\ \sigma^{\ \mu}_{ \alpha \dot{\beta}} \ \overline{\vartheta}^{\ \dot{\beta}}
\ \left ( \ \frac{-i}{2} \ \partial_{\ \mu} \ C \ \right )
\ + \ \color{red} \overline{\vartheta}^{\ 2} \ 0
\vspace*{0.3cm} \\
+ \ \color{red} \overline{\vartheta}^{\ \dot{\beta}} \ 0
\color{blue} \ + \ \vartheta^{\ \alpha} \ B_{\ \alpha}
\vspace*{0.3cm} \\
+ \ C
\end{array}
\right \rbrace
\end{array}
\end{equation}

}

\newpage

\color{blue}

{\large \bf

\noindent
We multiply $\color{red}  \overline{q}_{\ \dot{\beta}}$ with an infinitesimal Grassmann
spinor $\color{magenta} \overline{\tau}^{\ \dot{\beta}}$ from the left and recast
eq. (\ref{eq:158}) into the form

\vspace*{-0.5cm}
\begin{equation}
\label{eq:160}
\begin{array}{l}
\delta_{\ 2} \ = \ \color{magenta} \overline{\tau}^{\ \dot{\delta}}
\ \color{red}  \overline{q}_{\ \dot{\beta}}
\vspace*{0.2cm} \\
\delta_{\ 2} \ \color{blue} {\cal{L}} \ = 
\vspace*{0.3cm} \\
\left \lbrace
\begin{array}{c}
\vartheta^{\ 2} \ \overline{\vartheta}^{\ 2} \ \color{red} 0
\vspace*{0.3cm} \\
\vartheta^{\ 2} \ \overline{\vartheta}_{\ \dot{\delta}}
\ \color{magenta} \overline{\tau}^{\ \dot{\delta}}
\ \color{blue} \left ( \ - \ \frac{1}{2} \ \mbox{\fbox{\rule{0mm}{1.0mm}}} \ C
\ \right )
\ + \color{red} \ \overline{\vartheta}^{\ 2} \ \vartheta^{\ \gamma} \ 0
\vspace*{0.3cm} \\
\color{blue}
\vartheta^{\ 2} \ \color{magenta} \overline{\tau}^{\ \dot{\delta}}
\ \color{blue} \left ( \ -i \ \partial_{\ \gamma \dot{\delta}} \ B^{\ \gamma}
\ \right )
\ + \  \vartheta^{\ \gamma} 
\ \sigma^{\ \mu}_{ \gamma \dot{\delta}} \ \overline{\vartheta}^{\ \dot{\delta}}
\ \color{red} 0
\ \color{blue} + \ \color{red} \overline{\vartheta}^{\ 2} \ 0
\vspace*{0.3cm} \\
+ \ \color{red} \overline{\vartheta}^{\ \dot{\beta}} \ 0
\color{blue} \ + \ \vartheta^{\ \alpha} \ \sigma^{\ \mu}_{\ \alpha \dot{\delta}}
\ \color{magenta} \overline{\tau}^{\ \dot{\delta}}
\ \color{blue} \left ( \ - \ i \ \partial_{\ \mu} \ C \ \right ) 
\vspace*{0.3cm} \\
+ \ \color{red} 0
\end{array}
\right \rbrace
\end{array}
\end{equation}

\noindent
Comparing with eq. (\ref{eq:159}) we obtain, proceedeing slowly, term by term

\vspace*{-0.5cm}
\begin{equation}
\label{eq:161}
\begin{array}{l}
\delta_{\ 2} \ C \ = \ 0
\vspace*{0.2cm} \\
\delta_{\ 2} \ \left ( \ \frac{-i}{2} \ \partial^{\ \dot{\delta} \alpha} \ B_{\ \alpha} \ \right )
\ = \ \color{magenta} \overline{\tau}^{\ \dot{\delta}}
\ \color{blue} \left ( \ - \ \frac{1}{2} \ \mbox{\fbox{\rule{0mm}{1.0mm}}} \ C
\ \right ) \ \color{red} \rightarrow
\vspace*{0.2cm} \\
\delta_{\ 2} \ \left ( \ i \ \partial_{\ \mu} \ \sigma^{\ \mu \ \dot{\delta} \alpha} \ B_{\ \alpha}
\ \right )
\ = \ \partial_{\ \mu} \ \left ( \ \color{magenta} \overline{\tau}^{\ \dot{\delta}}
\ \color{blue} \partial^{\ \mu} \ C \ \right )
\end{array}
\end{equation}

\noindent
We rewrite the last relation in eq. (\ref{eq:161})

}

\newpage

\color{blue}

{\large \bf

\vspace*{-0.5cm}
\begin{equation}
\label{eq:162}
\begin{array}{l}
i \ \partial^{\ \dot{\delta} \alpha} \ \delta_{\ 2} \ B_{\ \alpha} \ =
\ \color{magenta} \overline{\tau}^{\ \dot{\delta}} \ \color{blue} \mbox{\fbox{\rule{0mm}{1.0mm}}} \ C
\vspace*{0.2cm} \\
\partial_{\ \gamma \dot{\delta}} \ \partial^{\ \dot{\delta} \alpha} \ = 
\ \delta_{\ \gamma}^{\ \ \alpha} \ \mbox{\fbox{\rule{0mm}{1.0mm}}}
\end{array}
\end{equation}

\noindent
Further we obtain

\vspace*{-0.5cm}
\begin{equation}
\label{eq:163}
\begin{array}{l}
\delta_{\ 2} \ A \ = \ \color{magenta} \overline{\tau}_{\ \dot{\delta}}
\ \color{blue} \left ( \ -i \ \partial^{\ \dot{\delta} \alpha} \ B_{\ \alpha} \ \right )
\vspace*{0.2cm} \\
\delta_{\ 2} \ \partial_{\ \mu} \ C \ = \ 0 \ \color{red} \leftarrow
\ \color{blue} \delta_{\ 2} \ C \ = \ 0
\vspace*{0.2cm} \\
\delta_{\ 2} \ B_{\ \alpha} \ = \ \color{magenta} \overline{\tau}^{\ \dot{\delta}}
\ \color{blue} \left ( \ - \ i \ \partial_{\ \alpha \dot{\delta}} \ C \ \right )
\end{array}
\end{equation}

\noindent
Finally -- within the completion of $\delta_{\ 2}$ we operate with
$i \ \partial^{\ \dot{\gamma} \ \alpha}$ on the last relation in eq. (\ref{eq:163})

\vspace*{-0.5cm}
\begin{equation}
\label{eq:164}
\begin{array}{l}
i \ \partial^{\ \dot{\gamma} \ \alpha} \ \delta_{\ 2} \ B_{\ \alpha} \ = 
\ \color{magenta} \overline{\tau}^{\ \dot{\delta}}
\ \color{blue} \left ( \ \partial^{\ \dot{\gamma} \ \alpha} \ \partial_{\ \alpha \dot{\delta}} \ C 
\ \right )
\vspace*{0.2cm} \\
\color{red} \rightarrow
\ \color{blue} i \ \partial^{\ \dot{\delta} \ \alpha} \ \delta_{\ 2} \ B_{\ \alpha} \ =
\ \color{magenta} \overline{\tau}^{\ \dot{\delta}} \ \color{blue} \mbox{\fbox{\rule{0mm}{1.0mm}}} \ C
\end{array}
\end{equation}

\color{red}
\begin{center}
7a) Susy transformations of the Lagrangean multiplet -- $\delta_{\ 2}$
\end{center}

\color{blue}

\noindent
Thus we summarize \color{red} results \color{blue} (eqs. \ref{eq:158} - \ref{eq:164})

\vspace*{-0.1cm}
\begin{equation}
\label{eq:165}
\begin{array}{lll}
\delta_{\ 2} \ A & = & \color{magenta} \overline{\tau}_{\ \dot{\delta}}
\ \color{blue} \left ( \ -i \ \partial^{\ \dot{\delta} \alpha} \ B_{\ \alpha} \ \right )
\vspace*{0.2cm} \\
\delta_{\ 2} \ B_{\ \alpha} & = & \color{magenta} \overline{\tau}^{\ \dot{\delta}}
\ \color{blue} \left ( \ - \ i \ \partial_{\ \alpha \dot{\delta}} \ C \ \right )
\vspace*{0.2cm} \\
\delta_{\ 2} \ C & = & 0
\end{array}
\end{equation}

}

\newpage

\color{blue}

{\large \bf

\color{red}
\begin{center}
7b) Susy transformations of the Lagrangean multiplet -- 
$\delta_{\ 1} \ = \ \color{magenta} \tau^{\ \delta} \ \color{red} q_{\ \delta}$
\end{center}

\color{blue}

\noindent
We repeat from eq. (\ref{eq:156}) the form of $\color{red} q_{\ \alpha}$

\vspace*{-0.5cm}
\begin{equation}
\label{eq:166}
\begin{array}{l}
\color{red}
q_{\ \alpha} \ = \ ( \ \partial_{\ \vartheta} \ )_{\ \alpha}
\ + \ \frac{i}{2} \ \overline{\vartheta}^{\ \dot{\beta}} \ ( \ \partial_{\ x} \ )_{\ \alpha \dot{\beta}}
\end{array}
\end{equation}

\noindent
The first part of $\delta_{\ 1} \ {\cal{L}}$ , 
$\color{red} ( \ \partial_{\ \vartheta} \ )_{\ \alpha} \ \color{blue} {\cal{L}}$ is displayed
in eq. (\ref{eq:157}) . 

\noindent
We now turn to the second part --

\begin{center} 
$\color{red} \frac{i}{2} \ \overline{\vartheta}^{\ \dot{\beta}} \ \partial_{\ \alpha \dot{\beta}}
\ \color{blue} {\cal{L}}$
\end{center}

\noindent
-- remembering the
form of ${\cal{L}}$ displayed in eq. (\ref{eq:154}) repeated for clarity below

\vspace*{-0.5cm}
\begin{equation}
\label{eq:167} 
\begin{array}{l}
{\cal{L}}  \ = 
\vspace*{0.3cm} \\
\left \lbrace
\begin{array}{c}
\vartheta^{\ 2} \ \overline{\vartheta}^{\ 2} 
\ \left ( \ - \ \frac{1}{4} \ \mbox{\fbox{\rule{0mm}{1.0mm}}} \ C
\ \right )
\vspace*{0.3cm} \\
\vartheta^{\ 2} \ \overline{\vartheta}_{\ \dot{\delta}}
\ \left ( \ \frac{-i}{2} \ \partial^{\ \dot{\delta} \alpha} \ B_{\ \alpha}
\ \right )
\ + \color{red} \ \overline{\vartheta}^{\ 2} \ \vartheta^{\ \alpha} \ 0
\vspace*{0.3cm} \\
\color{blue}
\vartheta^{\ 2} \ A \ + \  \vartheta^{\ \alpha} 
\ \sigma^{\ \mu}_{ \alpha \dot{\beta}} \ \overline{\vartheta}^{\ \dot{\beta}}
\ \left ( \ \frac{-i}{2} \ \partial_{\ \mu} \ C \ \right )
\ + \ \color{red} \overline{\vartheta}^{\ 2} \ 0
\vspace*{0.3cm} \\
+ \ \color{red} \overline{\vartheta}^{\ \dot{\beta}} \ 0
\color{blue} \ + \ \vartheta^{\ \alpha} \ B_{\ \alpha}
\vspace*{0.3cm} \\
+ \ C
\end{array}
\right \rbrace
\end{array}
\end{equation}

}

\newpage

\color{blue}

{\large \bf

\vspace*{-0.5cm}
\begin{equation}
\label{eq:168}
\begin{array}{l}
\color{red}
\frac{i}{2} \ \overline{\vartheta}^{\ \dot{\beta}} \ ( \ \partial_{\ x} \ )_{\ \alpha \dot{\beta}}
\ \color{blue} {\cal{L}} \ =
\vspace*{0.3cm} \\
\left \lbrace
\begin{array}{c}
\vartheta^{\ 2} \ \overline{\vartheta}^{\ 2} 
\ \left ( \ - \ \frac{1}{4} \ \mbox{\fbox{\rule{0mm}{1.0mm}}} \ B_{\ \alpha}
\ \right )
\vspace*{0.3cm} \\
\vartheta^{\ 2} \ \overline{\vartheta}^{\ \dot{\delta}}
\ \left ( \ \frac{i}{2} \ \partial_{\ \alpha \dot{\delta}} \ A
\ \right ) 
\ + \color{red} \ \overline{\vartheta}^{\ 2} \ \vartheta_{\ \alpha} 
\ \left ( \ \frac{1}{4} 
\ \mbox{\fbox{\rule{0mm}{1.0mm}}} 
\ C \ \right ) 
\vspace*{0.3cm} \\
\color{blue}
\vartheta^{\ 2} \ \color{red} 0  \ \color{blue} + 
\ \vartheta^{\ \gamma} \ \overline{\vartheta}^{\ \dot{\beta}}
\ \left ( \ \frac{-i}{2} \ \partial_{\ \alpha \dot{\beta}} \ B_{\ \gamma} \ \right )
\ + \ \color{red} \overline{\vartheta}^{\ 2} \ 0
\vspace*{0.3cm} \\
\color{blue} +
\ \color{red} \overline{\vartheta}^{\ \dot{\beta}} 
\ \frac{i}{2} \ \partial_{\ \alpha \dot{\beta}} \ C
\ \color{blue} + \ \vartheta^{\ \alpha} \ \color{red} 0
\vspace*{0.3cm} \\
+ \ \color{red} 0
\end{array}
\right \rbrace
\end{array}
\end{equation} 

\noindent
Next we repeat the first part
$\color{red} ( \ \partial_{\ \vartheta} \ )_{\ \alpha} \ \color{blue} {\cal{L}}$
(eq. \ref{eq:157})

\vspace*{-0.5cm}
\begin{equation}
\label{eq:169}
\begin{array}{l}
\color{red} ( \ \partial_{\ \vartheta} \ )_{\ \alpha} \ \color{blue} {\cal{L}}  \ = 
\vspace*{0.3cm} \\
\left \lbrace
\begin{array}{c}
\vartheta^{\ 2} \ \overline{\vartheta}^{\ 2} \ \color{red} 0
\vspace*{0.3cm} \\
\vartheta^{\ 2} \ \overline{\vartheta}_{\ \dot{\delta}} \ \color{red} 0
\ \color{blue} + \color{red} \ \overline{\vartheta}^{\ 2} \ \vartheta_{\ \alpha} 
\ \left ( \ - \ \frac{1}{4} \ \mbox{\fbox{\rule{0mm}{1.0mm}}} \ C
\ \right )
\vspace*{0.3cm} \\
\color{blue}
\vartheta^{\ 2} \ \color{red} 0 \ \color{blue} + \  \vartheta_{\ \alpha} 
\ \sigma^{\ \mu \ \dot{\delta} \gamma} \ \overline{\vartheta}_{\ \dot{\delta}}
\ \left ( \ \frac{-i}{2} \ \partial_{\ \mu} \ B_{\ \gamma} \ \right )
\ + \ \color{red} \overline{\vartheta}^{\ 2} \ 0
\vspace*{0.3cm} \\
+ \ \color{red} \sigma^{\ \mu}_{\ \alpha \dot{\delta}} \ \overline{\vartheta}^{\ \dot{\delta}}
\ \left ( \ \frac{-i}{2} \ \partial_{\ \mu} \ C \ \right )
\color{blue} \ + \ \vartheta_{\ \alpha} \ A
\vspace*{0.3cm} \\
+ \ B_{\ \alpha}
\end{array}
\right \rbrace
\end{array}
\end{equation}

\noindent
and add the two parts to obtain $\color{red} q_{\ \alpha} \ \color{blue} {\cal{L}}$

} 

\newpage

\color{blue}

{\large \bf

\vspace*{-0.5cm}
\begin{equation}
\label{eq:170}
\begin{array}{l}
\color{red}
q_{\ \alpha} \ \color{blue} {\cal{L}} \ =
\vspace*{0.3cm} \\
\left \lbrace
\begin{array}{c}
\vartheta^{\ 2} \ \overline{\vartheta}^{\ 2} 
\ \left ( \ - \ \frac{1}{4} \ \mbox{\fbox{\rule{0mm}{1.0mm}}} \ B_{\ \alpha}
\ \right )
\vspace*{0.3cm} \\
\vartheta^{\ 2} \ \overline{\vartheta}^{\ \dot{\delta}}
\ \left ( \ \frac{i}{2} \ \partial_{\ \alpha \dot{\delta}} \ A
\ \right ) 
\ + \color{red} \ \overline{\vartheta}^{\ 2} \ \vartheta_{\ \alpha} 
\ 0 
\vspace*{0.3cm} \\
\color{blue}
\vartheta^{\ 2} \ \color{red} 0  \ \color{blue} + 
\ \vartheta^{\ \gamma} \ \overline{\vartheta}^{\ \dot{\beta}}
\left ( \begin{array}{c} 
\frac{-i}{2} \ \partial_{\ \alpha \dot{\beta}} \ B_{\ \gamma}
\vspace*{0.2cm} \\
+ \ \varepsilon_{\ \alpha \gamma} 
\ \frac{i}{2} \ \partial_{\ \kappa \dot{\beta}} \ B^{\ \kappa}
\end{array} 
\right )
\ + \ \color{red} \overline{\vartheta}^{\ 2} \ 0
\vspace*{0.3cm} \\
\color{blue} +
\ \color{red} \overline{\vartheta}^{\ \dot{\beta}} 
\ 0
\ \color{blue} + \ \vartheta_{\ \alpha} \ A
\vspace*{0.3cm} \\
+ \ B_{\ \alpha}
\end{array}
\right \rbrace
\end{array}
\end{equation}

\noindent
We remark, as a check, that the chiral structure is indeed maintained.
Finally we mulyiply with $\color{magenta} \tau^{\ \alpha}$ from the left
to obtain $\delta_{\ 1} \ = \ \color{magenta} \tau^{\ \alpha} \ \color{red} q_{\ \alpha}$ \\
\color{blue}
( acting on ${\cal{L}} \ \color{red} \rightarrow$ )

}

\newpage

\color{blue}

{\large \bf

\vspace*{-0.5cm}
\begin{equation}
\label{eq:171}
\begin{array}{l}
\color{blue}
\delta_{\ 1} \ {\cal{L}} \ = \ \color{magenta} \tau^{\ \alpha} 
\ \color{red} q_{\ \alpha} \ \color{blue} {\cal{L}} \ =
\vspace*{0.3cm} \\
\left \lbrace
\begin{array}{c}
\vartheta^{\ 2} \ \overline{\vartheta}^{\ 2} 
\ \left ( \ - \ \color{magenta} \tau^{\ \alpha} 
\ \color{blue} \frac{1}{4} \ \mbox{\fbox{\rule{0mm}{1.0mm}}} \ B_{\ \alpha}
\ \right )
\vspace*{0.3cm} \\
\vartheta^{\ 2} \ \overline{\vartheta}^{\ \dot{\delta}}
\ \left ( \ - \ \color{magenta} \tau^{\ \alpha} \ \color{blue} \frac{i}{2} 
\ \partial_{\ \alpha \dot{\delta}} \ A
\ \right ) 
\ + \color{red} \ \overline{\vartheta}^{\ 2} \ \color{magenta} \tau^{\ \alpha} 
\ \color{red} \vartheta_{\ \alpha} 
\ 0 
\vspace*{0.3cm} \\
\color{blue}
\vartheta^{\ 2} \ \color{red} 0  \ \color{blue} + 
\ \vartheta^{\ \gamma} \ \overline{\vartheta}^{\ \dot{\beta}}
\left ( \begin{array}{c} 
- \ \color{magenta} \tau^{\ \alpha} \ \color{blue} \frac{i}{2} 
\ \partial_{\ \alpha \dot{\beta}} \ B_{\ \gamma}
\vspace*{0.2cm} \\
+ \ \color{magenta} \tau_{\ \gamma} \ \color{blue}
\frac{i}{2} \ \partial_{\ \kappa \dot{\beta}} \ B^{\ \kappa}
\end{array} 
\right )
\ + \ \color{red} \overline{\vartheta}^{\ 2} \ 0
\vspace*{0.3cm} \\
\color{blue} +
\ \color{red} \overline{\vartheta}^{\ \dot{\beta}} 
\ 0
\ \color{blue} + \ \vartheta^{\ \alpha} 
\ \color{magenta} \tau_{\ \alpha} \ \color{blue} A
\vspace*{0.3cm} \\
+ \ \color{magenta} \tau^{\ \alpha} \ \color{blue} B_{\ \alpha}
\end{array}
\right \rbrace
\end{array}
\end{equation}

\noindent
We compare with ${\cal{L}}$ (eq. \ref{eq:167}) reproduced below

\vspace*{-0.5cm}
\begin{equation}
\label{eq:172} 
\begin{array}{l}
{\cal{L}}  \ = 
\vspace*{0.3cm} \\
\left \lbrace
\begin{array}{c}
\vartheta^{\ 2} \ \overline{\vartheta}^{\ 2} 
\ \left ( \ - \ \frac{1}{4} \ \mbox{\fbox{\rule{0mm}{1.0mm}}} \ C
\ \right )
\vspace*{0.3cm} \\
\vartheta^{\ 2} \ \overline{\vartheta}_{\ \dot{\delta}}
\ \left ( \ \frac{-i}{2} \ \partial^{\ \dot{\delta} \alpha} \ B_{\ \alpha}
\ \right )
\ + \color{red} \ \overline{\vartheta}^{\ 2} \ \vartheta^{\ \alpha} \ 0
\vspace*{0.3cm} \\
\color{blue}
\vartheta^{\ 2} \ A \ + \  \vartheta^{\ \alpha} 
\ \sigma^{\ \mu}_{ \alpha \dot{\beta}} \ \overline{\vartheta}^{\ \dot{\beta}}
\ \left ( \ \frac{-i}{2} \ \partial_{\ \mu} \ C \ \right )
\ + \ \color{red} \overline{\vartheta}^{\ 2} \ 0
\vspace*{0.3cm} \\
+ \ \color{red} \overline{\vartheta}^{\ \dot{\beta}} \ 0
\color{blue} \ + \ \vartheta^{\ \alpha} \ B_{\ \alpha}
\vspace*{0.3cm} \\
+ \ C
\end{array}
\right \rbrace
\end{array}
\end{equation}

}

\newpage

\color{blue}

{\large \bf

\noindent
We compare component by component

\vspace*{-0.5cm}
\begin{equation}
\label{eq:173} 
\begin{array}{l}
\delta_{\ 1} \ \mbox{\fbox{\rule{0mm}{1.0mm}}} \ C \ =
\ \color{magenta} \tau^{\ \alpha} 
\ \color{blue}  \ \mbox{\fbox{\rule{0mm}{1.0mm}}} \ B_{\ \alpha}
\ \color{red} \leftarrow
\ \color{blue} \delta_{\ 1} \ C \ =
\ \color{magenta} \tau^{\ \alpha} 
\ \color{blue} B_{\ \alpha}
\vspace*{0.3cm} \\
\delta_{\ 1} \ i \ \partial^{\ \dot{\delta} \alpha} \ B_{\ \alpha}
\ = \ \color{magenta} \tau_{\ \alpha} \ \color{blue} i \ \partial^{\ \dot{\delta} \alpha} \ A
\vspace*{0.3cm} \\
\color{red} \uparrow \hspace*{0.7cm} \color{blue}
\delta_{\ 1} \ B_{\ \alpha} \ = \ \color{magenta} \tau_{\ \alpha} \ \color{blue} A
\vspace*{0.3cm} \\
\delta_{\ 1} \ A \ = \ 0
\vspace*{0.3cm} \\
\delta_{\ 1} \ \partial_{\ \gamma \dot{\beta}} \ C
\ = \ \left ( \ \begin{array}{c}
\color{magenta} \tau^{\ \alpha} \ \color{blue} \ \partial_{\ \alpha \dot{\beta}} \ B_{\ \gamma}
\vspace*{0.2cm} \\
- \ \color{magenta} \tau_{\ \gamma} \ \color{blue}
\partial_{\ \kappa \dot{\beta}} \ B^{\ \kappa} 
\end{array}
\ \right )
\end{array}
\end{equation}

\noindent
The last relation in eq. (\ref{eq:173}) can be transformed, using the identity

\vspace*{-0.5cm}
\begin{equation}
\label{eq:174} 
\begin{array}{l}
\partial_{\ \alpha \dot{\beta}} \ B_{\ \gamma} \ - \ \partial_{\ \gamma \dot{\beta}} \ B_{\ \alpha}
\ = \ \varepsilon_{\ \gamma \alpha} \ X \ \color{red} \rightarrow
\vspace*{0.2cm} \\
\delta_{\ \alpha}^{\ \gamma} \ X \ =
\ \partial_{\ \alpha \dot{\beta}} \ B^{\ \gamma} \ - \ \partial^{\ \gamma}_{\ \dot{\beta}} \ B_{\ \alpha}
\ \color{red} \rightarrow
\vspace*{0.2cm} \\
X \ = \ \partial_{\ \kappa \dot{\beta}} \ B^{\ \kappa}
\end{array}
\end{equation}

\noindent
From the identity (eq. \ref{eq:174}) it follows \footnote{\hspace*{0.1cm} \color{red}
Check all relations in eqs. (\ref{eq:173} - \ref{eq:175}) .}

\vspace*{-0.5cm}
\begin{equation}
\label{eq:175} 
\begin{array}{l}
\delta_{\ 1} \ \partial_{\ \gamma \dot{\beta}} \ C \ = 
\ \partial_{\ \gamma \dot{\beta}} \ \color{magenta} \tau^{\ \alpha}
\ \color{blue} B_{\ \alpha}
\vspace*{0.2cm} \\
\color{red} \uparrow \ \color{blue}
\delta_{\ 1} \ C \ =
\ \color{magenta} \tau^{\ \alpha} 
\ \color{blue} B_{\ \alpha}
\end{array}
\end{equation}

}

\newpage

\color{blue}

{\large \bf

\color{red}
\begin{center}
{\small
Results for $\delta_{\ 1} \ \color{blue} {\cal{L}}$ }
\end{center}

\color{blue}

\noindent
We summarize the susy variations $\delta_{\ 1} \ {\cal{L}}$ (eqs. \ref{eq:173} - \ref{eq:175}) :

\vspace*{-0.5cm}
\begin{equation}
\label{eq:176} 
\begin{array}{lll}
\delta_{\ 1} \ A & = & 0
\vspace*{0.2cm} \\
\delta_{\ 1} \ B_{\ \alpha} & = & \color{magenta} \tau_{\ \alpha} \ \color{blue} A
\vspace*{0.2cm} \\
\delta_{\ 1} \ C & = & \color{magenta} \tau^{\ \alpha} \ \color{blue} B_{\ \alpha}
\end{array}
\end{equation}

\noindent
and repeat $\delta_{\ 2} \ {\cal{L}}$ (eq. \ref{eq:165}) below

\vspace*{-0.5cm} 
\begin{equation}
\label{eq:177}
\begin{array}{lll}
\delta_{\ 2} \ A & = & \color{magenta} \overline{\tau}_{\ \dot{\delta}}
\ \color{blue} \left ( \ -i \ \partial^{\ \dot{\delta} \alpha} \ B_{\ \alpha} \ \right )
\vspace*{0.2cm} \\
\delta_{\ 2} \ B_{\ \alpha} & = & \color{magenta} \overline{\tau}^{\ \dot{\delta}}
\ \color{blue} \left ( \ - \ i \ \partial_{\ \alpha \dot{\delta}} \ C \ \right )
\vspace*{0.2cm} \\
\delta_{\ 2} \ C & = & \color{magenta} \tau^{\ \alpha} 
\ \color{blue} B_{\ \alpha}
\end{array}
\end{equation}

\color{red}
\begin{center}
7c) Susy transformations of the Lagrangean multiplet --
$\delta \ = \ \delta_{\ 1} \ + \ \delta_{\ 2}
\ = \ \color{magenta} \tau^{\ \gamma} \ \color{red} q_{\ \gamma}
\ + \ \color{magenta} \overline{\tau}^{\ \dot{\delta}} \ \color{red} \overline{q}_{\ \dot{\delta}}$
\end{center}

\color{blue}

\noindent
Thus the full susy variations of ${\cal{L}}$ are \\

\begin{center}
$\begin{array}{c}
\delta \ = \ \delta \ \left ( \ \color{magenta} \tau \ , \ \overline{\tau} 
\ \color{blue} \right ) 
\end{array}$
\end{center}
\vspace*{0.2cm}  

\vspace*{-0.8cm}
\begin{equation}
\label{eq:178}
\begin{array}{|@{\hspace*{0.5cm}}lll ll@{\hspace*{0.4cm}}|}
\hline \vspace*{-0.7cm} \\
  & & & & \vspace*{-0.3cm} \\
\delta \ A & = & & & \color{magenta} \overline{\tau}^{\ \dot{\delta}}
\ \color{blue} \left ( \ -i \ \partial_{\ \gamma \dot{\delta}} \ B^{\ \gamma} \ \right )
\vspace*{-0.3cm} \\
 & & & &
\vspace*{-0.2cm} \\
\delta \ B_{\ \alpha} & = & \color{magenta} \tau^{\ \gamma} \ \color{blue} 
\varepsilon_{\ \alpha \gamma} \ A
& + & \color{magenta} \overline{\tau}^{\ \dot{\delta}}
\ \color{blue} \left ( \ - \ i \ \partial_{\ \alpha \dot{\delta}} \ C \ \right )
\vspace*{-0.3cm} \\
 & & & &
\vspace*{-0.2cm} \\
\delta \ C & = & \color{magenta} \tau^{\ \gamma} 
\ \color{blue} B_{\ \gamma} & &
\vspace*{-0.3cm} \\
 & & & &
\vspace*{-0.0cm} \\
\hline
\end{array}
\end{equation}

}

\newpage

\color{blue}

{\large \bf

\color{red}
\begin{center}
8) Indirect derivation of the anomaly of the susy current
\end{center}

\color{blue}

\noindent
We return to the course of the Vienna seminar and the search for a {\it minimum} of
the effective action $\Gamma \ ( \ {\cal{L}} \ , \ \overline{{\cal{L}}} \ )$ 
( eqs. \ref{eq:136} - \ref{eq:143} ) , within the minimal source extension described in
the previous section .

\noindent
Here we include the boundary values given in eq. (\ref{eq:145}) 

\vspace*{-0.5cm}
\begin{equation}
\label{eq:179}
\begin{array}{l}
\color{red} \frac{1}{g^{\ 2}_{\ \infty}} \ , 
\ \Theta_{\ \infty} \ ,
\ m_{\ \infty}
\end{array}
\end{equation}

\color{blue}

\noindent
in the definition of the arguments of $\Gamma$ with the identification
(eq. \ref{eq:136})

\vspace*{-0.5cm}
\begin{equation}
\label{eq:180}
\begin{array}{c}
{\cal{L}} \ \rightarrow \ \Phi
\ \color{red} \leftrightarrow 
\ \color{blue} \Gamma \ ( \ {\cal{L}} \ , \ \overline{\cal{L}} \ ) \ \rightarrow
\ \Gamma \ ( \ \Phi \ , \ \overline{\Phi} \ )
\vspace*{0.2cm} \\
\mbox{with}
\hspace*{0.3cm}
\Phi \ \rightarrow
\hspace*{0.2cm}
\left \lbrace \begin{array}{l}
\vartheta^{\ 2} \ A
\vspace*{0.2cm} \\
+ \ \vartheta^{\ \alpha} \ B_{\ \alpha}
\vspace*{0.2cm} \\
+ \ C  
\end{array}
\right \rbrace \ ( \ x^{\ -} \ )  
\end{array}
\end{equation}

\noindent
and -- following ref. \cite{PMtheta} -- setting

\vspace*{-0.5cm}
\begin{equation}
\label{eq:181}
\begin{array}{c}
\color{red} \Theta_{\ \infty} \ \rightarrow \ \Theta_{\ \infty} \ - \ \Theta^{\ '}
\ \rightarrow \ 0
\end{array}
\end{equation}

\noindent
We keep $\color{red} m_{\ \infty} \ \rightarrow \ 0$ \color{blue} as an infrared regulater mass, to be
set zero in the susy approaching limit. 

}

\newpage

\color{blue}

{\large \bf

\noindent
The thermodynamic limit has been carefully described in ref. \cite{LeibMink} and
in the thesis of L. Bergamin \cite{Bergathesis} .

\color{red}
\begin{center}
3 ({\small of topics}) On the road of conserved susy 
\end{center}

\color{blue}

\noindent
We pursue the road along the logical possibility $b_{\ +}$ as defined in sections 1 and 2,
hence assuming strict conservation of the susy (super)current 

\vspace*{-0.5cm}
\begin{equation}
\label{eq:182}
\begin{array}{c}
j_{\ \beta \alpha}^{\ \dot{\gamma}} \ = \ \color{red} k \ \color{blue} \Lambda^{\ * \ \dot{\gamma} \ A} 
\ f_{\ \left \lbrace \beta \alpha \right \rbrace}^{\ A} 
\vspace*{0.2cm} \\
j_{\ \mu \ \alpha} \ = \ \left ( \ \sigma_{\ \mu} \ \right )^{\ \beta}_{\ \dot{\gamma}}
\ j_{\ \beta \alpha}^{\ \dot{\gamma}} 
\vspace*{0.2cm} \\
\mbox{assuming \color{red} as regula falsi \color{blue} :} 
\ \partial^{\ \mu} \ j_{\ \mu \ \alpha} \ = \ 0 
\vspace*{0.3cm} \\
f_{\ \left \lbrace \beta \alpha \right \rbrace}^{\ A} \ =
\ \frac{1}{2} \ \left ( \ \sigma^{\ \mu \nu} \ \right )_{\ \left \lbrace \ \gamma \alpha
\ \right \rbrace} \ {\cal{F}}_{\ \mu \nu}
\vspace*{0.3cm} \\
{\cal{F}}_{\ \mu \nu}^{\ A} \ = \ \partial_{\ \mu} \ {\cal{W}}_{\ \nu}^{\ A}
\ - \  \partial_{\ \nu} \ {\cal{W}}_{\ \mu}^{\ A} \ + 
\ {\cal{W}}_{\ \mu}^{\ B} \ {\cal{W}}_{\ \nu}^{\ C}
\ i \ f_{\ A B C}
\vspace*{0.2cm} \\
{\cal{F}}_{\ \mu \nu}^{\ A} \ = \ \frac{1}{i} \ F_{\ \mu \nu}^{\ A}
\hspace*{0.2cm} ; \hspace*{0.2cm}
{\cal{W}}_{\ \mu}^{\ A} \ = \ i \ v_{\ \mu}^{\ A}
\end{array}
\end{equation} 

\noindent
The numerical value of the constant \color{red} k \color{blue} in eq. (\ref{eq:182}) ,
which ensures the normalisation condition in the equal time commutator in eq. (\ref{eq:1})
is here immaterial.

\noindent
We recall the definitions
in eqs. (\ref{eq:74} - \ref{eq:76}) \color{red} $\rightarrow$

}

\newpage

\color{blue}

{\large \bf

\vspace*{-0.5cm}
\begin{equation}
\label{eq:183}
\begin{array}{l}
\left ( \ \sigma^{\ \mu \nu} \ \right )_{\ \gamma}^{\ \hspace*{0.3cm} \delta}
\ = \ \left ( \ \varepsilon^{\ '} \ \right )^{\ \delta \alpha}
\ \left ( \ \sigma^{\ \mu \nu} \ \right )_{\ \left \lbrace \ \gamma \alpha \ \right \rbrace}
\vspace*{0.2cm} \\
\left ( \ \sigma^{\ \mu} \ \right )_{\ \gamma \dot{\beta}}
\ \left ( \ \sigma^{\ \nu} \ \right )^{\ \dot{\beta} \delta}
\ = \ \eta^{\ \mu \nu} \ \delta_{\ \gamma}^{\ \hspace*{0.3cm} \delta}
\ + \ \left ( \ \sigma^{\ \mu \nu} \ \right )_{\ \gamma}^{\ \hspace*{0.3cm} \delta}
\vspace*{0.2cm} \\
\eta^{\ \mu \nu} \ = \ diag \ ( \ 1 \ , \ - 1 \ , \ - 1 \ , \ - 1 \ )
\vspace*{0.2cm} \\
\left (
\ \sigma^{\ \mu \nu} \ = 
\ \left \lbrace \ \begin{array}{ll}
\sigma_{\ r} & \mbox{for  } ^{\mu} \ = 0 \ , \ ^{\nu} \ = \ r
\vspace*{0.2cm} \\
\frac{1}{i} \ \varepsilon_{\ s t r} \ \sigma_{\ r} & \mbox{for  } ^{\mu} \ = s \ , \ ^{\nu} \ = \ t
\end{array} \right \rbrace
\ \right )_{\ \gamma}^{\ \hspace*{0.3cm} \delta}
\end{array}
\end{equation}

\noindent
This implies the symmetric property -- with respect to the spinor indices $_{\ \beta \alpha}$ --
of the current $j_{\ \beta \alpha}^{\ \dot{\gamma}}$ .

\noindent
Further the right chiral components project on the duality related combination
(eq. \ref{eq:93} - \ref{eq:94}) \footnote{\hspace*{0.1cm} \color{red} In the Euclidean transcription this
becomes the anti-selfdual combination 
$\left ( \ \vec{B}^{\ A} \ - \ \vec{E}^{\ A} \ \right )_{\ Eucl.}$ ( show ! ) .} 

\vspace*{-0.5cm}
\begin{equation}
\label{eq:184}
\begin{array}{l}
F_{\ \mu \nu}^{\ + \ A} \ = \ F_{\ \mu \nu}^{\ A} \ - \ i 
\ \widetilde{F}_{\ \mu \nu}^{\ A}
\ \color{red} \leftrightarrow \ \vec{B}^{\ A} \ - i \ \vec{E}^{\ A}
\vspace*{0.3cm} \\
\color{blue}
\widetilde{\sigma}_{\ \mu \nu} \ = \ \frac{1}{2} \ \varepsilon_{\ \mu \nu \varrho \tau}
\ \sigma^{\ \varrho \tau}
\vspace*{0.2cm} \\
\widetilde{\sigma}^{\ \mu \nu} \ =
\ \left \lbrace \ \begin{array}{ll}
i \ \sigma_{\ r} & \mbox{for  } ^{\mu} \ = 0 \ , \ ^{\nu} \ = \ r
\vspace*{0.2cm} \\
\varepsilon_{\ s t r} \ \sigma_{\ r} & \mbox{for  } ^{\mu} \ = s \ , \ ^{\nu} \ = \ t
\end{array} \right \rbrace                                    
\vspace*{0.2cm} \\
\color{red} \rightarrow
\hspace*{0.3cm} \color{blue}
\widetilde{\sigma}^{\ \mu \nu} \ = \ i \ \sigma^{\ \mu \nu}
\end{array}
\end{equation}

}

\newpage

\color{blue}

{\large \bf

\noindent
The logical deductions are shown step by step :
\vspace*{-0.3cm}

\begin{description}
\item 1 susy covariance

It follows from the assumption in eq. (\ref{eq:182}) that
the effective potential $\Gamma \ ( \ \Phi \ , \ \overline{\Phi} \ )$ in eq. (\ref{eq:180})
inherits full susy covariance.

\item 2 Restriction of $\Phi$ to bosonic arguments -- \color{red} and eventual
'would be' goldstino condensates.

\color{blue}
\item 3 The effective potential must be bounded from below, yet exhibit a true
minimum, corresponding to a (the) stable ground state.

\item 4 The two bosonic arguments of $\Phi \ \leftrightarrow \ A \ , \ C$
characterizing the minimum correspond to vacuum expected values of the associated
operators; a priori these values -- any one of the two or both  -- can well be 0

\item 5 Absence of \color{red} spontaneous \color{blue} susy breaking requires
$A \ = \ 0$ , conversely $A \ \neq \ 0$ implies \color{red} spontanous 
\color{blue} susy breaking and implies indeed the appearance of a \color{red} 
Goldstone-fermion \color{blue} the goldstino. 

\item 6 The goldstino implies a \color{red} positive vacuum energy density 
\color{blue} by the deduction in section 2.

\item 7 The stability of the gauge boson condensate 
${\cal{B}}^{\ 2} \ = \ \left \langle \ \Omega \ \right |
\ \frac{1}{4} \ F_{\ \mu \nu}^{\ A} \ F^{\ \mu \nu \ A} \ \left | \ \Omega \ \right \rangle$
requires a \color{red} non-positive vacuum energy density

\end{description}

}

\newpage

\color{blue}

{\large \bf

\begin{description}

\item 8 This apparently leaves only the logical possibility of no spontaneous breaking at all
($b_{\ 0}$) -- yet this is not given .

\end{description}

\color{red}
\begin{center}
8a) Ad 1 , 2 : susy covariance and the fermionic arguments of $\Phi$
\end{center}

\color{blue}
From the discussion of the $D^{\ A}$ eliminated Lagrangean multiplet in section 6) , 6a) 
( eqs. \ref{eq:98} - \ref{eq:135} ) we infer, 

\vspace*{-0.5cm}
\begin{equation}
\label{eq:185}
\begin{array}{l}
\Phi \ \rightarrow \ \Phi \ = \ \Phi_{\ orig} \ \color{red} J_{\ \infty} 
\ \color{blue} = \ \left . {\cal{L}} \ \color{red} \right |_{\  \color{red} J \ = \ 0} 
\vspace*{0.2cm} \\
\color{blue}
\Phi \ = \ \left ( \ A \ , \ B_{\ \alpha} \ , \ C \ \right )
\end{array}
\end{equation}

\noindent
on the 'classical' fermionic fields $B_{\ \alpha}$ can involve -- in the
thermodynamic limit, to be discussed below -- constant (in space-time) nontrivial
associated condensates, for bilinears corresponding to $B_{\ \alpha}$ associated operators

\vspace*{-0.5cm}
\begin{equation}
\label{eq:186}
\begin{array}{l}
\underline{B}_{\ \alpha} \ ( \ x \ ) \ \rightarrow 
\vspace*{0.2cm} \\
\eta^{\ 2} \ = \ \left \langle \ \Omega \ \right |
\ : \ \underline{B}^{\ \alpha}
\ \underline{B}_{\ \alpha} \ ( \ x \ ) \ : \ \left | \ \Omega \ \right \rangle \ = \ constant
\end{array}
\end{equation}

\noindent
The : : in eq. (\ref{eq:186}) denotes a normal ordering with respect to an unstable ground state.

}

\newpage

\color{blue}

{\large \bf

\noindent
The latter condensates ( $\eta^{\ 2} \ , \ \overline{\eta}^{\ 2} \ = \ ( \ \eta^{\ 2} \ )^{\ *}$ ) ,
if not zero, do modify the minimum conditions. They must be distinguished from the condensates
of the base fields $\Lambda^{\ A \ \alpha} \ \Lambda^{\ A}_{\ \alpha}$ and its hermitian conjugate.
 
\noindent
The arguments of $\Gamma$ can be reduced -- in the 'static' or thermodynamic limit
to the \color{red} dependent \color{blue} chiral pair \cite{shore}

\vspace*{-0.5cm}
\begin{equation}
\label{eq:187}
\begin{array}{l}
\Phi \ = \ \Phi_{\ 0} \ =
\ \left \lbrace \ \begin{array}{c}
\vartheta^{\ 2} \ A 
\vspace*{0.2cm} \\
+ \ \vartheta^{\ \alpha} \ B_{\ \alpha}
\vspace*{0.2cm} \\
+ \ C
\end{array} \ \right \rbrace
\ ( \ x_{\ 0} \ )
\vspace*{0.2cm} \\
\Psi \ = \ \Phi_{\ 1} \ =
\ \left \lbrace \ \begin{array}{c}
\vartheta^{\ 2} \ \color{red} 0 
\vspace*{0.2cm} \\
\color{blue}
+ \ \vartheta^{\ \alpha} \ \color{red} 0
\vspace*{0.2cm} \\
+ \ A^{\ *}
\end{array} \ \right \rbrace
\ ( \ x_{\ 0} \ )
\end{array}
\end{equation}

\noindent
The common and arbitrary space time argument $x_{\ 0}$ for all classical fields is
the result of a Taylor expansion of all quantities around this base point, whereby
all derivatives (momenta) are set zero.

\noindent
In this limit -- which can well be dangerous because of infrared singularities --
the static effectice potentials is constrained by susy covariance.

}

\newpage

\color{blue}

{\large \bf

It is determined from a K\"{a}hler-like potential $K$ 
and a complex conjugate pair of superpotentials 
$S \ , \ S^{\ *}$ and takes the form

\vspace*{-0.5cm}
\begin{equation}
\label{eq:188}
\begin{array}{l}
\color{red} 
K \ = \ K \ \left ( \ \Phi \ , \ \Psi \ ; \ \Phi^{\ *} \ , \ \Psi^{\ *} \ \right )
\vspace*{0.2cm} \\
S \ = \ S \ \left ( \ \Phi \ \right ) \ , \ S^{\ *} \ = \ \left \lbrack 
\ S \ \left ( \ \Phi \ \right )
\ \right \rbrack^{\ *}
\vspace*{0.2cm} \\
b \ = \ \frac{1}{2} \ B^{\ \alpha} \ B_{\ \alpha} \ ,
\ \mbox{and} \ b \ \rightarrow \ b^{\ *}  
\vspace*{0.2cm} \\
\color{blue}
\Gamma \ \left ( \ \Phi \ , \ \Phi^{\ *} \ \right ) \ =
\vspace*{0.2cm} \\
\ = \ \left \lbrack \ \begin{array}{c} 
\left (
\ K_{\ 1 \ \overline{1}} \ A^{\ *} \ A
\ + \ S_{\ 1} \ A \ + \ S^{\ *}_{\ \overline{1}} \ A^{\ *} \ \right )
\vspace*{0.2cm} \\
\ + \ \left ( \ \begin{array}{c}
- \ ( \ K_{\ 1 \ 1 \ \overline{1}} \ A^{\ *} \ + \ S_{\ 1 \ 1} \ ) \ \color{red} b 
\vspace*{0.2cm} \\
\color{blue}
- \ ( \ K_{\ 1 \ \overline{1} \ \overline{1}} \ A \ + \ S^{\ *}_{\ \overline{1} \ \overline{1}} \ )
\ \color{red} b^{\ *}
\vspace*{0.2cm} \\
\color{blue}
+ \ K_{\ 1 \ 1 \ \overline{1} \ \overline{1}} \ \color{red} b^{\ *} \ b
\end{array}
\color{blue}
\ \right )
\end{array}
\right \rbrack
\end{array}
\end{equation}

\noindent
In eq. (\ref{eq:188}) the symbols $_{\ 1} \ , \ _{\ \overline{1}}$ mean

\vspace*{-0.5cm}
\begin{equation}
\label{eq:189}
\begin{array}{l}
F_{\ 1} \ = \ \left ( \ \partial \ / \ \partial \ C \ \right ) \ F
\hspace*{0.2cm} , \hspace*{0.2cm}
F_{\ \overline{1}} \ = \ \left ( \ \partial \ / \ \partial \ C^{\ *} \ \right ) \ F
\end{array}
\end{equation}

\noindent
With the notation $K_{\ 1 \ \overline{1}} \ = \ g$ and $S_{\ 1} \ = \ s$
we note the associated reduced arguments

\vspace*{-0.5cm}
\begin{equation}
\label{eq:190}
\begin{array}{l}
g \ = \ g \ \left ( \ C \ , \ A^{\ *} \ ; \ C^{\ *} \ , \ A \ \right )
\vspace*{0.2cm} \\
s \ = \ s \ ( \ C \ ) 
\end{array}
\end{equation}

}

\newpage

\color{blue}

{\large \bf

\noindent
$\Gamma$ in eq. (\ref{eq:188}) then takes the form

\vspace*{-0.5cm}
\begin{equation}
\label{eq:191}
\begin{array}{l}
\Gamma \ \left ( \ \Phi \ , \ \Phi^{\ *} \ \right ) \ =
\vspace*{0.2cm} \\
\ = \ \left \lbrack \ \begin{array}{c} 
\left (
\ g \ A^{\ *} \ A
\ + \ s \ A \ + \ s^{\ *} \ A^{\ *} \ \right )
\vspace*{0.2cm} \\
\ + \ \left ( \ \begin{array}{c}
- \ ( \ g_{\ 1} \ A^{\ *} \ + \ s_{\ 1} \ ) \ \color{red} b 
\vspace*{0.2cm} \\
\color{blue}
- \ ( \ g_{\ \overline{1}} \ A \ + \ s^{\ *}_{\ \overline{1}} \ )
\ \color{red} b^{\ *}
\vspace*{0.2cm} \\
\color{blue}
+ \ g_{\ 1 \ \overline{1}} \ \color{red} b^{\ *} \ b
\end{array}
\color{blue}
\ \right )
\end{array}
\right \rbrack
\vspace*{0.3cm} \\
g \ = \ g \ \left ( \ C \ , \ A^{\ *} \ ; \ C^{\ *} \ , \ A \ \right )
\vspace*{0.2cm} \\
s \ = \ s \ ( \ C \ ) \ , \ s^{\ *} \ = \ s^{\ *} \ ( \ C^{\ *} \ )
\end{array}
\end{equation}

\noindent
It was proven by S. Portmann \cite{Portdipl} that the restricted functional form

\vspace*{-0.5cm}
\begin{equation}
\label{eq:192}
\begin{array}{l}
s \ = \ s \ ( \ C \ , \ A^{\ *} \ ) \ \rightarrow \ s \ ( \ C \ )
\end{array}
\end{equation} 

\noindent
is no losss of generality \footnote{\hspace*{0.1cm} \color{red} This should be checked
when the quadratic fermionic quantities $b \ , \ b^{\ *}$ are included .}

\noindent
The quantities $\color{red} b \ , \ b^{\ *}$ only occur in quadratic order as shown in
eqs. (\ref{eq:188} , \ref{eq:191}) .

\noindent
It follows that $\Gamma$ is bounded from below exactly if

\vspace*{-0.5cm}
\begin{equation}
\label{eq:193}
\begin{array}{l}
\Gamma \ > \ - \ \infty \ \color{red} \leftrightarrow \ \color{blue} g_{\ 1 \ \overline{1}} \ \geq \ 0
\end{array}
\end{equation}

}

\newpage

\color{blue}

{\large \bf

\color{red}
\begin{center}
8a 1) Ad 3 : Trivial (in)dependence of $\Gamma$ on the fermionic arguments
\end{center}

\color{blue}

\noindent
The case where $\Gamma$ is identically independent of the arguments
$\color{red} b \ , \ b^{\ *}$ corresponds to the functional identities

\vspace*{-0.5cm}
\begin{equation}
\label{eq:194}
\begin{array}{l}
g_{\ 1} \ , \ g_{\ \overline{1}} \ \equiv \ 0
\ \color{red} \rightarrow \ \color{blue} g \ = \ g \ ( \ A \ , \ A^{\ *} \ )
\vspace*{0.2cm} \\
s_{\ 1} \ , \ s^{\ *}_{\ \overline{1}} \ \equiv \ 0
\ \color{red} \rightarrow \ \color{blue} s \ , \ s^{\ *} \ = \ constant 
\end{array}
\end{equation} 

\noindent
This is not acceptable, since as a consequence the effective potential minimum 
\color{red} does not determine the spontaneous parameter

\color{blue}

\vspace*{-0.5cm}
\begin{equation}
\label{eq:195}
\begin{array}{l}
C \ \sim \ \left \langle \ \Omega \ \right | \ \Lambda^{\ A \gamma} \ \Lambda^{\ A}_{\ \gamma}
\ \left | \ \Omega \ \right \rangle 
\end{array}
\end{equation}

\noindent
rendering it a 'modulus'. We do not discuss this case further.

\color{red}
\begin{center}
8a 2) Ad 3 : General dependence of $\Gamma$ on the fermionic arguments
\end{center}

\color{blue}

\noindent
By (hermitian) bilinear  completion and assuming $g \ , \ g_{\ 1 \ \overline{1}} \ 
\not \hspace*{-0.1cm} \equiv \ 0$ , $\Gamma$ (eq. \ref{eq:191}) takes the form

}

\newpage

\color{blue}

{\large \bf

\vspace*{-0.5cm}
\begin{equation}
\label{eq:196}
\begin{array}{l}
\Gamma \ \left ( \ \Phi \ , \ \Phi^{\ *} \ \right ) \ =
\vspace*{0.2cm} \\
\ = \ \left \lbrack \ \begin{array}{c} 
\ g \ ( \ A^{\ *} \ + \ \sigma^{\ *} \ ) \ ( \ A \ + \ \sigma \ )
\vspace*{0.2cm} \\
\ - \ g \ \sigma^{\ *} \ \sigma
\vspace*{0.2cm} \\
\ + \ \left ( \ \begin{array}{c}
g_{\ 1 \ \overline{1}} \ ( \ \color{red} b^{\ *} \ \color{blue} - \ \tau^{\ *} \ )
\ ( \ \color{red} b \ \color{blue} - \ \tau \ )
\vspace*{0.2cm} \\
- \ g_{\ 1 \ \overline{1}} \ \tau^{\ *} \ \tau
\end{array}
\color{blue}
\ \right )
\end{array}
\right \rbrack
\vspace*{0.3cm} \\
\sigma \ = \ s^{\ *} \ / \ g
\hspace*{0.2cm} , \hspace*{0.2cm} \tau \ = 
\ ( \ g_{\ \overline{1}} \ A \ + \ s^{\ *}_{\ \overline{1}} \ ) 
\ / \ g_{\ 1 \ \overline{1}} 
\vspace*{0.2cm} \\
\mbox{and} \ \sigma \ \rightarrow \ \sigma^{\ *} 
\hspace*{0.2cm} , \hspace*{0.2cm} \tau \ \rightarrow \ \tau^{\ *}
\end{array}
\end{equation}

\noindent
Thus as a first result we obtain the reduced effective potential 

\vspace*{-0.5cm}
\begin{equation}
\label{eq:197}
\begin{array}{l}
\color{red} \left . b \ \color{blue} \right |_{\ \color{blue} min} \ \color{blue} = \ \tau 
\hspace*{0.2cm} \mbox{and c.c.} \ \color{red} \rightarrow
\vspace*{0.3cm} \\
\color{blue}
\Gamma \ \left ( \ \Phi \ , \ \Phi^{\ *} \ \right ) \ \rightarrow
\vspace*{0.2cm} \\
\ \rightarrow \ \left \lbrack \ \begin{array}{c} 
g \ ( \ A^{\ *} \ + \ \sigma^{\ *} \ ) \ ( \ A \ + \ \sigma \ )
\vspace*{0.2cm} \\
\ - \ g \ \sigma^{\ *} \ \sigma \ - \ g_{\ 1 \ \overline{1}} \ \tau^{\ *} \ \tau
\end{array}
\right \rbrack
\vspace*{0.3cm} \\
\sigma \ = \ s^{\ *} \ / \ g
\hspace*{0.2cm} , \hspace*{0.2cm} \tau \ = 
\ ( \ g_{\ \overline{1}} \ A \ + \ s^{\ *}_{\ \overline{1}} \ ) 
\ / \ g_{\ 1 \ \overline{1}} 
\vspace*{0.2cm} \\
\mbox{and} \ \sigma \ \rightarrow \ \sigma^{\ *} 
\hspace*{0.2cm} , \hspace*{0.2cm} \tau \ \rightarrow \ \tau^{\ *}
\end{array}
\end{equation}

}

\newpage

\color{blue}

{\large \bf

\color{red}
\begin{center}
8b) Special values for the arguments of $\Gamma$
\end{center}

\color{blue}

\begin{description}
\item 1) $A \ , \ A^{\ *} \ = \ 0$

\noindent
For $A \ ( \ A^{\ *} \ ) \ =  \ 0$ the effective potential becomes

\vspace*{-0.5cm}
\begin{equation}
\label{eq:198}
\begin{array}{l}
\left . \ \Gamma \ \right |_{\ A \ = \ 0} \ \rightarrow
\ - \ \left | \ s_{\ 1} \ \right |^{\ 2} \ / \ \left . \ g_{\ 1 \ \overline{1}} \ \right|_{\ A \ = \ 0}
\end{array}
\end{equation}

\noindent
As in section 8a 1) we discard the possibility that $s_{\ 1} \ , \ s^{\ *}_{\ \overline{1}}$
are identically zero, for all of its arguments $C \ , \ C^{\ *}$ respectively.

\noindent
It then follows that the overall minimum value of $\Gamma$ is negative.
Since the effective potential is a highest component in its associated susy multiplet this
fact in itself proves the spontaneous part of susy breaking.

\noindent
This is independent of the arguments $A \ , \ A^{\ *}$ for which this minimum is attained.

\noindent
The effective potential beeing bounded from below, a fortiori for restricted values of the
arguments, now implies, remembering the condition $g_{\ 1 \ \overline{1}} \ > \ 0$ (eq. \ref{eq:193})

\vspace*{-0.5cm}
\begin{equation}
\label{eq:199}
\begin{array}{l}
Max_{\ C \ , \ C^{\ *}} 
\ \left ( \ \left | \ s_{\ 1} \ \right |^{\ 2} \ / 
\ \left . \ g_{\ 1 \ \overline{1}} \ \right|_{\ A \ = \ 0} \ \right )
\ < \ \infty
\end{array}
\end{equation}

\end{description}

}

\newpage

\color{blue}

{\large \bf

\begin{description}
\item 2) $\left | \ A \ \right | \ \rightarrow \ \infty$

\noindent
For $\left | \ A \ \right | \ \rightarrow \ \infty$ , $\Gamma$ asymptotically becomes

\vspace*{-0.5cm}
\begin{equation}
\label{eq:200}
\begin{array}{l}
\Gamma \ \sim 
\ \left ( \ \left | \ A \ \right |^{\ 2} \ / \ g_{\ 1 \ \overline{1}} \ \right )
\ \left ( \ g \ g_{\ 1 \ \overline{1}} \ - \ \left | \ g_{\ 1} \ \right |^{\ 2} \ \right ) 
\end{array}
\end{equation}

\noindent
Also in the above limit the boundedness of $\Gamma$ from below imposes nontrivial conditions.

\end{description}

\color{red}
\begin{center}
9) Last but not least -- the main problem :
   inconsistency of only sponteneous susy breaking
\end{center}

\color{blue}

\noindent
We return to the topics of the seminar in Vienna. 

\noindent
The identification of the gauge boson condensate :

\noindent
The quantity $F \ = \ - \ \color{red} g_{\ \infty}^{\ 2} 
\ \color{blue} ( \ A \ + \ A^{\ *} \ )$ , related to the arguments $A \ , \ A^{\ *}$
of $\Gamma$ for which $\Gamma \ = \ minimum$ is attained, 
is associated with the gauge boson condensate

\vspace*{-0.5cm}
\begin{equation}
\label{eq:201}
\begin{array}{l}
F \ \color{red} \leftrightarrow \ \color{blue} \underline{F} 
\ \begin{array}[t]{ll}  = \ - &
\ \left ( \begin{array}{c}
\frac{1}{2} \ \underline{D}^{\ A} \ \underline{D}^{\ A} \ \color{magenta} (a)
\vspace*{0.2cm} \\
\color{blue}
+ \ \underline{\Lambda}^{\ *} \ i \ \sigma^{\ \mu} \ 
\stackrel{\leftharpoondown \hspace*{-0.3cm} \rightharpoonup}{D}_{\ \mu} \ \underline{\Lambda}
\ \color{magenta} (b)
\vspace*{0.2cm} \\
\color{blue}
- \ \frac{1}{4} \ \underline{F}_{\ \mu \nu}^{\ A} \ \underline{F}^{\ A \ \mu \nu}
\end{array}
\ \right )
\vspace*{0.3cm} \\
  \rightarrow &
\left \langle \ \Omega \ \right | \ \underline{F} \ \left | \ \Omega \ \right \rangle
\end{array}
\end{array}
\end{equation} 

\footnote{\hspace*{0.1cm} \color{magenta} (a) $\rightarrow \ 0$ for vanishing sources,
(b) : gauge invariant kinetic energy density for base fermions .}

}

\newpage

\color{blue}

{\large \bf

\noindent
The quantities \color{magenta} (a) \color{blue} and \color{magenta} (b)
\color{blue} in eq. (\ref{eq:201}) are thought not to contribute to
the vacuum expected value 
$\left \langle \ \Omega \ \right | \ \underline{F} \ \left | \ \Omega \ \right \rangle$ .

\color{red}
\begin{center}
9a) Connection with trace anomaly (N=1 susy) and the sign of the vacuum energy density
\end{center}

\color{blue}

\noindent
The trace anomaly \cite{trace} retains its unique operator identity form 
specifically for the N=1 super Yang-Mills system, using the definitions in eq. (\ref{eq:201})

\vspace*{-0.5cm}
\begin{equation}
\label{eq:202}
\begin{array}{l}
\underline{\vartheta}^{\ \mu}_{\ \mu} \ = 
\ \color{red} \left ( \ 2 \ \beta \ / \ g^{\ 3} \ \right ) \ \color{blue} \underline{F}_{\ \color{red} g}
\vspace*{0.2cm} \\
\underline{F} _{\ \color{red} g} \ = 
\ \frac{1}{4} \ \underline{F}{\ \mu \nu}^{\ A} \ \underline{F}^{\ A \ \mu \nu}
\end{array}
\end{equation}

\noindent
The suffix in $\underline{F}_{\ \color{red} g}$ \color{blue} as defined in eq. (\ref{eq:202})
indicates that the normalization of the operator $\underline{F}_{\ \color{red} g}$
is implicitely dependent on the corresponding normalization of the running coupling constant
$g \ = \ g \ ( \ \mu \ )$ as discussed in \cite{trace} , while the energy momentum tensor
components are -- by definition -- renormalization group invariant.

\noindent
While eq. (\ref{eq:202}) is valid to all orders in the coupling constant, the
lowest order perturbative contribution to the $\beta$ function is

\vspace*{-0.5cm}
\begin{equation}
\label{eq:203}
\begin{array}{l}
\color{red} \beta \ / \ g^{\ 3} \ \sim \ b_{\ 0} \ \color{blue} / \ ( \ 16 \pi^{\ 2} \ )
\vspace*{0.2cm} \\
\color{red} b_{\ 0} \ = \ \color{blue}  - \ 3 \ C_{\ 2} \ ( \ G \ )
\end{array}
\end{equation}

}

\newpage

\color{blue}

{\large \bf

\noindent
$C_{\ 2} \ ( \ G \ )$ in eq. (\ref{eq:203}) denotes the second Casimir operator of the 
underlying gauge group G, with the conventional $SUN$ (embedding-) normalization

\vspace*{-0.5cm}
\begin{equation}
\label{eq:204}
\begin{array}{l}
C_{\ 2} \ ( \ SUN \ ) \ = \ N
\end{array}
\end{equation} 

\noindent
This sets up the clash of signs for the vacuum expectation value

\vspace*{-0.5cm}
\begin{equation}
\label{eq:205}
\begin{array}{l}
\left \langle \ \Omega \ \right | \ \underline{\vartheta}_{\ \mu \nu} 
\ \left | \ \Omega \ \right \rangle \ = \ \varepsilon \ g_{\ \mu \nu} 
\vspace*{0.2cm} \\
\varepsilon \ = \ \color{red} ( \ \beta \ / \ 2 \ g^{\ 3} \ )
\ \color{blue} \left \langle \ \Omega \ \right | \ \underline{F}_{\ \color{red} g} 
\ \color{blue} \left | \ \Omega \ \right \rangle
\end{array}
\end{equation}

\noindent
since the vacuum expected value of 

\vspace*{-0.5cm}
\begin{equation}
\label{eq:206}
\begin{array}{l}
\underline{F} \ = \ \frac{1}{2} \ \left ( 
\ \left ( \ \vec{\underline{B}}^{\ A} \ \right )^{\ 2} \ -
\ \left ( \ \vec{\underline{E}}^{\ A} \ \right )^{\ 2}
\ \right )
\end{array}
\end{equation}

\noindent
requires the "Watt-less" positive sign for \cite{Yildiz} 

\vspace*{-0.5cm}
\begin{equation}
\label{eq:207}
\begin{array}{c}
F_{\ \color{red} g} \ \color{blue} = \ \left \langle \ \Omega \ \right | 
\ \underline{F}_{\ \color{red} g} 
\ \color{blue} \left | \ \Omega \ \right \rangle \ \geq \ 0
\ \color{red} \rightarrow
\vspace*{0.2cm} \\
\color{blue}
\varepsilon \ \leq \ 0
\end{array}
\end{equation}

\noindent
whereas the Goldstino situation ( for $\varepsilon \ \neq \ 0$ ) requires
according to section 2 ( eq. \ref{eq:7} )

\vspace*{-0.5cm}
\begin{equation}
\label{eq:208}
\begin{array}{c}
\varepsilon \ = \ \left | \ f_{\ g} \ \right |^{\ 2} \ \geq \ 0
\end{array}
\end{equation}

\noindent
The clash of signs between eq. (\ref{eq:207} ) and eq. (\ref{eq:208}) is the final result
of this section \footnote{\hspace*{0.1cm} \color{red}
We have come a long way $\cdots$ }.

}

\newpage

\color{blue}

{\large \bf

\noindent
Superficially this might be interpreted as showing the correctness of no susy breaking
of any kind \\
( case $b_{\ 0}$ in section 1 ) \cite{Burgess} , but really it constitutes an indirect proof,
that the susy current must develop an anomalous divergence \cite{Casher} .
 
\color{red}
\begin{center}
10) Consequence and conjectured structure of the anomalous susy current divergence
\end{center}

\color{blue}

\begin{description}
\item a) Consequence

Assuming the existence of a genuine anomalous susy current the appearance of a massles
goldstino mode is no more warranted, to the contrary the super Yang-Mills system
develops -- like QCD with one massless quark flavor -- a mass gap.

The 'would be goldstino' mass is then generated similarly as the corresponding
pseudoscalar $\eta^{\ '}$ like mass.

\item b) Conjectured anomaly structure

It was realized after the seminar beeing summarized here, that the susy current 
anomalous divergence may not be represented by a local operator, rather a local
relation may involve a minimal (nonnegative) power p of the d'Alembert operator 
in the form

\end{description}

}

\newpage

\color{blue}

{\large \bf

\begin{description}
\item

\vspace*{-0.5cm}
\begin{equation}
\label{eq:209}
\begin{array}{c}
\left ( \ \mbox{\fbox{\rule{0mm}{1.0mm}}} \ \right )^{\ p} 
\ \partial_{\ \mu} \ j^{\ \mu}_{\ \alpha} \ ( \ x \ ) \ = \ \delta_{\ \alpha}^{\ (p)} \ ( \ x \ ) 
\hspace*{0.2cm} \mbox{and h.c.}
\vspace*{0.2cm} \\
p \ = \ (0) \ , \ 1 \ , \cdots
\end{array}
\end{equation}

The local spinor operator $\delta_{\ \alpha}^{\ (p)}$ and the minimal power p in eq. (\ref{eq:209})
are to be considered as unknowns.

The naive structure of the susy current in spinor basis is ( eq. \ref{eq:97} )

\vspace*{-0.5cm}
\begin{equation}
\label{eq:210}
\begin{array}{c}
j^{\ \dot{\gamma}}_{\ \beta \alpha} \ = \ k \ \Lambda^{\ * \ A \ \dot{\gamma}} 
\ {\cal{F}}_{\ \beta \alpha}^{\ A}
\hspace*{0.2cm} ; \hspace*{0.2cm} \color{red} dim \ j \ = \ \frac{7}{2}
\vspace*{0.4cm} \\ 
\hline \vspace*{-0.4cm} \\
j^{\ \mu}_{\ \alpha} \ = \ \left ( \ \sigma^{\ \mu} \ \right )_{\ \dot{\gamma}}^{\ \beta}
\ j^{\ \dot{\gamma}}_{\ \beta \alpha} 
\vspace*{0.2cm} \\
{\cal{F}}_{\ \alpha}^{\ A \ \gamma} \ = \ \frac{1}{2} \ {\cal{F}}_{\ \mu \nu}
\ \left ( \ \sigma^{\ \mu \nu} \ \right )_{\ \alpha}^{\hspace*{0.3cm} \gamma}
\end{array}
\end{equation}

The normalization constant k in eq. (\ref{eq:210}) is here of no interest.

It follows for the dimension of $\delta_{\ \alpha}^{\ (p)}$ 

\vspace*{-0.5cm}
\begin{equation}
\label{eq:211}
\begin{array}{c}
\color{red} dim \ \delta_{\ \alpha}^{\ (p)} \ = \ ( \ 9 \ + \ 4 \ p \ ) \ / \ 2
\end{array}
\end{equation}

While no candidate operators can be constructed for $p \ = \ 0$, for $p \ = \ 1$ the structure of
$\delta_{\ \alpha}^{\ (1)}$ is

\end{description}

}

\newpage

\color{blue}

{\large \bf

\begin{description}
\item

\vspace*{-0.5cm}
\begin{equation}
\label{eq:212}
\begin{array}{c}
\delta_{\ \alpha}^{\ (1)} \ \sim \ \Lambda^{\ * \ A}_{\ \dot{\gamma}}
\ \Lambda^{\ * \ B \ \dot{\gamma}} 
\ \Lambda^{\ C \ \beta} \ {\cal{F}}_{\ \beta \alpha}^{\ D} \ c_{\ A B C D}
\vspace*{0.3cm} \\
c_{\ A B C D} \ = \ c_{\ B A C D}
\end{array}
\end{equation}

In eq. (\ref{eq:212}) $c_{\ A B C D}$ denotes an invariant coupling coefficient for
a product of 4 adjoint representations, with the symmetry constraint imposed by Fermi statistics.

Simple choice are

\vspace*{-0.5cm}
\begin{equation}
\label{eq:213}
\begin{array}{c}
\left ( \ c_{\ 1} \ \right )_{\ A B C D} \ = \ \delta_{\ A B} \ \delta_{\ C D}
\vspace*{0.3cm} \\
\left ( \ c_{\ 2} \ \right )_{\ A B C D} \ = \ \delta_{\ A C} \ \delta_{\ B D}
\ + \ \delta_{\ B C} \ \delta_{\ A D}
\end{array}
\end{equation}

We just note here the interesting combination of fields corresponding to $c_{\ 1}$
in eq. (\ref{eq:213}) , where the 'would be goldstino' field $g_{\ \alpha}$ is locally
multiplied by the (anti) fermion bilinear $\Lambda^{\ *} \ \Lambda^{\ *}$

\vspace*{-0.3cm}
\begin{equation}
\label{eq:214}
\begin{array}{c}
\left ( \delta_{\ \alpha}^{\ (1)} \ \right )_{\ 1} \ \sim 
\ \left ( \ \Lambda^{\ * \ A}_{\ \dot{\gamma}}
\ \Lambda^{\ * \ A \ \dot{\gamma}} \ \right ) \ g_{\ \alpha}
\vspace*{0.2cm} \\
g_{\ \alpha} \ = \ \Lambda^{\ B \ \beta} \ {\cal{F}}_{\ \beta \alpha}^{\ B}
\end{array}
\end{equation}

\end{description}

}

\newpage

\color{red}

{\large \bf

\begin{center}
11)  Outlook (conclusions)
\end{center}

\color{blue}

\noindent
The following conclusions are taken from the actual seminar. They do not reflect the
entire body of deductions included here.

\color{red}

\begin{description}
\item 1) The clear case consistent with all analogous derivations in QCD , 
for $N \ = \ 1$ super Yang-Mills systems exhibits the central property :

the susy curents $j^{\ \mu}_{\ \alpha} \ , \ j^{\ * \ \mu}_{\ \dot{\alpha}}$
are anomalous \\
( {\small \color{magenta} in the generalized sense of eq. \ref{eq:209}} 
\color{red} ) .

Precursor ideas have been defended by Aharony Casher \cite{Casher} .

{\small \color{magenta} Luzi Bergamin and Elisabeth Kraus were near .}

In addition susy is also spontaneously broken and develops a finite mass gap.
As a consequence the 'would be goldstino'
acquires a mass analogously to $\eta^{\ '}$ in $N_{\ fl} \ = \ 1$ QCD. 

\item 2) Let me express my gratitude and high esteem for my collaborators ( since 1996 ) :

\begin{tabular}{c}
Markus Leibundgut \\
Luzi Bergamin \\
Bernhard Scheuner {\color{black} $\dagger$}\\
{\small \color{magenta} He took his life a few months after this seminar} \\
Samuel Portmann
\end{tabular}
\end{description}

}

\newpage

\color{red}

{\large \bf

\begin{center}
\vspace*{1.0cm}

3) Next ...  $N \ = \ 4$ ?
\vspace*{0.5cm}

Lets be open minded.
\vspace*{3.5cm}

Thank you.
\end{center}

}

\newpage

\color{blue}

{\large \bf

}

\end{PSlide}

%\begin{PSlide}{}
%{\large \bf

%\color{red}
%\begin{center}
%{ \Large \bf 5 Conclusions and outlook}
%\\
%\end{center}
%\hspace*{0.3cm} , \hspace*{0.3cm}

%\color{blue}

%}
%\end{PSlide}

\end{document}